\documentclass[11pt]{article}
\usepackage[margin=1in]{geometry}
\usepackage[utf8]{inputenc}
\usepackage[T1]{fontenc}
\usepackage{amsmath, amssymb, amsfonts, amsthm}
\usepackage{xcolor}
\usepackage{multirow}
\usepackage{authblk}
\usepackage{booktabs}
\usepackage{graphicx}
\usepackage{booktabs}
\usepackage{microtype}
\usepackage{newtxtext}
\usepackage[subscriptcorrection]{newtxmath}
\usepackage{natbib}
\usepackage{hyperref}
\hypersetup{colorlinks=true, allcolors=blue}

\usepackage[plain,noend]{algorithm2e}
\usepackage[section]{placeins}

\graphicspath{{./art/}}

\makeatletter
\renewcommand{\algocf@captiontext}[2]{#1\algocf@typo. \AlCapFnt{}#2} % text of caption
\def\@algocf@capt@plain{top}
\renewcommand{\algocf@makecaption}[2]{%
  \addtolength{\hsize}{\algomargin}%
  \sbox\@tempboxa{\algocf@captiontext{#1}{#2}}%
  \ifdim\wd\@tempboxa >\hsize%     % if caption is longer than a line
    \hskip .5\algomargin%
    \parbox[t]{\hsize}{\algocf@captiontext{#1}{#2}}% then caption is not centred
  \else%
    \global\@minipagefalse%
    \hbox to\hsize{\box\@tempboxa}% else caption is centred
  \fi%
  \addtolength{\hsize}{-\algomargin}%
}
\makeatother

\newcommand{\tbl}[2]{\caption{#1} #2}
\newenvironment{tabnote}{\begin{quote}\small}{\end{quote}}

\begin{document}

\title{Flexible and Scalable Bayesian Modelling Of Spatio-Temporal Hawkes Processes}

\author[1]{Wenqing Liu\thanks{Emails: \texttt{wenqing.liu@usi.ch}, \texttt{deborah.sulem@usi.ch}}}
\author[2]{Xenia Miscouridou\thanks{Email: \texttt{miscouridou.xenia@ucy.ac.cy}}}
\author[1]{Déborah Sulem\protect\footnotemark[1]}

\affil[1]{Faculty of Informatics, Università della Svizzera italiana, Lugano, Switzerland}
\affil[2]{Department of Mathematics and Statistics, University of Cyprus, Nicosia, Cyprus}
\date{}
\maketitle
\begin{abstract}
Existing spatio-temporal Hawkes process models typically rely on either parametric or semiparametric assumptions, limiting the model's ability to capture complex endogenous and exogenous event dynamics. We propose a fully Bayesian nonparametric framework for spatio-temporal Hawkes processes using additive Gaussian processes for the prior distributions on the background rate and the triggering kernel. This additive structure enhances interpretability by decoupling temporal and spatial effects while maintaining high modelling flexibility across the entire spatio-temporal domain. To address scalability, we develop a sparse variational inference scheme based on the Gaussian variational family. Synthetic experiments demonstrate that the proposed method accurately recovers background and triggering structures, achieving superior performance compared to existing alternatives. When applied to real-world datasets, it achieves higher held-out log-likelihoods and reveals interpretable spatio-temporal structures of the self-excitation mechanism. Overall, the framework provides a flexible, scalable, interpretable, and uncertainty-aware approach for modelling complex excitation patterns in spatio-temporal event data.
\end{abstract}


\section{Introduction}

Many real-world phenomena exhibit contagious behaviour, where events cluster in both time and space. In seismology, major earthquakes are typically followed by aftershock sequences concentrated near the epicenter and shortly after the main shock~\citep{ogata1988statistical,ogata1998space}. In criminology, burglary incidents often exhibit ``near-repeat" victimisation, where one crime triggers additional offenses in nearby areas within a short period~\citep{mohler2011self}. Similar spatio-temporal clustering has been observed in terrorism and armed conflict data, where initial attacks may precipitate cycles of follow-up violence~\citep{wang2023modeling}. These examples all share the feature that the occurrence of one event increases the likelihood of subsequent events nearby in space and time. This characteristic, often referred to as self-excitation, means that past events increase the probability of future events. Hawkes processes~\citep{hawkes1971spectra} provide a general framework for modelling such history-dependent dynamics of events occurrences. They represent the conditional intensity as the sum of a background rate, capturing exogenous influences, and a triggering kernel that describes how past events increase the likelihood of a future event. By extending this formulation to incorporate spatial dependence, spatio-temporal Hawkes processes model how both the timing and the locations of past events jointly shape the risk of future events, and have become a standard tool for studying event data that exhibit clustering and contagion effects.

While the Hawkes model is conceptually flexible, practical implementations typically impose strong assumptions on the forms of its main components, the background rate and the triggering kernel. As a result, most spatio-temporal Hawkes processes rely on specified parametric forms for these components. Temporal triggering kernels are typically assumed to be exponential or power-law~\citep{bernabeu2025spatio} whereas spatial kernels are often taken to be Gaussian~\citep{miscouridou2023cox}. Furthermore, most Hawkes processes in literature assume a constant background rate~\citep{bernabeu2025spatio,hawkes1971spectra}, thereby ignoring the temporal or spatial variation coming from exogenous factors. All the aforementioned assumptions facilitate inference but restrict the flexibility of the model, and may prevent it from capturing complex temporal trends, spatial heterogeneity, and interactions between temporal and spatial coordinates that are often present in real-world event data.

To overcome the limitations of parametric spatio-temporal Hawkes processes, the background rate and the triggering kernel should ideally be modelled with flexible, data-driven functional forms rather than restricted to fixed parametric shapes. A Bayesian nonparametric formulation naturally provides this flexibility through priors that allow the functions to vary freely while quantifying uncertainty in the estimated spatio-temporal structure. At the same time, fully nonparametric Bayesian models require posterior inference over a vast latent space that scales with both data size and dimensionality, rendering them computationally prohibitive in large-scale spatio-temporal settings, which makes scalability an essential practical consideration. Furthermore, our framework is computationally scalable to large spatio-temporal point patterns thanks to a variational approximation. Therefore, we propose a novel fully Bayesian nonparametric spatio-temporal Hawkes process that satisfies both flexibility and scalability requirements with the following contributions:

(1) We model both the background rate and the triggering kernel in a fully Bayesian nonparametric way using Gaussian process (GP) priors. Existing spatio-temporal Hawkes processes typically impose parametric forms on at least one of these components; this entirely nonparametric Bayesian framework allows the data to determine temporal and spatial structure while providing uncertainty quantification.

(2) We employ an additive GP representation for both components of the Hawkes process. This additive structure increases interpretability by modelling temporal, spatial, and cross-dimensional interaction effects separately rather than forcing them into a single joint kernel. 
Beyond interpretability, we empirically observe that the additive formulation typically enhances numerical stability and performs at least as well as its non-additive alternative.
%To the best of our knowledge, ours is the first work to incorporate a complete additive GP structure that explicitly accounts for cross-dimensional interactions within the Hawkes process framework.

(3) We perform posterior inference using a variational Gaussian approximation (VGA) built on a sparse GP representation of the  latent components of the Hawkes process. VGA along with the sparse GP enables efficient optimisation and computational scalability.
%preserves the covariance structure of the latent GP, avoiding restrictive independence assumptions, while its Gaussian form enables efficient optimisation. This, along with the sparse GP, ensures computational scalability.

(4) We evaluate our framework using various link functions and GP kernels across both synthetic and real-world datasets, whereas existing Bayesian nonparametric temporal Hawkes methods are typically restricted to specific link functions for conjugacy. This demonstrates the versatility of our framework in supporting diverse model specifications.

The remainder of this paper is organised as follows. Section~\ref{sec:related_work} reviews related work, Section~\ref{sec:method} presents the proposed methodology, Section~\ref{sec:experiments} validates the performance on both synthetic and real-world data, and Section~\ref{sec:conclusion} offers a discussion and conclusion.

\section{Related methods}
\label{sec:related_work}

In purely temporal settings, Hawkes processes have been used to model complex point patterns~\citep{hawkes1971spectra,sulem2022scalable,sulem2024bayesian}. Their extension to the spatio-temporal domain has received increasing attention in recent years due to their ability to capture self-exciting phenomena across both dimensions. This modelling framework has been applied in various fields, including finance~\citep{bacry2015hawkes,chen2022hawkes}, neuroscience~\citep{johnson1996point}, criminology~\citep{mohler2011self}, epidemiology~\citep{chiang2022hawkes}, and seismology~\citep{ogata1998space,kwon2023flexible}.

Several recent surveys provide broader overviews of these developments. In particular, \cite{bernabeu2024spatio} focuses on recent methodological advances, while \cite{reinhart2018review} covers foundational aspects along with applications across multiple domains.

Most classical spatio-temporal Hawkes processes adopt a fully parametric specification, in which both the background rate and the triggering kernel are described by fixed functional forms. The background rate is often modeled as constant~\citep{schoenberg2023estimating,bernabeu2025spatio}, while the triggering kernel is typically taken to be separable, with exponential decay over time and Gaussian spread over space, such as in~\citep{miscouridou2023cox,bernabeu2024spatio,mohler2014marked}. Such parametric formulations offer interpretability and computational tractability, but their form can be inadequate for capturing complex or irregular spatio-temporal patterns often observed in real-world data.

Relaxing these assumptions, a line of work has explored nonparametric modelling of the intensity within a Bayesian framework, most notably via log-Gaussian Cox processes. In these models, a GP prior is placed on a latent log-intensity, which is then mapped to the positive domain through an exponential or sigmoid link function to ensure non-negativity~\citep{moller1998log,samo2015scalable}. In the context of Hawkes processes, \cite{miscouridou2023cox} proposed the Cox–Hawkes model, which uses a log-Gaussian Cox process prior for the background rate while retaining parametric exponential and Gaussian kernels for the triggering kernel. This hybrid design introduces nonparametric flexibility for the baseline but still relies on strong structural assumptions for the excitation mechanism. Posterior inference is performed using MCMC, aided by pre-trained GP generators to accelerate sampling.

\cite{alaimo2025semi} proposed a semi-parametric Hawkes process model, which treats the background component nonparametrically by estimating its spatio-temporal varying intensity directly on the road network, while preserving a structured form for the excitation mechanism. This framework utilises a stochastic-reconstruction algorithm that iteratively updates the background risk through empirical smoothing of events, rather than assuming a fixed parametric baseline. By combining this flexible background with a parametric triggering component, the model effectively balances local data adaptability with the interpretability of self-excitation patterns.

A different approach is taken by~\cite{zhou2020efficient}, who proposed a fully nonparametric temporal Hawkes model by placing GP priors on both the background rate and the triggering kernel. A sigmoid link function is used not only to enforce positivity but also, when combined with suitable latent variables, to achieve conjugacy that enables a range of inference schemes, including EM algorithms, Gibbs sampling, and variational inference (VI). While this framework avoids parametric assumptions on the triggering kernel and captures rich temporal dynamics, it is limited to purely temporal data and is tightly coupled to the sigmoid link function. Extending it to a general spatio-temporal setting would require latent-variable updates over a high-dimensional space, causing the computational cost to grow prohibitively. Overall, the method is difficult to extend directly to the spatio-temporal domain or to accommodate alternative link functions.

While the method of~\cite{zhou2020efficient} represents a Bayesian nonparametric direction, a substantial literature on frequentist nonparametric estimation for Hawkes processes has also emerged. Maximum-likelihood-based methods have been developed to estimate triggering kernels and, in some cases, background rates in a flexible manner~\citep{marsan2008extending,lewis2011nonparametric,zhou2013learning}. These approaches are attractive due to their efficiency and asymptotic properties, but they yield point estimates rather than full distributions, and therefore do not provide uncertainty quantification for the inferred intensity components.

From a Bayesian perspective, posterior inference is often carried out using Markov Chain Monte Carlo (MCMC) algorithms~\citep{adams2009tractable,donnet2020nonparametric,miscouridou2023cox}, which offer principled uncertainty quantification but can be computationally demanding, particularly for large spatio-temporal datasets. VI has emerged as a scalable alternative, providing approximate posterior distributions at a fraction of the computational cost~\citep{malem2021nonlinear,sulem2022scalable,zhou2020efficient}. Theoretical guarantees for such Bayesian nonparametric formulations in the spatio-temporal domain have also been established, including results on posterior contraction rates~\citep{miscouridou2026posterior}. However, many existing VI approaches rely on independence assumptions that factorise the posterior over latent variables, potentially limiting the ability to capture complex posterior dependencies. This motivates the use of more expressive approximations, such as the Gaussian variational family, which we adopt in this work.

Overall, these factors indicate the need to further relax parametric assumptions while maintaining scalability in Hawkes process modelling. However, existing methods either (i) maintain parametric structure in at least one component, (ii) remain confined to purely temporal settings, or (iii) lack a fully Bayesian treatment that is both nonparametric and scalable in space and time. Our work fills these gaps by proposing a novel fully Bayesian nonparametric spatio-temporal Hawkes model in which both the background rate and the triggering kernel are modelled with GP priors, and a scalable VI algorithm is used to approximate the posterior.
%This enables joint learning of spatial and temporal excitation patterns without restrictive parametric forms, while retaining uncertainty quantification and computational feasibility for high-dimensional settings.

\section{Methodology}
\label{sec:method}
\subsection{Problem Setup}

We assume that we observe a spatio-temporal point process $N$ on a bounded domain
$\mathcal{W}=\mathcal{T}\times\mathcal{S}$, where $\mathcal{T}=[0,T]$ is the temporal window, $T > 0$ is the time horizon, and $\mathcal{S}\subset\mathbb{R}^d$ is the spatial domain. Typical choices include $d=1$ for a single spatial coordinate, $d=2$ for planar 
settings, and $d=3$ for applications such as latitude-longitude-depth. The data consists of one or multiple realisations of this point process $N$. For clarity, we first consider that we only observe one sequence of points
$\{(t_i,s_i)\}_{i=1}^{n}$ with $t_i\in\mathcal{T}$ and $s_i\in\mathcal{S}$. Let
$\mathcal{H}_t=\{(t_i,s_i):\,t_i<t\}$ denote the history up to time $t$.

The law of this process is specified by its conditional intensity.
For each $(t,s)\in\mathcal{W}$, the conditional intensity, given the history 
$\mathcal{H}_t$, of a Hawkes process is
\begin{equation}
\lambda(t,s\mid\mathcal{H}_t)
 \;=\;
 \mu(t,s)
 \;+\;
 \sum_{t_i<t}
 \phi\!\bigl(t-t_i,\, s-s_i\bigr),
\label{eq:sthp_intensity}
\end{equation}
where $\mu(t, s) \ge 0$ is the \textit{background rate} representing exogenous arrivals, and $\phi(\Delta t, \Delta s) \ge 0$ is the \textit{triggering kernel} characterizing the endogenous self-exciting effect, where $\Delta t = t-t_i$ and $\Delta s = s-s_i$ denote the relative time and spatial lags from a past event, respectively. The cumulative impact of these two components can be characterised by their respective $L_1$-norms. Specifically, the total expected number of background events within the observation domain $\mathcal{W}$ is given by:
\begin{equation}
    \|\mu\|_1 = \int_{\mathcal{S}} \int_{\mathcal{T}} \mu(t, s) \, \mathrm{d}t \, \mathrm{d}s,
\end{equation}
Meanwhile, the overall strength of the self-excitation is summarised by the branching ratio, defined as the $L_1$-norm of $\phi$ over its support:
\begin{equation}
    \|\phi\|_1 = \int_{\mathcal{S}} \int_{0}^{\infty} \phi(\Delta t, \Delta s) \, \mathrm{d}\Delta t \, \mathrm{d}\Delta s.
\end{equation}
The branching ratio $\|\phi\|_1$ represents the expected number of offspring events generated by a single event. To ensure the process is stable, i.e., does not explode, we assume the stationarity condition $\|\phi\|_1 < 1$~\citep{bernabeu2024spatio,siviero2024flexible,reinhart2018review,miscouridou2023cox}. Under this condition, the total expected number of events in the bounded domain $\mathcal{W}$ can be approximated as $\mathbb{E}[N(\mathcal{W})] \approx \|\mu\|_1 / (1 - \|\phi\|_1)$~\citep{miscouridou2023cox} with $N(\mathcal{W}$ the number of events in $\mathcal{W}$. A detailed derivation of this approximation is provided in Appendix~\ref{app:expected_events}.

\subsection{Gaussian Process Prior}
\label{sec:gp_prior}

To flexibly model both the background rate $\mu$ and the triggering kernel $\phi$, we introduce
latent functions $f_\mu$ and $f_\phi$ and apply link functions to ensure
non-negativity. Specifically, the conditional intensity of a Hawkes process in~\eqref{eq:sthp_intensity} is reformulated as
\begin{equation}
\lambda(t,s\mid\mathcal{H}_t)
=
h_\mu\!\bigl(f_\mu(t,s)\bigr)
+
\sum_{t_i<t}
h_\phi\!\bigl(f_\phi(t-t_i,s-s_i)\bigr)
\label{eq:latent_gp_prior}
\end{equation}
where $h_\mu,h_\phi:\mathbb{R}\to\mathbb{R}_{\ge 0}$ are fixed, non-negative, bijective and non-decreasing link functions, such as the softplus~\citep{mei2017neural,sulem2022scalable}, sigmoid~\citep{zhou2020efficient,malem2021nonlinear}, or exponential functions~\citep{moller1998log,diggle2013spatial,miscouridou2023cox}.

The latent functions $f_{\mu}$ and $f_{\phi}$ are given independent Gaussian 
process priors,
\[
f_{\mu}(t,s)\sim\mathcal{GP}(\eta_{\mu},K_{\mu}),\qquad
f_{\phi}(\Delta t,\Delta s)\sim\mathcal{GP}(\eta_{\phi},K_{\phi})
\]
where $f_{\mu}$ is defined on absolute coordinates $(t,s)$, while 
$f_{\phi}$ is defined on temporal and spatial lags 
$(\Delta t,\Delta s)=(t-t_i,\,s-s_i)$ with $\Delta t\ge 0$. Above, $\eta_{\mu}$ and $\eta_{\phi}$ denote the GP mean functions, which are assumed to be zero without loss of generality~\citep{zhou2019generalized}.
The covariance kernels $K_{\mu}$ and $K_{\phi}$ encode prior beliefs about the smoothness and correlation structures of the latent functions. We primarily employ stationary families such as the squared-exponential (RBF) or Mat\'{e}rn~\citep{williams2006gaussian}, parameterised by a signal variance $\sigma^2$ (controlling amplitude) and lengthscales $\ell$ (controlling correlation ranges). For the spatio-temporal domain, we consider both traditional separable kernels which factorise into independent spatial and temporal components
\begin{equation}
    K_{\text{separable}}\bigl((t,s),(t',s')\bigr) = \sigma^2k_{t}(t,t' \mid 1, \ell_t)\, k_{s}(s,s' \mid 1, \ell_s),
\end{equation}
and a more interpretable additive framework~\citep{plate1999accuracy,duvenaud2011additive}, where marginal and interaction effects are explicitly modelled to capture cross-dimensional dependencies
\begin{equation}
    K_{\text{additive}}\bigl((t,s),(t',s')\bigr) = k_t(t, t' \mid \sigma_t^2, \ell_t) + k_s(s, s' \mid \sigma_s^2, \ell_s) + k_{t,s}\bigl((t,s),(t',s') \mid \sigma_{t,s}^2, \ell_{t,s}\bigr).
\end{equation}
Mathematical definitions for all kernels are provided in Appendix~\ref{sec:appendix_kernels}.

\subsection{Model Likelihood}
We now specialise to the two-dimensional spatial case ($d=2$). 
For simplicity, let $\mathcal{W} = [0,T] \times [0,X] \times [0,Y]$ with $X, Y > 0$ and 
$\mathcal{D}=\{(t_i,x_i,y_i)\}_{i=1}^n$ denote the observed sequence of events on $\mathcal{W}$. Following \cite{daley2008introduction}, the likelihood of a spatio-temporal Hawkes process is:
\begin{equation}
p(\mathcal{D}\mid \mu,\phi) = \left( \prod_{i=1}^{n} \lambda(t_i,x_i,y_i \mid \mathcal{H}_{t_i}) \right) \exp\!\left( - \iiint_{\mathcal{W}} \lambda(t,x,y\mid \mathcal{H}_t)\, \rm{d}t\, \rm{d}x\, \rm{d}y \right).
\label{eq:likelihood_1seq}
\end{equation}
For $M$ independent sequences $\{\mathcal{D}^{(m)} = \{(t_i^{(m)}, x_i^{(m)}, y_i^{(m)})\}_{i=1}^{n_m}\}_{m=1}^M$, the joint likelihood is the product of the likelihoods of each sequence:
\[
p(\{\mathcal{D}^{(m)}\}_{m=1}^M \mid \mu, \phi) = \prod_{m=1}^M \left[ \left( \prod_{i=1}^{n_m} \lambda\big(t_i^{(m)},x_i^{(m)},y_i^{(m)} \mid \mathcal{H}^{(m)}_{t_i}\big) \right) \exp\!\left( - \iiint_{\mathcal{W}} \lambda(t,x,y\mid \mathcal{H}^{(m)}_t)\, \rm{d}t\, \rm{d}x\, \rm{d}y \right) \right].
\]

Evaluating the likelihood in~\eqref{eq:likelihood_1seq} is computationally demanding for nonparametric models. Unlike parametric triggering kernels, which often admit closed-form integrals, a nonparametric triggering kernel modelled by a GP requires explicit computation over history. Without structural restrictions, each event depends on all previous events, resulting in quadratic complexity in the number of observations.

To reduce this burden, we follow~\citep{sulem2024bayesian,zhou2020efficient} and assume the triggering kernel $\phi$ has finite support $\Omega_\phi = [0, T_\phi] \times [-X_\phi, X_\phi] \times [-Y_\phi, Y_\phi]$, with $T_\phi, X_\phi, Y_\phi > 0$, vanishing elsewhere. Under this assumption, only recent events within the support contribute to the intensity and the integral, reducing both computations to linear complexity. This assumption is empirically justified, as influence is typically localised in domains like neuroscience~\citep{staerman2023fadin,eichler2017graphical}, finance~\citep{bacry2015hawkes,carreira2021exponential}, and criminology~\citep{zhou2020efficient}.

Under this assumption, the likelihood~\eqref{eq:likelihood_1seq} simplifies to the form derived in Appendix~\ref{app:LH_derivation}:
\begin{equation}
\begin{aligned}
p(\mathcal{D}\mid \mu,\phi)
&=
\exp\!\left(
- \int_0^T \int_0^X \int_0^Y \mu(t,x,y)\,\rm{d}t\,\rm{d}x\,\rm{d}y
\right) \\
&\quad \times \exp\!\left(
- \sum_{i=1}^{n}
\int_{0}^{\min(T_\phi,\,T-t_i)}
\int_{\max(-X_\phi,\,-x_i)}^{\min(X_\phi,\,X-x_i)}
\int_{\max(-Y_\phi,\,-y_i)}^{\min(Y_\phi,\,Y-y_i)}
\phi(\Delta t, \Delta x, \Delta y)\,\rm{d}\Delta t\,\rm{d}\Delta x\,\rm{d}\Delta y
\right) \\
&\quad \times \prod_{i=1}^{n}
\left(
\mu(t_i,x_i,y_i)
+ \sum_{j:\, t_j<t_i}
\phi(\Delta t_{ij}, \Delta x_{ij}, \Delta y_{ij})
\right).
\end{aligned}
\label{eq:full_likelihood}
\end{equation}
where $\Delta t_{ij} = t_i - t_j$, $\Delta x_{ij} = x_i - x_j$, and $\Delta y_{ij} = y_i - y_j$.

\subsection{Inference}
\label{sec:VI}
Having established the expression of the likelihood, our goal is to perform Bayesian inference for the background rate $\mu$ and the triggering kernel $\phi$ from the observed events $\mathcal{D}$. Following the nonparametric framework established in Section~\ref{sec:gp_prior}, we do not place priors directly on $\mu$ and $\phi$. Instead, these components are modelled as transformations of latent Gaussian processes, i.e., $\mu=h_\mu(f_\mu)$ and $\phi=h_\phi(f_\phi)$. Consequently, the inference problem translates to estimating these latent functions $f=\{f_\mu, f_\phi\}$ and hyperparameters $\theta=\{\theta_\mu, \theta_\phi\}$ of the GP kernels. The specific joint posterior distribution over these latent functions is:
\begin{equation}
p(f_\mu, f_\phi, \theta_\mu,\theta_\phi \mid \mathcal{D})
\;\propto\;
p(\mathcal{D} \mid f_\mu, f_\phi)\,
p(f_\mu \mid \theta_\mu)\,
p(f_\phi \mid \theta_\phi)\,
p(\theta_\mu)\,
p(\theta_\phi).
\label{eq:posterior_specific}
\end{equation}
where $p(f \mid \theta)$ denotes the GP prior and $p(\theta)$ the hyperprior (for background or triggering kernel components).

For practical computation, the latent functions $f_\mu$ and $f_\phi$ are evaluated on a finite discretisation grid. 
Specifically, for $f_\mu$, the grid has $n_{\mu,t}$ points in time and $n_{\mu,x}, n_{\mu,y}$ points in space, giving a total number of points of $N_\mu = n_{\mu,t} \times n_{\mu,x} \times n_{\mu,y}$. 
The grid for $f_\phi$ has $n_{\phi,t}, n_{\phi,x}, n_{\phi,y}$ points in each dimension, resulting in $N_\phi = n_{\phi,t} \times n_{\phi,x} \times n_{\phi,y}$ grid points.

Direct inference on this specific posterior~\eqref{eq:posterior_specific} is intractable due to the non-conjugate relationship between the Hawkes likelihood and the Gaussian process priors. While Markov Chain Monte Carlo (MCMC) methods can theoretically yield exact samples from the posterior, they are computationally prohibitive due to the dimensionality of the discretised latent functions. Consequently, we adopt variational inference~\citep{jordan1999introduction,blei2017variational} to achieve scalability and efficiency.

The goal of variational inference is to approximate the intractable posterior 
$p(f_\mu, f_\phi, \theta_\mu, \theta_\phi \mid \mathcal{D})$ in~\eqref{eq:posterior_specific}. 
A variational family $\mathcal{Q}$ is introduced to provide a tractable set of candidate 
densities over the latent variables, defined on the same domain as the posterior distribution. 
The variational distribution 
$q(f_\mu, f_\phi, \theta_\mu, \theta_\phi) \in \mathcal{Q}$ is chosen to minimise the 
Kullback--Leibler (KL) divergence to the true posterior. 
Finding this optimal $q$ is equivalent to maximising the Evidence Lower Bound (ELBO):
\begin{equation}
\mathcal{L}(q) =
\mathbb{E}_{q} \Bigl[ \log p(f_\mu, f_\phi, \theta_\mu, \theta_\phi,\mathcal{D}) \Bigr]
- \mathbb{E}_{q} \Bigl[ \log q(f_\mu, f_\phi, \theta_\mu, \theta_\phi) \Bigr].
\label{eq:elbo_general}
\end{equation}
A detailed derivation demonstrating this equivalence is provided in Appendix~\ref{app:VI_derivation}.

The choice of the variational family $\mathcal{Q}$ is critical for the quality of posterior 
approximation. In the present setting, the latent functions $f_\mu$ and $f_\phi$ are modelled as GPs, and it is 
therefore essential that the variational family preserves the correlation structure within each 
latent process while remaining computationally tractable. We adopt a Variational Gaussian Approximation (VGA)~\citep{opper2009variational,zhou2019generalized} with the block-wise factorisation:
\begin{equation}
q(f_\mu, f_\phi, \theta_\mu, \theta_\phi)
=
q(f_\mu,\theta_\mu)\,
q(f_\phi,\theta_\phi)
\label{eq:VGA_block}
\end{equation}
which assumes independence between the background and triggering components. Within each block, an additional factorisation is used:
\begin{equation}
q(f_\mu,\theta_\mu)=q(f_\mu)\,q(\theta_\mu),
\qquad
q(f_\phi,\theta_\phi)=q(f_\phi)\,q(\theta_\phi).
\label{eq:VGA_inner}
\end{equation}
This independence between the latent function and its hyperparameters is introduced for efficiency. 
The variational distributions over the latent functions are chosen as multivariate Gaussians:
\begin{equation}
q(f_\mu)=\mathcal{N}(m_\mu,\Sigma_\mu), 
\qquad 
q(f_\phi)=\mathcal{N}(m_\phi,\Sigma_\phi).
\end{equation}
The full covariance matrices $\Sigma_\mu$ and $\Sigma_\phi$ preserve the dependence between 
function values, allowing the variational distributions to reflect the smoothness structure induced 
by the GP priors. It is important to retain dependencies within the hyperparameters of each latent process. Since the hyperparameters $\theta$ of the GP kernels are positive, we define the variational approximation in the log-domain. Specifically, we assign multivariate Gaussian distributions to the log-transformed hyperparameters $\alpha_\mu = \log \theta_\mu$ and $\alpha_\phi = \log \theta_\phi$:
\begin{equation}
q(\alpha_\mu)=\mathcal{N}(m_{\theta_\mu},\Sigma_{\theta_\mu}),
\qquad
q(\alpha_\phi) =\mathcal{N}(m_{\theta_\phi},\Sigma_{\theta_\phi})
\label{eq:VGA_theta}
\end{equation}
implying that $q(\theta_\mu)$ and $q(\theta_\phi)$ follow multivariate Log-normal distributions:
\begin{equation}
q(\theta_\mu)=\mathcal{LN}(m_{\theta_\mu},\Sigma_{\theta_\mu}),
\qquad
q(\theta_\phi) =\mathcal{LN}(m_{\theta_\phi},\Sigma_{\theta_\phi})
\label{eq:VGA_theta_ln}
\end{equation}

The use of full covariance matrices $\Sigma_{\theta_\mu}$ and $\Sigma_{\theta_\phi}$ allows the variational distribution to capture potential correlations between hyperparameters~\citep{flaxman2015fast}, enabling the model to represent dependencies that may arise in the posterior.

\subsection{Sparse GP Approximation}
\label{sec:sparse_gp}

Under the VGA structure defined in Equations~\eqref{eq:VGA_block}--\eqref{eq:VGA_inner}, calculating the KL divergence between the
variational distributions $q(f_\mu), q(f_\phi)$ and the priors $p(f_\mu), p(f_\phi)$
requires the inversion of the kernel matrices $K_{\mu}$ and $K_{\phi}$. In our spatio-temporal setting, the sizes of these matrices are respectively \(N_\mu \times N_\mu\) and \(N_\phi \times N_\phi\) and are typically so large that these matrix operations, which scale as $\mathcal{O}(N_\mu^3 + N_\phi^3)$, become computationally
infeasible. To ensure scalability, a sparse GP approximation is adopted 
following~\cite{titsias2009variational}, replacing each full GP with a lower-rank 
inducing-point representation and substantially reducing the cost of ELBO optimisation. For the background component, let 
\(Z_\mu=\{z_m\}_{m=1}^{M_\mu}\subset[0,T]\times[0,X]\times[0,Y]\) be a set of 
inducing locations, and define the inducing variables 
\(\mathbf u_\mu := f_\mu(Z_\mu) \in \mathbb R^{M_\mu}\).
Under the GP prior
\[
p(\mathbf u_\mu) = \mathcal{N}(0,\,
K_{Z_\mu Z_\mu}),
\]
where \(K_{Z_\mu Z_\mu}\in \mathbb R^{M_\mu \times M_\mu}\) is the kernel matrix evaluated at the inducing locations.
The triggering component is handled analogously using 
inducing locations $Z_\phi = \{\mathbf{z}_n\}_{n=1}^{M_\phi} \subset [0,T_\phi] \times [-X_\phi, X_\phi] \times [-Y_\phi, Y_\phi]$ and the corresponding inducing variables \(\mathbf u_\phi=f_\phi(Z_\phi)\).

To select the inducing variables, we adopt the approach of~\citep{zhou2020efficient}, where the inducing locations $Z_\mu$ and $Z_\phi$ are placed on equidistant grids over their respective domains. Given the inducing variables, the latent functions at the grid locations 
%locations $Z$ 
are recovered via Gaussian 
conditioning~\citep{titsias2009variational,zhou2019generalized}. For the background latent function \(f_\mu\),
\[
p(f_\mu \mid \mathbf u_\mu)
= \mathcal{N}(m^{f \mid u}_\mu,\Sigma^{f \mid u}_\mu),
\]
with
\begin{align}
m^{f \mid u}_\mu 
&= K_{ZZ_\mu}\,
   K_{ Z_\mu Z_\mu}^{-1}\,
   \mathbf u_\mu, \\
\Sigma^{f \mid u}_\mu
&= K_{ZZ} -
    K_{ZZ_\mu}\,
   K_{ Z_\mu  Z_\mu}^{-1}\,
k_{ Z_\mu Z}.
\end{align}

%To obtain variational approximations, we place multivariate Gaussian distributions over the  inducing variables:
Under our variational Gaussian approximation, the variational posterior on the inducing variables is
\[
q(\mathbf{u}_\mu) = \mathcal{N}({\mu}_\mu, {S}_\mu), 
\qquad 
q(\mathbf{u}_\phi) = \mathcal{N}({\mu}_\phi, {S}_\phi).
\]
The variational posteriors over the latent functions are obtained by marginalizing out the inducing variables:
\[
q(f_\mu) = 
\int p(f_\mu \mid \mathbf u_\mu)\,
q(\mathbf u_\mu)\, d\mathbf u_\mu,
\qquad
q(f_\phi) = 
\int p(f_\phi \mid \mathbf u_\phi)\,
q(\mathbf u_\phi)\, d\mathbf u_\phi.
\]
Since both terms are Gaussian and conjugate, the marginals remain Gaussian with closed-form expressions.  
For \(f_\mu\),
\[
q(f_\mu)=\mathcal{N}\bigl(\tilde m_\mu,\tilde \Sigma_\mu\bigr),
\]
where $\tilde m_\mu$ and $\tilde \Sigma_\mu$ are given by \citep{lloyd2015variational}, with the derivation provided in Appendix~\ref{app:sparse_gp_derivation},
\begin{align}
\tilde m_\mu 
&= K_{ZZ_\mu}\, K_{Z_\mu Z_\mu}^{-1}\,\mu_\mu, \\
\tilde \Sigma_\mu
&= K_{ZZ} 
- K_{ZZ_\mu}\, K_{Z_\mu Z_\mu}^{-1}\, K_{ Z_\mu Z} 
+ K_{Z Z_\mu}\, K_{Z_\mu Z_\mu}^{-1}\, 
S_\mu\, K_{Z_\mu Z_\mu}^{-1}\, K_{Z_\mu Z}.
\end{align}

We can interpret the form of $\tilde{\Sigma}_\mu$ as follows: the first term 
$K_{ZZ}$ represents the original GP prior uncertainty, while the second term,
$K_{ZZ_\mu}\, K_{ Z_\mu  Z_\mu}^{-1}\, K_{ Z_\mu Z}$,
corresponds to the part of this uncertainty that can be explained by the inducing variables, 
so their difference captures the remaining uncertainty that the inducing variables cannot represent~\citep{quinonero2005unifying}. The final term,
$K_{ZZ_\mu}\, K_{Z_\mu  Z_\mu}^{-1}\, 
    S_\mu\, K_{ Z_\mu  Z_\mu}^{-1}\, K_{ Z_\mu Z}$,
adds back the uncertainty arising from the fact that the inducing variables themselves are not known exactly but have an approximate posterior with covariance $ S_\mu$. The expressions for \(q(f_\phi)\) are analogous.

Although the sparse construction provides closed-form variational posteriors
$q(f_\mu)$ and $q(f_\phi)$, these distributions are still defined over all latent
function evaluations, and their covariance matrices remain dense with sizes 
$N_\mu \times N_\mu$ and $N_\phi \times N_\phi$.  
Consequently, computing ELBO terms that require sampling from 
$q(f_\mu)$ or $q(f_\phi)$ involves operations on these dense covariance matrices, 
necessitating a Cholesky decomposition with 
$\mathcal{O}(N_\mu^{3})$ or $\mathcal{O}(N_\phi^{3})$ complexity. This computational burden motivates the additional approximations introduced below.

We adopt the
Deterministic Training Conditional (DTC) approximation
\citep{quinonero2005unifying, hensman2018variational}, in which the latent
function $f_\mu^{\ast}$ is based on the projection:
\[
f_\mu^{\ast}
= K_{ZZ_\mu} K_{ Z_\mu  Z_\mu}^{-1} \mathbf u_\mu,
\qquad 
\mathbf u_\mu \sim q(\mathbf u_\mu)
= \mathcal N({\mu}_\mu, S_\mu).
\]

The posterior mean follows directly from the linear mapping:
\[
m_\mu^{\ast}
= \mathbb{E}[f_\mu^{\ast}]
= K_{ZZ_\mu} K^{-1}_{ Z_\mu  Z_\mu}{\mu}_\mu.
\]

The covariance is
\[
\Sigma_\mu^{\ast}
=
\mathrm{Cov}\!\left(f_\mu^{\ast}\right)
=
K_{ZZ_\mu}
  K^{-1}_{ Z_\mu  Z_\mu} S_\mu
  K^{-1}_{ Z_\mu  Z_\mu}
  K_{ Z_\mu Z}.
\]

In DTC, the prior residual term
\(
K_{ZZ} - k_{Z Z_\mu} K^{-1}_{ Z_\mu  Z_\mu} K_{Z_\mu Z}
\)
present in the full sparse posterior is omitted. As this residual term
vanishes when the inducing set becomes the original grid, the DTC approximation improves
with larger numbers of inducing variables, consistent with the classical Nyström
approximation~\citep{williams2000using}. The same approximation is applied to the posterior on the triggering kernel's latent function. 
%construction is applied to the GP governing the triggering kernel.

In this sparse approximation setting, the posterior distribution in
\eqref{eq:posterior_specific} can be written as
\begin{equation}
\begin{aligned}
p(f_\mu^{\ast}, f_\phi^{\ast}, \mathbf u_\mu, \mathbf u_\phi, \theta_\mu, \theta_\phi \mid \mathcal{D})
&\propto p\!\left(\mathcal{D} \mid \mu = h_\mu(f_\mu^{\ast}), \, \phi = h_\phi(f_\phi^{\ast})\right)\,
   p(f_\mu^{\ast} \mid \mathbf u_\mu, \theta_\mu)\,
   p(f_\phi^{\ast} \mid \mathbf u_\phi, \theta_\phi) \\
&\quad \times p(\mathbf u_\mu)\,
   p(\mathbf u_\phi)\,
   p(\theta_\mu)\,
   p(\theta_\phi).
\end{aligned}
\end{equation}
Since $f_\mu^{\ast}$ and $f_\phi^{\ast}$ are conditionally recovered from the inducing variables, 
the variational distribution only needs to be defined over the inducing values 
and the hyperparameters,
\(q(\mathbf u_\mu, \mathbf u_\phi, \theta_\mu, \theta_\phi)\).
The corresponding ELBO in~\eqref{eq:elbo_general} is
\begin{equation}
\mathcal{L}(q)
= \mathbb{E}_{q}\!\left[\log p(f_\mu^{\ast}, f_\phi^{\ast}, \mathbf{u}_\mu, \mathbf{u}_\phi, 
\theta_\mu, \theta_\phi, \mathcal{D})\right]
 - \mathbb{E}_{q}\!\left[\log q(\mathbf{u}_\mu, \mathbf{u}_\phi, \theta_\mu, \theta_\phi)\right].
\label{eq:sparse_elbo_fully_factorized}
\end{equation}

In summary, the sparse GP approximation introduces inducing variables to obtain a
low-rank representation of each GP prior, while the DTC approximation further simplifies the
variational posterior by projecting all latent function evaluations onto these inducing
variables. This combination yields a scalable variational inference scheme for spatio-temporal
Hawkes processes.

\section[Experiments]{Experiments\protect\footnote{The model implementation can be found at \url{https://github.com/Wenqing-Liu-1010/Flexible-and-Scalable-Bayesian-Modeling-Of-Spatio-Temporal-Hawkes-Processes}.}}
\label{sec:experiments}

Detailed hyperparameter settings, grid resolutions, and computational configurations used in our experiments are provided in Appendix~\ref{app:experimental_settings}.

\subsection{Baselines and Metrics}
\label{sec:baselines_metrics}

We compare our model against several spatio-temporal Hawkes process baselines, which are estimated using variational inference for consistency with our approach. 

\paragraph{Parametric Hawkes Process~\citep{bernabeu2025spatio}:} The background rate $\mu$ is assumed constant and the triggering kernel takes the parametric form:
\begin{equation}
    \frac{\alpha\beta}{2\pi \sigma_x \sigma_y} \exp\!\bigl(-\beta (t - t_i)\bigr) \exp\!\left( -\frac{(x - x_i)^2}{2\sigma_x^2} -\frac{(y - y_i)^2}{2\sigma_y^2} \right)
    \label{eq:param_triggerirng}
\end{equation}
with hyperparameters $\alpha, \beta, \sigma_x, \sigma_y > 0$. 

\paragraph{Cox--Hawkes Process~\citep{miscouridou2023cox}:} The background rate $\mu(t,x,y)$ is modelled using a log-Gaussian Cox process, while the triggering kernel retains the same parametric form as~\eqref{eq:param_triggerirng}. Although their GP prior consists of an additive RBF kernel with separate temporal and spatial components, here for consistency we adopt the interaction-aware additive structure described in Section~\ref{sec:gp_prior}. 

A key limitation of both baselines is their reliance on triggering kernels with exponential decay and Gaussian forms, which are inherently unsuitable for capturing non-monotonic triggering patterns. To evaluate model performance, we employ the following metrics, with detailed mathematical formulations provided in Appendix \ref{sec:metrics}:

\paragraph{Posterior Mean Error ($PM_{\mathrm{mse}}$):} The mean squared error between the posterior mean estimates $\hat{\mu}$, $\hat{\phi}$ and the ground truth.
\paragraph{Posterior Expected Error ($PE_{\mathrm{mse}}$):} The expected mean squared error calculated under the variational posterior distribution of $\mu$ and $\phi$.
\paragraph{Expected Log-Likelihood ($ELL$):} The expected log-likelihood of the observed data under the variational posterior. This metric serves as a measure of model fit for real-world datasets.

\subsection{Synthetic Data Experiments}

For synthetic data experiments, events are generated using a cluster-based simulation~\citep{miscouridou2023cox,bernabeu2025spatio} 
in a spatio-temporal domain $(T,X,Y)=(10,10,10)$. For the  triggering kernel, we set  $(T_\phi,X_\phi,Y_\phi)=(0.5,0.3,0.3)$. We consider the following three synthetic scenarios, with increasing complexity:
\begin{align*}
    \textbf{Scenario 1.}\quad 
    &\mu(t,x,y)=4.5, \\
    &\phi(\Delta t,\Delta x,\Delta y)
    = 5\,\exp\!\bigl(-(\Delta x^{2}+\Delta y^{2})/0.02\bigr)\exp(-\Delta t).
    \\[0.4em]
    \textbf{Scenario 2.}\quad 
    &\mu(t,x,y)
    = (1.5+\sin(2\pi t/T)) + (1.5+\sin(2\pi x/X)) + (1.5+\sin(2\pi y/Y)), \\
    &\phi(\Delta t,\Delta x,\Delta y)
    = 5\,\exp\!\bigl(-(\Delta x^{2}+\Delta y^{2})/0.02\bigr)\exp(-\Delta t).
    \\[0.4em]
    \textbf{Scenario 3.}\quad
    &\mu(t,x,y)
    = (1.5+\sin(2\pi t/T)) 
    + (1.5+\sin(2\pi x/X)) 
    + (1.5+\sin(2\pi y/Y))  \\
    &\qquad
    + \sin(2\pi t/T)\sin(2\pi x/X)\sin(2\pi y/Y), \\
    &\phi(\Delta t,\Delta x,\Delta y)
    = (1+10(\Delta t/T_\phi)(1-\Delta t/T_\phi))
    \exp\!(-(\Delta x^{2}/X_\phi^{2}+\Delta y^{2}/Y_\phi^{2})).
    \end{align*}

The first scenario corresponds to a parametric Hawkes process with constant background rate. The second 
introduces an additive nonstationary background rate that varies independently in time 
and space. The third has 
a background rate which is nonstationary, with additive and interaction terms and a triggering kernel which is non-monotonic. These scenarios 
allow us to evaluate our model under classical settings, time-varying background 
rate, and the most flexible scenarios involving both nonstationary $\mu(t,x,y)$ and 
non-monotonic $\phi(\Delta t,\Delta x,\Delta y)$.

To assess the impact of prior specifications on the inference, we evaluate our model across various configurations, including separable and additive RBF and Mat\'{e}rn kernels (with smoothness level $\nu=2.5$), as well as different link functions (softplus, exponential, and sigmoid\footnote{The sigmoid link function is implemented as $\sigma(x) = \alpha / (1 + \exp(-x))$, where $\alpha$ is set to a relatively large constant to accommodate the scale of the target latent intensity.}). We present the results for these combinations for Scenarios 1 to 3 in Tables~\ref{tab:scenario1_comparison}, \ref{tab:scenario2_comparison}, and \ref{tab:scenario3_comparison}, which report $PM_{\mathrm{mse}}(\mu)$, $PM_{\mathrm{mse}}(\phi)$, $PE_{\mathrm{mse}}(\mu)$, and $PE_{\mathrm{mse}}(\phi)$ computed against the ground-truth background rate and triggering kernel. Detailed results regarding different smoothness levels for the Mat\'{e}rn kernels ($\nu=0.5, 1.5$) across all scenarios are provided in Tables~\ref{tab:scenario1_full}, \ref{tab:scenario2_full}, and \ref{tab:scenario3_full} of Appendix~\ref{app:full_results}.

\begin{figure}
\centering
\includegraphics[width=1.0\textwidth]{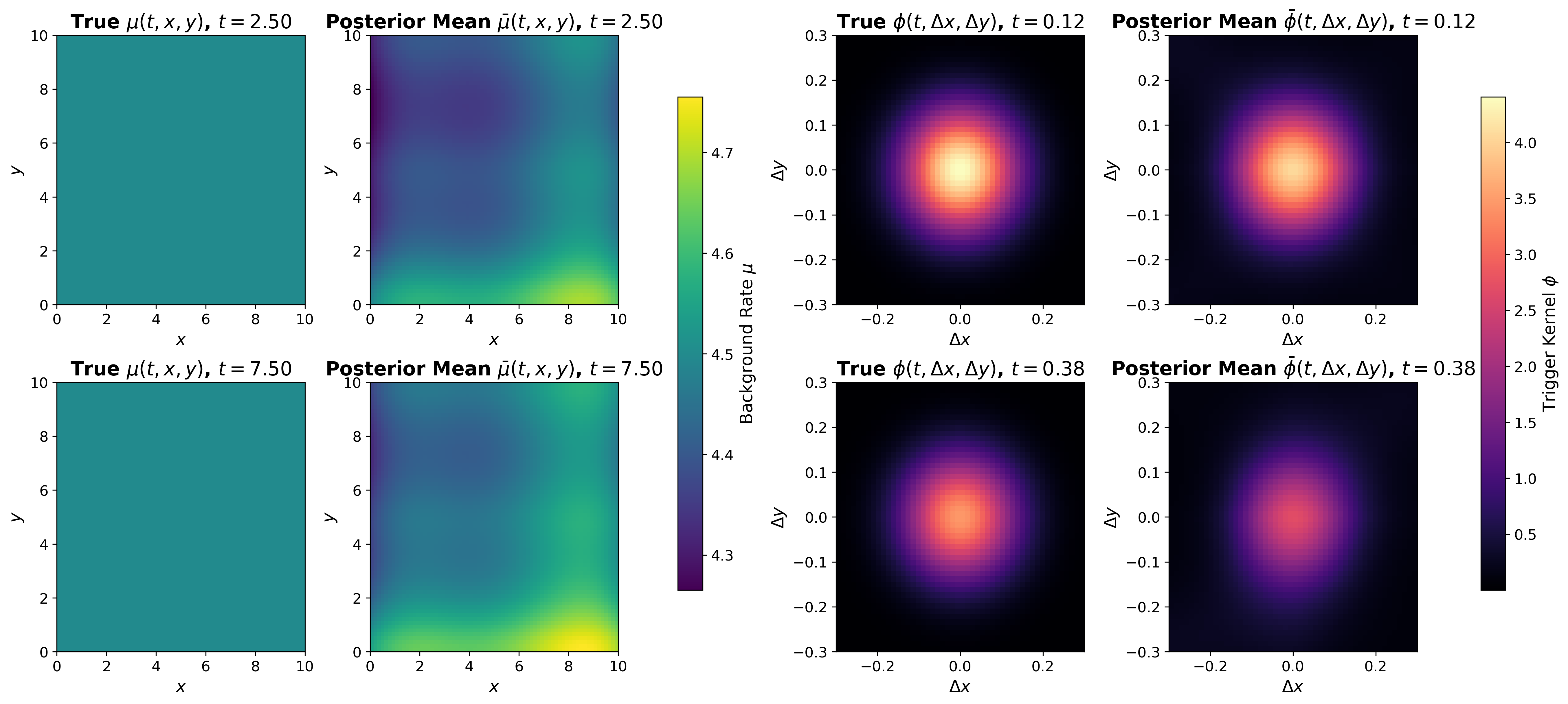}
\caption{Estimated Hawkes process components for Scenario 1. The figure is divided into two main panels: the background rate $\mu(t,x,y)$ (left) and the triggering kernel $\phi(\Delta t, \Delta x, \Delta y)$ (right), each shown at two different time snapshots. For each component, the ground truth is on the left and the posterior mean estimate is on the right.
%The model successfully recovers the constant background rate and the point-source excitation patterns.
}
\label{fig:intensity_case1}
\end{figure}

\begin{figure}
\centering
\includegraphics[width=1.0\textwidth]{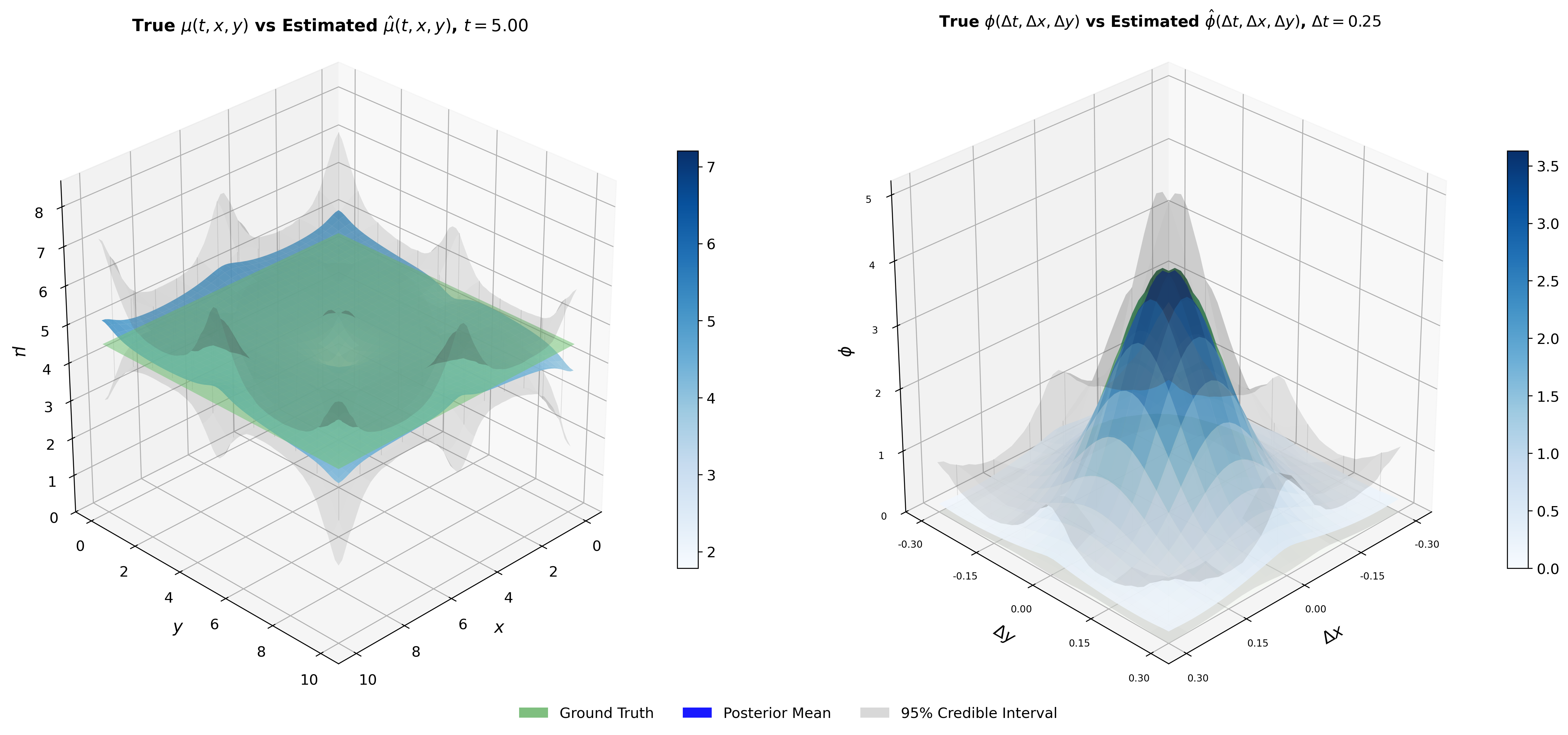}
\caption{Three-dimensional visualisation of the estimated Hawkes process components for Scenario 1, including posterior uncertainty. Left: background rate $\mu(t,x,y)$ at $t=5$. Right: triggering kernel $\phi(\Delta t,\Delta x,\Delta y)$ at $\Delta t=0.25$. The plots show the variational posterior mean (blue), ground truth (green), and the 95\% CIs (grey shaded area).}
\label{fig:intensity_3d_case1}
\end{figure}

\begin{figure}
\centering

\begin{minipage}{0.48\textwidth}
\centering
\includegraphics[width=\linewidth]{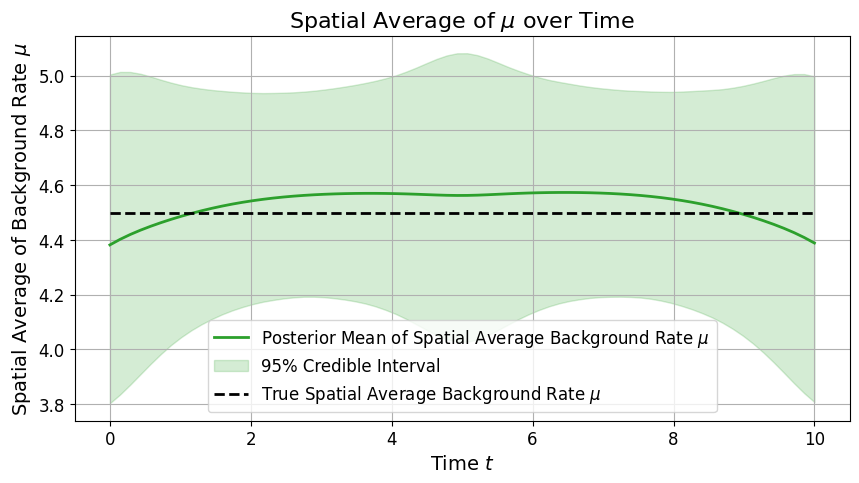}
\end{minipage}
\hfill
\begin{minipage}{0.48\textwidth}
\centering
\includegraphics[width=\linewidth]{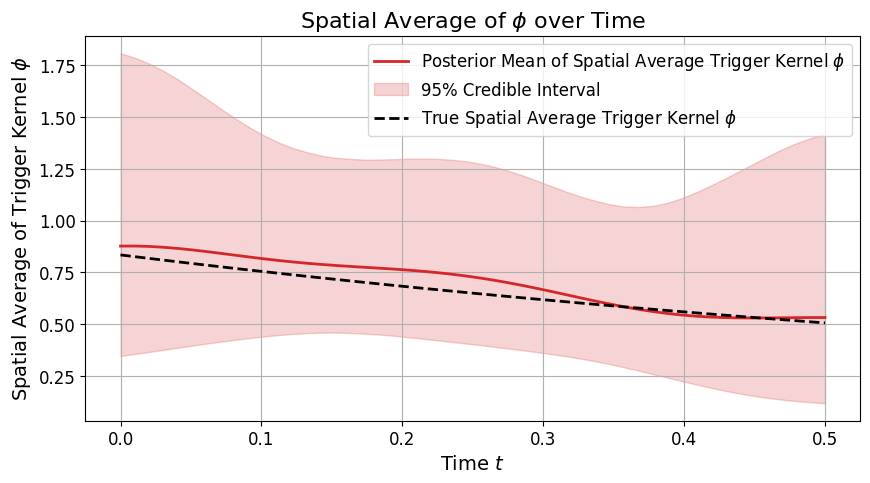}
\end{minipage}

\caption{Spatial averages of the Hawkes process components for Scenario 1. Left: spatial average of the background rate over its domain. Right: spatial average of the triggering kernel over its domain. The plots show the variational posterior mean (blue), ground truth (green), and 95\% CIs.}
\label{fig:case1_integrals}
\end{figure}

\begin{figure}
\centering
\includegraphics[width=1.0\textwidth]{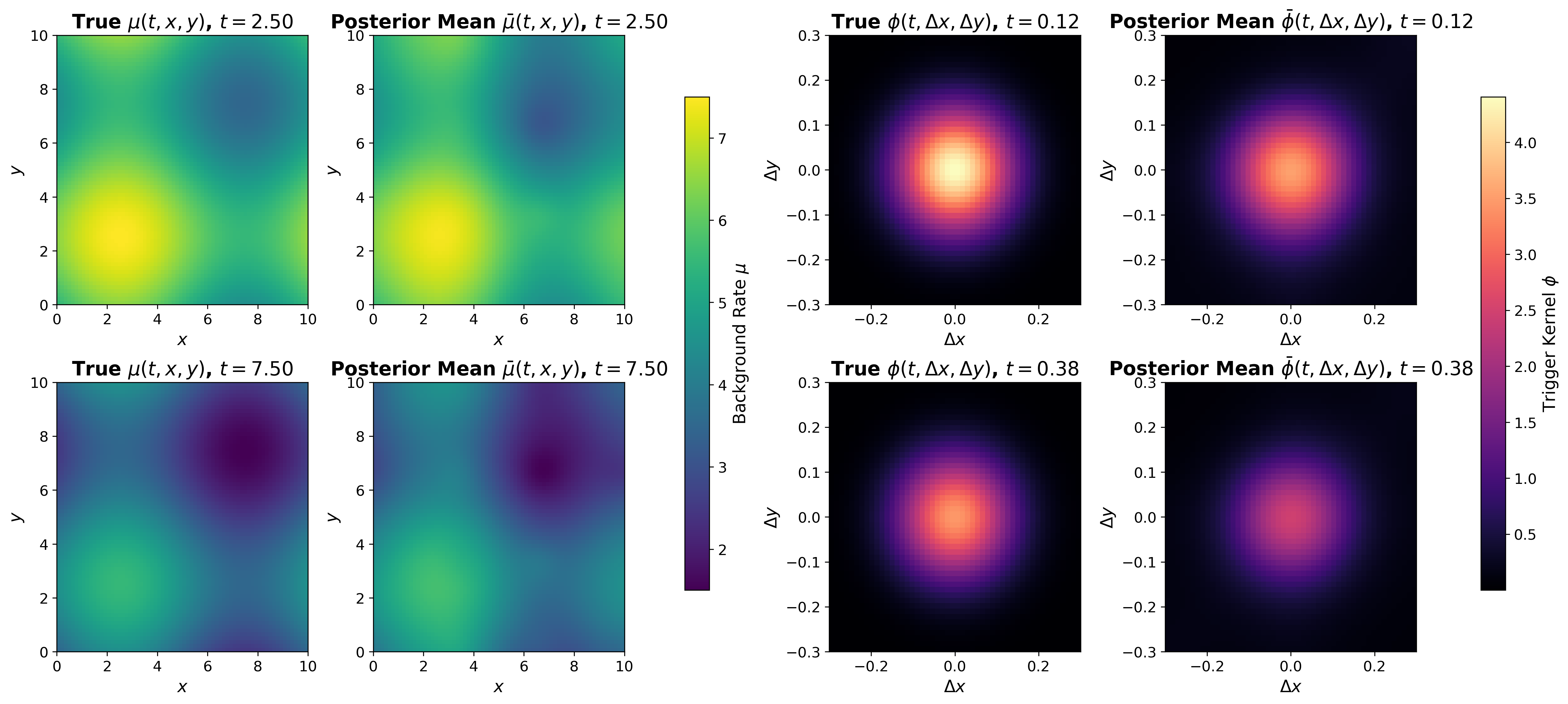}
\caption{Estimated Hawkes process components for Scenario 2. The background rate $\mu$ (left) and triggering kernel $\phi$ (right) are shown at different time snapshots. For each component, the ground truth is on the left and the variational posterior mean estimate is on the right. 
%The estimation accurately captures the non-stationary background rate which varies independently in space and time.
}
\label{fig:intensity_case2}
\end{figure}

\begin{figure}
\centering
\includegraphics[width=1.0\textwidth]{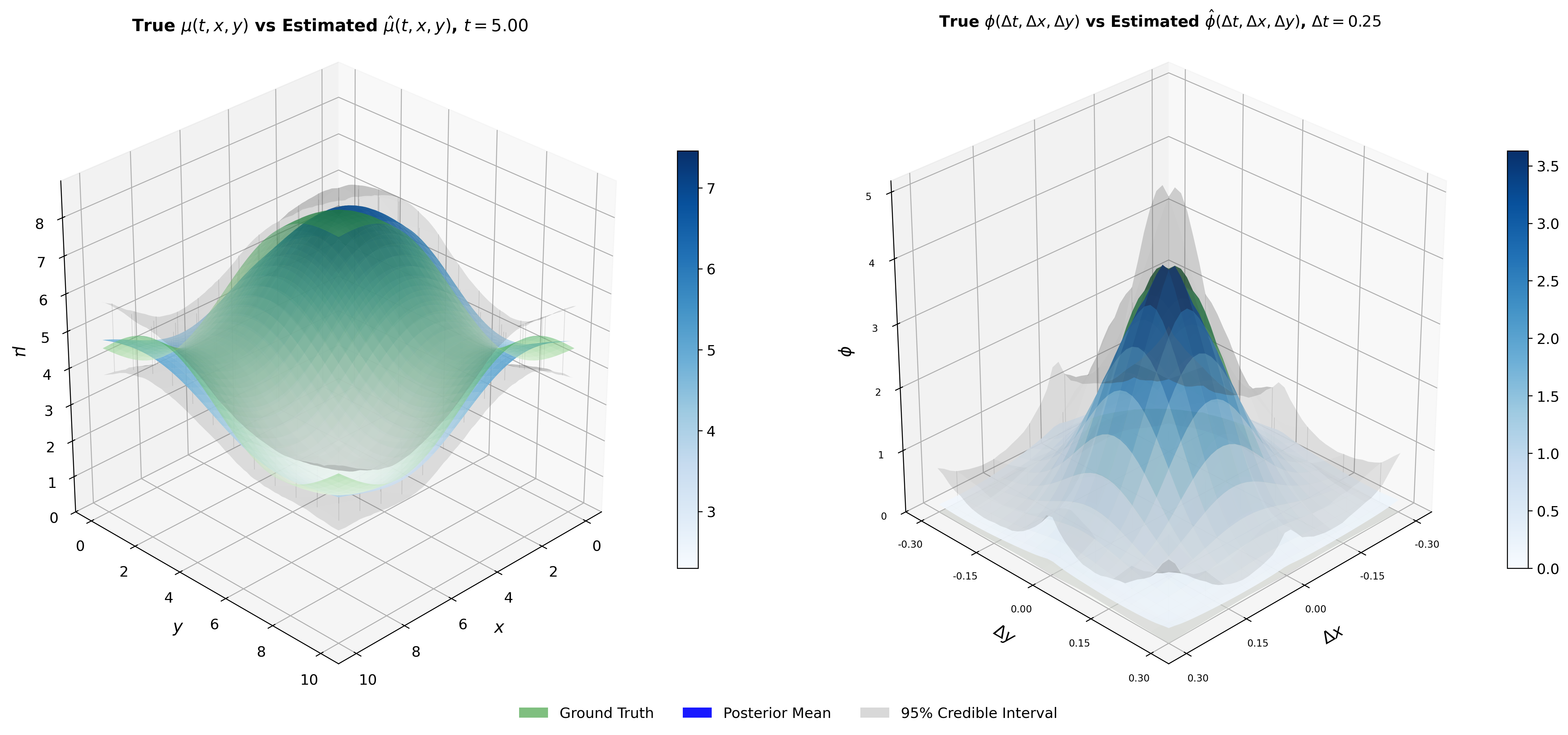}
\caption{Three-dimensional visualisation of the estimated Hawkes process components for Scenario 2, demonstrating captured posterior uncertainty. Left: background rate $\mu(t,x,y)$ at $t=5$. Right: triggering kernel $\phi(\Delta t,\Delta x,\Delta y)$ at $\Delta t=0.25$. The plots display the variational posterior mean (blue), ground truth (green), and the 95\% CIs (grey shaded area).}
\label{fig:intensity_3d_case2}
\end{figure}

\begin{figure}
\centering

\begin{minipage}{0.48\textwidth}
\centering
\includegraphics[width=\linewidth]{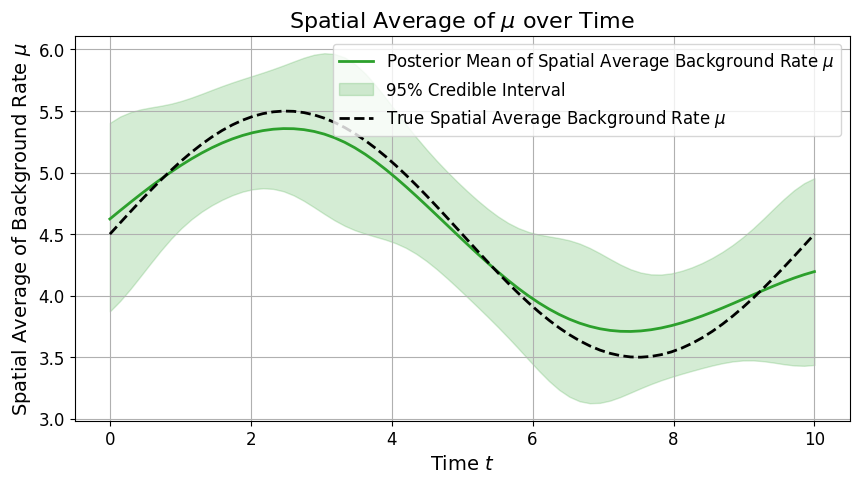}
\end{minipage}
\hfill
\begin{minipage}{0.48\textwidth}
\centering
\includegraphics[width=\linewidth]{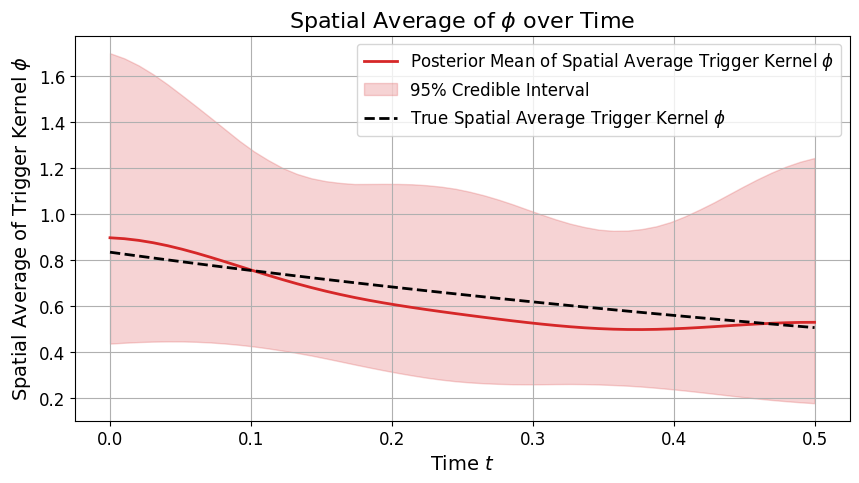}
\end{minipage}

\caption{Spatial averages of the Hawkes process components for Scenario 2. Left: spatial average of the background rate over its domain. Right: spatial average of the triggering kernel over its domain. The plots show the variational posterior mean (blue), ground truth (green), and 95\% CIs.}
\label{fig:case2_integrals}
\end{figure}

\begin{figure}
\centering
\includegraphics[width=1.0\textwidth]{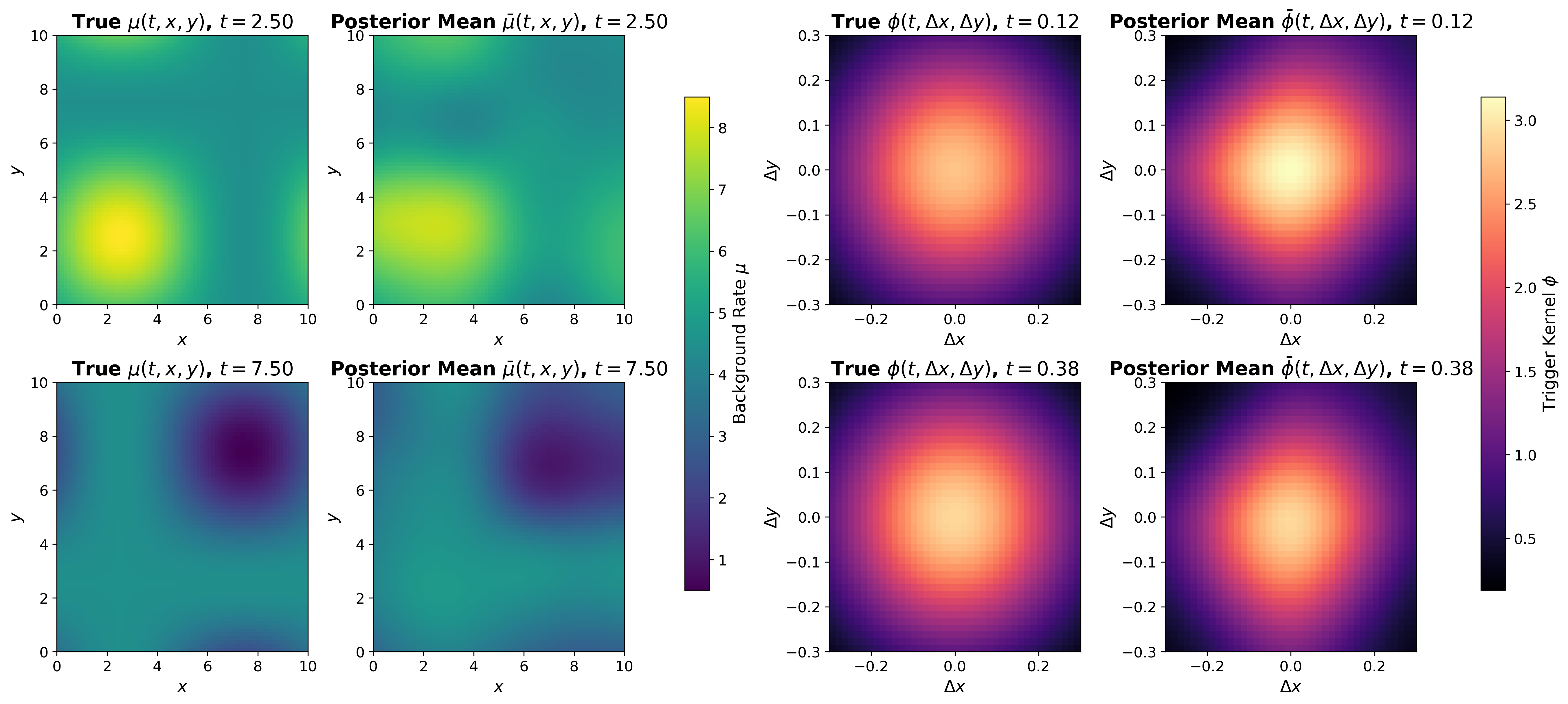}
\caption{Estimated Hawkes process components for Scenario 3. The background rate $\mu$ (left) and triggering kernel $\phi$ (right) are shown at different snapshots. For each component, the ground truth is on the left and the posterior mean estimate is on the right. %The model captures the complex interactive structures in the background rate and the non-monotonic temporal dynamics of the triggering kernel.
}
\label{fig:intensity_case3}
\end{figure}

\begin{figure}
\centering
\includegraphics[width=1.0\textwidth]{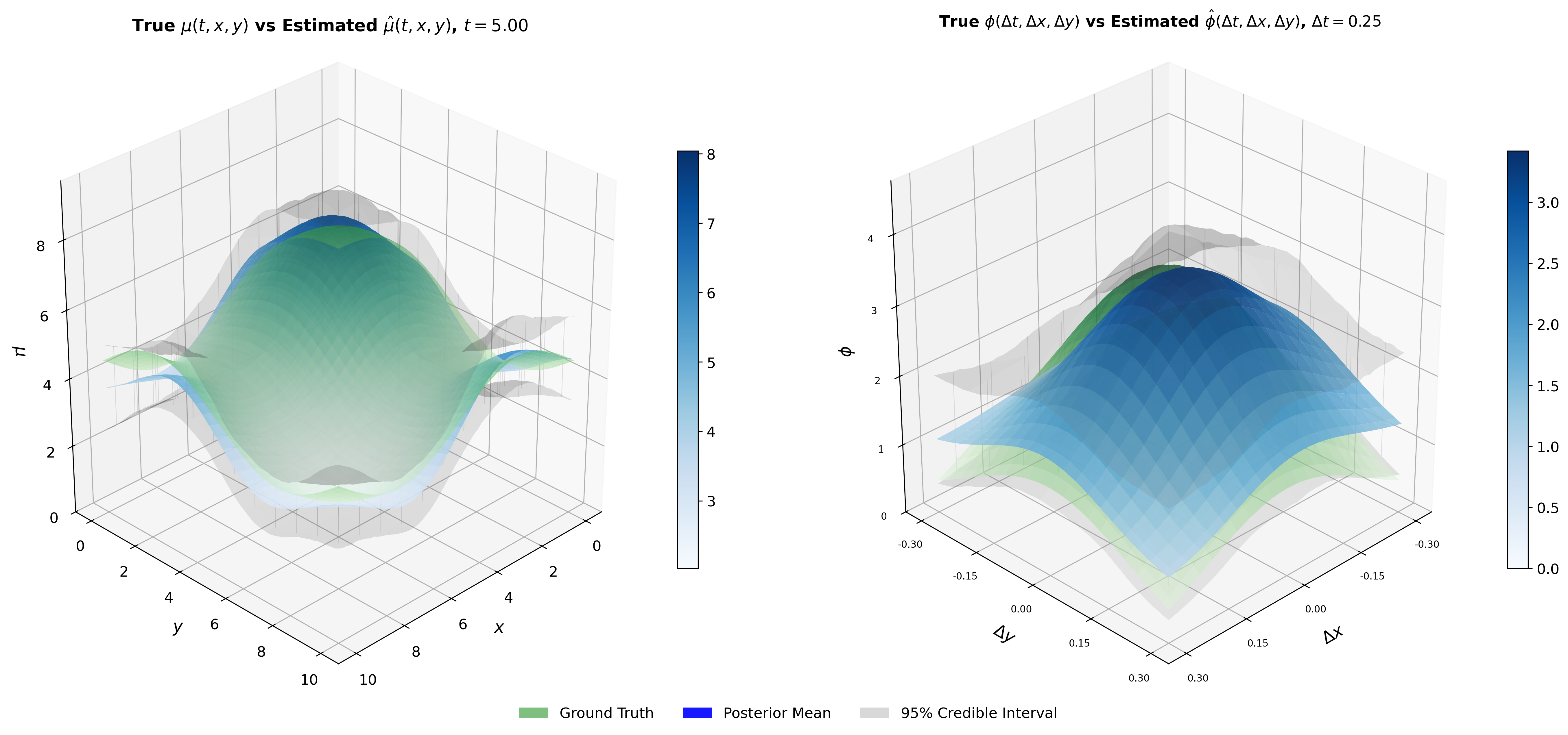}
\caption{Three-dimensional visualisation of the estimated Hawkes process components for Scenario 3, showing the posterior mean and associated uncertainty. Left: background rate $\mu(t,x,y)$ at $t=5$. Right: triggering kernel $\phi(\Delta t,\Delta x,\Delta y)$ at $\Delta t=0.25$. The results illustrate the variational posterior mean (blue), ground truth (green), and the 95\% CIs (grey shaded area).}
\label{fig:intensity_3d_case3}
\end{figure}

\begin{figure}
\centering

\begin{minipage}{0.48\textwidth}
\centering
\includegraphics[width=\linewidth]{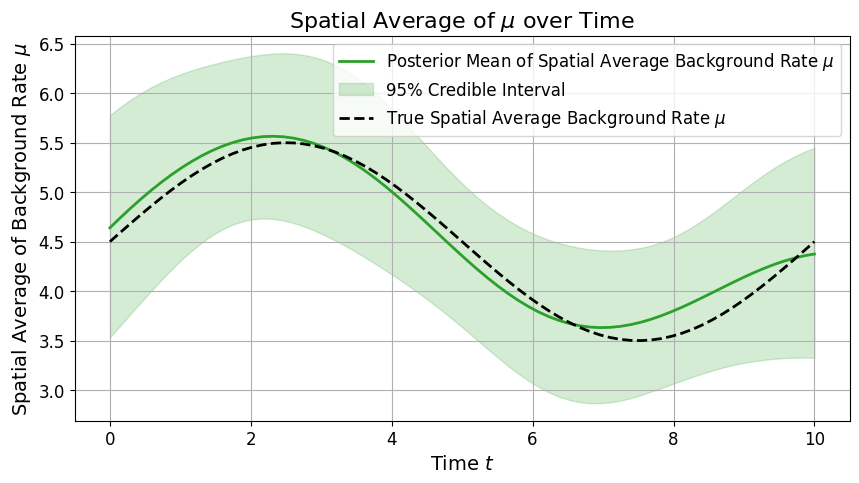}
\end{minipage}
\hfill
\begin{minipage}{0.48\textwidth}
\centering
\includegraphics[width=\linewidth]{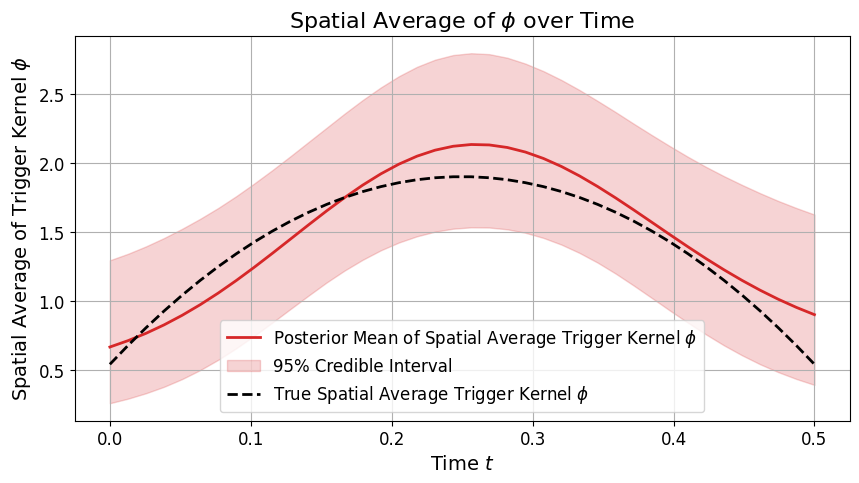}
\end{minipage}

\caption{Spatial averages of the Hawkes process components for Scenario 3. Left: spatial average of the background rate over its domain. Right: spatial average of the triggering kernel over its domain. The plots show the variational posterior mean (blue), ground truth (green), and 95\% CIs.}
\label{fig:case3_integrals}
\end{figure}

The variational posterior mean on the  background rate $\hat{\mu}(t,x,y)$ and triggering kernel $\hat{\phi}(\Delta t,\Delta x,\Delta y)$, along with their ground-truth counterparts, are visualised in Figures~\ref{fig:intensity_case1}, \ref{fig:intensity_case2}, and \ref{fig:intensity_case3} at selected timestamps. Furthermore, Figures~\ref{fig:intensity_3d_case1}, \ref{fig:intensity_3d_case2}, and \ref{fig:intensity_3d_case3} provide a 3D visualisation of the full posterior distributions for these components, showcasing the posterior mean, the ground truth, and the 95\% credible intervals (CIs) to illustrate the uncertainty of our inference. 
All results are obtained under the additive kernel configuration.
To quantify the temporal evolution of these components, we compute the spatial average for both the background rate and the triggering kernel. Specifically, for each variational posterior sample, we calculate the following spatial averages over their respective domains:
\begin{equation}
    \bar{\mu}(t) = \frac{1}{|\mathcal{S}_\mu|} \iint_{\mathcal{S}_\mu} \mu(t, x, y) \,dx dy, \quad \bar{\phi}(\Delta t) = \frac{1}{|\mathcal{S}_\phi|} \iint_{\mathcal{S}_\phi} \phi(\Delta t, \Delta x, \Delta y) \,d\Delta x d\Delta y,
\end{equation}
where $\mathcal{S}_\mu = [0, X] \times [0, Y]$ and $\mathcal{S}_\phi = [-X_\phi, X_\phi] \times [-Y_\phi, Y_\phi]$ denote the spatial support for the background rate and triggering kernel, respectively, with $|\mathcal{S}_\mu|$ and $|\mathcal{S}_\phi|$ representing their corresponding areas. Figures~\ref{fig:case1_integrals}, \ref{fig:case2_integrals}, and \ref{fig:case3_integrals} present the resulting posterior mean of these spatial averages, the 95\% CIs derived from the posterior samples, and the corresponding ground-truth curves.

The results in Tables~\ref{tab:scenario1_comparison}, \ref{tab:scenario2_comparison}, and \ref{tab:scenario3_comparison} summarise the performance of our framework. In Scenario~1, the Parametric Hawkes and Cox-Hawkes models achieve the best performance with lower $PM_{\mathrm{mse}}$ and $PE_{\mathrm{mse}}$ than our model, as their specifications match the data-generating processes more closely. However, our proposed model yields estimates close to the ground truth. As illustrated in Figures~\ref{fig:intensity_case1}, \ref{fig:intensity_3d_case1}, and~\ref{fig:case1_integrals}, our posterior mean estimates recover the steady, invariant background rate and point-source diffusion patterns (Fig.~\ref{fig:intensity_case1}), while 3D visualisations confirm that the true intensity surfaces remain within the 95\% CIs (Fig.~\ref{fig:intensity_3d_case1}). Notably, the model recovers the constant background rate and the temporal decay of the triggering kernel (Fig.~\ref{fig:case1_integrals}).

For Scenario 2, the results are presented in Table~\ref{tab:scenario2_comparison} and Figures~\ref{fig:intensity_case2}, \ref{fig:intensity_3d_case2}, and \ref{fig:case2_integrals}. In this complex setting, Parametric Hawkes model yields the poorest performance due to the limitations of its predefined functional forms. Although the Cox-Hawkes model is the most closely aligned with the true data-generating process, our proposed nonparametric framework demonstrates comparable or even slightly superior performance in certain metrics. 

As illustrated in Fig.~\ref{fig:intensity_case2}, the posterior mean estimates accurately capture the oscillating spatial intensity across different time slices. This precision is further confirmed by the 3D surface plots in Fig.~\ref{fig:intensity_3d_case2}, where the estimated surfaces (blue) closely overlap with the ground truth (green), with the true intensity remaining well within the 95\% CIs (grey shaded area). Fig.~\ref{fig:case2_integrals} characterises the temporal evolution of the spatial averages; the results clearly demonstrate that our model captures the significant fluctuations in the background rate $\mu$, with the posterior mean tightly tracking the trajectory of the ground truth while simultaneously recovering the decay structure of the triggering kernel $\phi$.

The results for Scenario 3 are presented in Table~\ref{tab:scenario3_comparison} and Figures~\ref{fig:intensity_case3}, \ref{fig:intensity_3d_case3}, and \ref{fig:case3_integrals}. In this scenario, our model outperforms all baselines, achieving the lowest errors across all metrics. This performance gap highlights the inherent limitations of parametric or semi-parametric baselines when confronted with complex dynamics. 

A key challenge in Scenario 3 is that the triggering kernel $\phi$ is non-monotonic over time, as clearly shown in the right panel of Fig.~\ref{fig:case3_integrals}. Such a complex temporal structure is extremely difficult to capture using standard parametric functional forms, which typically assume monotonic decay. In contrast, our fully nonparametric approach offers superior flexibility, allowing the posterior mean to accurately track this non-monotonic rise and fall. 

Furthermore, Fig.~\ref{fig:intensity_case3} and Fig.~\ref{fig:intensity_3d_case3} demonstrate that our model maintains high precision in estimating the spatio-temporal background rate $\mu$. The 3D surface plots confirm that even under these highly dynamic conditions, the true intensity surfaces remain consistently within the 95\% CIs, showcasing the robustness of our framework in recovering complex Hawkes process components.

In addition, the results across all scenarios in Tables~\ref{tab:scenario1_comparison}, 
\ref{tab:scenario2_comparison}, and \ref{tab:scenario3_comparison} demonstrate that 
our additive kernels generally outperform their separable counterparts, or at the very 
least exhibit comparable performance. This advantage is particularly evident in the 
background rate estimation ($PM_{\mathrm{mse}}(\mu)$ and $PE_{\mathrm{mse}}(\mu)$), 
while the performance on the triggering kernel $\phi$ remains comparable to that of the separable kernels. We attribute this gain to the enhanced numerical stability of the additive formulation. 
In sparse GP inference, the stability of inverting the inducing point covariance matrices 
$K_{{Z}_\mu {Z}_\mu}$ and $K_{{Z}_\phi {Z}_\phi}$ depends on their condition number
$\kappa(K) = \lambda_{\max}(K) / \lambda_{\min}(K)$. 
Here $\lambda_{\max}$ and $\lambda_{\min}$ denote the maximum and minimum eigenvalues, respectively.

As summarised in Table~\ref{tab:condition_number}, which reports the condition numbers of $K_{{Z}_\mu {Z}_\mu}$ and $K_{{Z}_\phi {Z}_\phi}$ for the separable and additive RBF kernels under the softplus link function, separable kernels generally exhibit 
higher condition numbers, leading to ill-conditioned matrices. In contrast, the additive 
structure brings lower condition numbers (or at least comparable ones), thereby ensuring 
more robust estimation.

\begin{table}[ht!]
\centering
\tbl{Comparison of different methods for Scenario 1 using $PM_{\mathrm{mse}}(\mu)$, $PM_{\mathrm{mse}}(\phi)$, $PE_{\mathrm{mse}}(\mu)$, $PE_{\mathrm{mse}}(\phi)$. \label{tab:scenario1_comparison}}
{   \small\addtolength{\tabcolsep}{-1.1pt}\begin{tabular}{crrr|crrr}
        Method & Link & $PM_{\mathrm{mse}}(\mu)$ & $PM_{\mathrm{mse}}(\phi)$ & 
        & $PE_{\mathrm{mse}}(\mu)$ & $PE_{\mathrm{mse}}(\phi)$ & \\ 
        \midrule
        Parametric Hawkes & {} & $\color{red}{1.21 \pm 1.90}$& $\color{blue}4.49 \pm 2.26$ & & $\color{red}{1.90 \pm 1.87}$ & $\color{red}{3.33 \pm 0.47}$ & \\ 
        Cox-Hawkes & {exp}  & $4.63 \pm 2.76$ & $\color{red}2.92 \pm 1.66$ && $13.84 \pm 4.90$ & $\color{blue}4.42 \pm 1.66$ & \\ 
        \midrule
Ours (separable RBF) & {softplus-softplus}  & $8.13 \pm 2.50$ & $9.82 \pm 2.31$ && $14.95 \pm 2.73$ & $14.64 \pm 2.42$ &\\ 
        Ours (separable RBF) & {sigmoid-sigmoid}   & $8.03 \pm 2.86$ & $8.16 \pm 2.26$ && $15.96 \pm 3.51$ & $15.23 \pm 2.12$ &\\ 
        Ours (separable RBF) & {exp-exp}   & $8.48 \pm 2.50$ & $8.57 \pm 1.97$ && $16.29 \pm 2.85$ & $14.39 \pm 2.12$ &\\ 
        \midrule
Ours (separable Mat\'{e}rn, $\nu=2.5$)& {softplus-softplus}      & $6.58 \pm 2.08$ & $7.36 \pm 2.69$ && $13.27 \pm 2.24$ & $12.03 \pm 2.64$ &\\ 
        Ours (separable Mat\'{e}rn, $\nu=2.5$)&{sigmoid-sigmoid}      & $7.34 \pm 2.61$ & $7.04 \pm 2.32$ && $15.13 \pm 3.16$ & $13.35 \pm 2.28$ &
\\ 
        Ours (separable Mat\'{e}rn, $\nu=2.5$)&{exp-exp}      & $8.11 \pm 2.69$ & $8.12 \pm 2.09$ && $16.18 \pm 3.19$ & $14.01 \pm 1.99$ &
\\ 
        \midrule
Ours (additive RBF) & {softplus-softplus}   & \color{blue}$2.88 \pm 0.88$ & $5.01 \pm 1.36$ && $\color{blue}7.63 \pm 0.82$ & $10.16 \pm 1.47$ &\\ 
        Ours (additive RBF) & {sigmoid-sigmoid}    & $3.43 \pm 1.20$ & $7.91 \pm 2.93$ && $9.75 \pm 1.50$ & $15.43 \pm 3.28$ &\\ 
        Ours (additive RBF) & {exp-exp}    & $4.78 \pm 2.22$ & $7.26 \pm 3.46$ && $13.69 \pm 4.23$ & $14.08 \pm 3.78$ &\\ 
        \midrule
Ours (additive Mat\'{e}rn, $\nu=2.5$)& {softplus-softplus}       & $3.14 \pm 1.10$ & $5.28 \pm 1.77$ && $8.44 \pm 1.45$ & $10.83 \pm 1.93$ &\\ 
        Ours (additive Mat\'{e}rn, $\nu=2.5$)&{sigmoid-sigmoid}       & $3.76 \pm 1.51$ & $7.55 \pm 2.93$ && $10.45 \pm 2.06$ & $14.62 \pm 3.41$ &
\\ 
        Ours (additive Mat\'{e}rn, $\nu=2.5$)&{exp-exp}       & $5.11 \pm 2.50$ & $7.36 \pm 2.31$ && $13.82 \pm 4.84$ & $13.97 \pm 2.31$ &
\\ 
        \bottomrule
    \end{tabular}}
    \begin{tabnote}
    The first and second terms in the Link column refer to the link functions for the background rate $\mu$ and the triggering kernel $\phi$, respectively. Results are averaged over 8 independent realisations. Mean values and standard deviations ($\pm$) are reported. Red and blue indicate the best and second-best performance, respectively. All values are scaled by $10^{2}$.
\end{tabnote}
\end{table}

\begin{table}[ht!]
\centering
\tbl{Comparison of different methods for Scenario 2 based on $PM_{\mathrm{mse}}(\mu)$, $PM_{\mathrm{mse}}(\phi)$, $PE_{\mathrm{mse}}(\mu)$, $PE_{\mathrm{mse}}(\phi)$. \label{tab:scenario2_comparison}}
{   \small\addtolength{\tabcolsep}{-1.1pt}\begin{tabular}{crrr|crrr}
        Method & Link & $PM_{\mathrm{mse}}(\mu)$ & $PM_{\mathrm{mse}}(\phi)$ & 
        & $PE_{\mathrm{mse}}(\mu)$ & $PE_{\mathrm{mse}}(\phi)$ & \\ 
        \midrule
        Parametric Hawkes & {} &$152.64 \pm 3.24$& $16.78 \pm 7.56$ & & $	153.37 \pm 3.26$ & $\color{blue}11.58 \pm 6.34$ & \\ 
        Cox-Hawkes & {exp}  & $16.08 \pm 4.72$ & $\color{red}4.21 \pm 2.33$ && $30.57 \pm 6.09$ & $\color{red}5.68 \pm 2.27$ & \\ 
        \midrule
Ours (separable RBF) & {softplus-softplus}   & $22.09 \pm 7.26$ & $13.24 \pm 1.68$ && $37.01 \pm 6.70$ & $18.04 \pm 1.94$ &
\\ 
        Ours (separable RBF) & {sigmoid-sigmoid}    & $19.56 \pm 5.97$ & $11.76 \pm 2.07$ && $36.59 \pm 5.88$ & $19.24 \pm 2.48$ &
\\ 
        Ours (separable RBF) & {exp-exp}   & $22.34 \pm 6.55$ & $11.58 \pm 1.74$ && $38.29 \pm 6.21$ & $17.71 \pm 2.21$ &
\\ 
        \midrule
Ours (separable Mat\'{e}rn, $\nu=2.5$)& {softplus-softplus}       & $18.16 \pm 6.48$ & $9.16 \pm 1.43$ && $31.73 \pm 5.80$ & $14.58 \pm 1.89$ &
\\ 
        Ours (separable Mat\'{e}rn, $\nu=2.5$)&{sigmoid-sigmoid} & $17.28 \pm 5.91$ & $10.14 \pm 1.66$ && $32.90 \pm 5.69$ & $16.91 \pm 2.08$ &
\\ 
        Ours (separable Mat\'{e}rn, $\nu=2.5$)&{exp-exp}       & $18.62 \pm 6.08$ & $10.87 \pm 1.36$ && $33.69 \pm 5.60$ & $17.27 \pm 1.80$ &
\\ 
        \midrule
Ours (additive RBF) & {softplus-softplus}   & $\color{red}11.14 \pm 6.21$ & $\color{blue}7.67 \pm 1.37$ && $\color{red}20.22 \pm 6.14$ & $13.44 \pm 2.74$ &
\\ 
        Ours (additive RBF) & {sigmoid-sigmoid}    & $11.95 \pm 6.18$ & $11.96 \pm 2.58$ && $22.88 \pm 6.12$ & $20.20 \pm 3.56$ &
\\ 
        Ours (additive RBF) & {exp-exp}    & $14.12 \pm 5.72$ & $9.87 \pm 1.91$ && $25.52 \pm 5.75$ & $17.37 \pm 3.16$ &\\ 
        \midrule
Ours (additive Mat\'{e}rn, $\nu=2.5$)& {softplus-softplus}       & $\color{blue}11.81 \pm 6.14$ & $8.08 \pm 1.26$ && $\color{blue}21.00 \pm 5.86$ & $14.48 \pm 1.71$ &
\\ 
        Ours (additive Mat\'{e}rn, $\nu=2.5$)&{sigmoid-sigmoid}       & $12.38 \pm 5.80$ & $10.02 \pm 1.85$ && $23.27 \pm 5.79$ & $16.98 \pm 2.55$ &
\\ 
        Ours (additive Mat\'{e}rn, $\nu=2.5$)&{exp-exp} & $14.43 \pm 6.22$ & $10.50 \pm 1.97$ && $25.50 \pm 6.44$ & $18.26 \pm 2.01$ &
\\ 
        \bottomrule
    \end{tabular}}
    \begin{tabnote}
Results are averaged over 8 independent realisations. Mean values and standard deviations ($\pm$) are reported. Red and blue indicate the best and second-best performance, respectively. All values are scaled by $10^{2}$.
\end{tabnote}
\end{table}

\begin{table}[ht!]
\centering
\tbl{Comparison of different methods for Scenario 3 based on $PM_{\mathrm{mse}}(\mu)$, $PM_{\mathrm{mse}}(\phi)$, $PE_{\mathrm{mse}}(\mu)$, $PE_{\mathrm{mse}}(\phi)$. \label{tab:scenario3_comparison}}
{   \small\addtolength{\tabcolsep}{-1.1pt}\begin{tabular}{crrr|crrr}
        Method & Link & $PM_{\mathrm{mse}}(\mu)$ & $PM_{\mathrm{mse}}(\phi)$ & 
        & $PE_{\mathrm{mse}}(\mu)$ & $PE_{\mathrm{mse}}(\phi)$ & \\ 
        \midrule
        Parametric Hawkes & {} & $169.55 \pm  5.35$ & $32.50 \pm  4.42$ & & $170.35 \pm  5.31$ & $31.11 \pm  3.81$ & \\ 
        Cox-Hawkes & {exp} & $24.39 \pm 5.60$ & $24.97 \pm 1.46$ && $47.32 \pm 5.61$ & $26.02 \pm 1.44$ &\\ 
        \midrule
Ours (separable RBF) & {softplus-softplus}    & $22.92 \pm 3.97$ & $10.72 \pm 5.46$ && $41.49 \pm 3.82$ & $18.92 \pm 4.71$ &
\\ 
        Ours (separable RBF) & {sigmoid-sigmoid}     & $25.24 \pm 5.24$ & $11.61 \pm 5.07$ && $47.67 \pm 4.93$ & $20.31 \pm 4.94$ &
\\ 
        Ours (separable RBF) & {exp-exp} & $28.08 \pm 5.25$ & $12.41 \pm 5.29$ && $48.23 \pm 5.34$ & $20.75 \pm 4.68$ &
\\ 
        \midrule
Ours (separable Mat\'{e}rn, $\nu=2.5$)& {softplus-softplus}      & $20.75 \pm 3.86$ & $10.23 \pm 4.39$ && $38.07 \pm 3.80$ & $18.74 \pm 3.77$ &
\\ 
        Ours (separable Mat\'{e}rn, $\nu=2.5$)&{sigmoid-sigmoid}      & $23.00 \pm 5.34$ & $10.57 \pm 5.20$ && $43.84 \pm 5.06$ & $18.39 \pm 4.84$ &
\\ 
        Ours (separable Mat\'{e}rn, $\nu=2.5$)&{exp-exp}      & $24.79 \pm 5.32$ & $11.75 \pm 4.36$ && $44.17 \pm 5.43$ & $20.18 \pm 3.93$ &
\\ 
        \midrule
Ours (additive RBF) & {softplus-softplus}   & $\color{blue}19.17 \pm 5.28$ & $10.14 \pm 4.27$ && $\color{red}37.11 \pm 4.42$ & $18.54 \pm 4.23$ &\\ 
        Ours (additive RBF) & {sigmoid-sigmoid}    & $25.50 \pm 4.14$ & $\color{blue}9.31 \pm 4.61$ && $40.57 \pm 8.14$ & $\color{red}17.49 \pm 4.42$ &\\ 
        Ours (additive RBF) & {exp-exp}    & $24.88 \pm 6.60$ & $\color{red}9.26 \pm 4.24$ && $43.98 \pm 6.32$ & $\color{blue}17.84 \pm 3.96$ &
\\ 
        \midrule
Ours (additive Mat\'{e}rn, $\nu=2.5$)& {softplus-softplus}      & $\color{red}18.27 \pm 3.56$ & $10.91 \pm 5.89$ && $37.76 \pm 3.77$ & $18.75 \pm 5.45$ &
\\ 
        Ours (additive Mat\'{e}rn, $\nu=2.5$)&{sigmoid-sigmoid}       & $26.22 \pm 4.27$ & $9.70 \pm 4.78$ && $41.47 \pm 8.80$ & $17.94 \pm 4.50$ &
\\ 
        Ours (additive Mat\'{e}rn, $\nu=2.5$)&{exp-exp}       & $25.32 \pm 6.85$ & $10.50 \pm 4.39$ && $45.91 \pm 7.83$ & $19.47 \pm 4.47$ &
\\ 
        \bottomrule
    \end{tabular}}
    \begin{tabnote}
Results are averaged over 8 independent realisations. Mean values and standard deviations ($\pm$) are reported. Red and blue indicate the best and second-best performance, respectively. All values are scaled by $10^{2}$.
\end{tabnote}
\end{table}

\begin{table}[ht!]
    \centering
    \tbl{Comparison of condition numbers across different synthetic scenarios. $\kappa({K}_{{Z}_\mu {Z}_\mu})$ and $\kappa({K}_{{Z}_\phi {Z}_\phi})$ denote the condition numbers for the background and triggering components, respectively. Lower condition numbers usually indicate better numerical stability.}
    {\small\addtolength{\tabcolsep}{-1.1pt}\begin{tabular}{clccc} 
    \toprule
        Scenario & Method & Link & $\kappa({K}_{{Z}_\mu {Z}_\mu})$ & $\kappa({K}_{{Z}_\phi {Z}_\phi})$ \\ 
    \midrule
    \multirow{2}{*}{Scenario 1}
        & Ours (separable RBF) & {softplus-softplus} & $1.84 \times 10^{4} \pm 1.02 \times 10^{4}$ & $1.79 \times 10^{2} \pm 1.02 \times 10^{2}$  \\
        & Ours (additive RBF)  & {softplus-softplus} & $5.67 \times 10^{1} \pm 5.09 \times 10^{0}$ & $4.84 \times 10^{1} \pm 6.75 \times 10^{0}$  \\ 
    \midrule
    \multirow{2}{*}{Scenario 2}
        & Ours (separable RBF) & {softplus-softplus} & $4.92 \times 10^{4} \pm 1.32 \times 10^{4}$ & $2.50 \times 10^{2} \pm 6.67 \times 10^{1}$  \\
        & Ours (additive RBF)  & {softplus-softplus} & $1.37 \times 10^{2} \pm 1.50 \times 10^{1}$ & $4.49 \times 10^{1} \pm 7.74 \times 10^{0}$  \\ 
    \midrule
    \multirow{2}{*}{Scenario 3}
        & Ours (separable RBF) & {softplus-softplus} & $1.77 \times 10^{4} \pm 1.05 \times 10^{4}$ & $1.96 \times 10^{2} \pm 3.17 \times 10^{2}$  \\
        & Ours (additive RBF)  & {softplus-softplus} & $4.37 \times 10^{2} \pm 3.48 \times 10^{2}$ & $3.51 \times 10^{2} \pm 5.14 \times 10^{2}$ \\ 
    \bottomrule
    \end{tabular}}
    \label{tab:condition_number}
    \begin{tabnote}
Mean and standard deviation ($\pm$) of the condition numbers for the induction points. Values are multiplied by $10^{2}$.
\end{tabnote}
\end{table}

\clearpage

\subsection{Real-world Data Experiments}
\label{sec:real_data}

\begin{figure}[h!]
\centering

\begin{minipage}{0.48\textwidth}
\centering
\includegraphics[width=\linewidth]{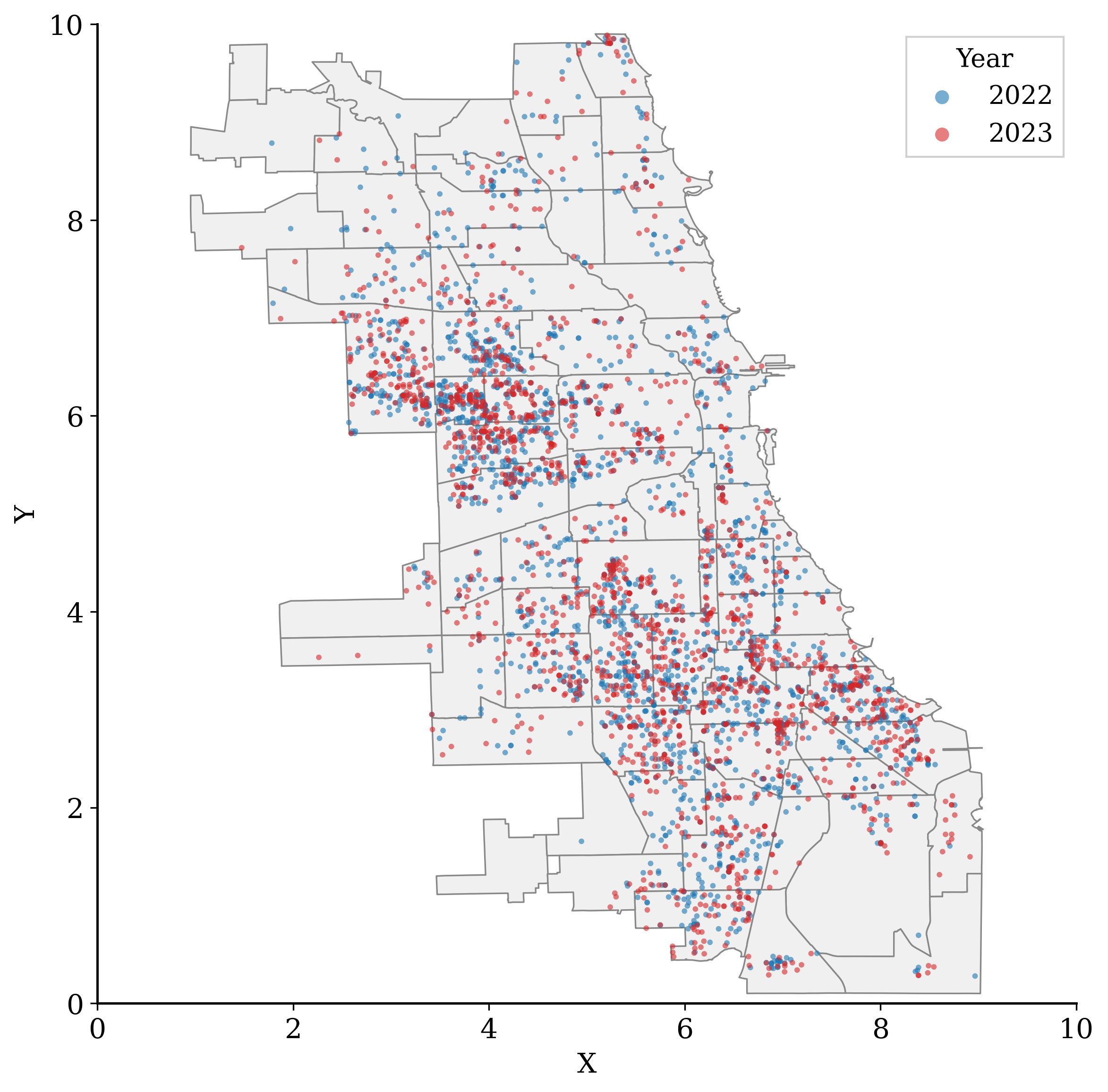}
\end{minipage}
\hfill
\begin{minipage}{0.48\textwidth}
\centering
\includegraphics[width=\linewidth]{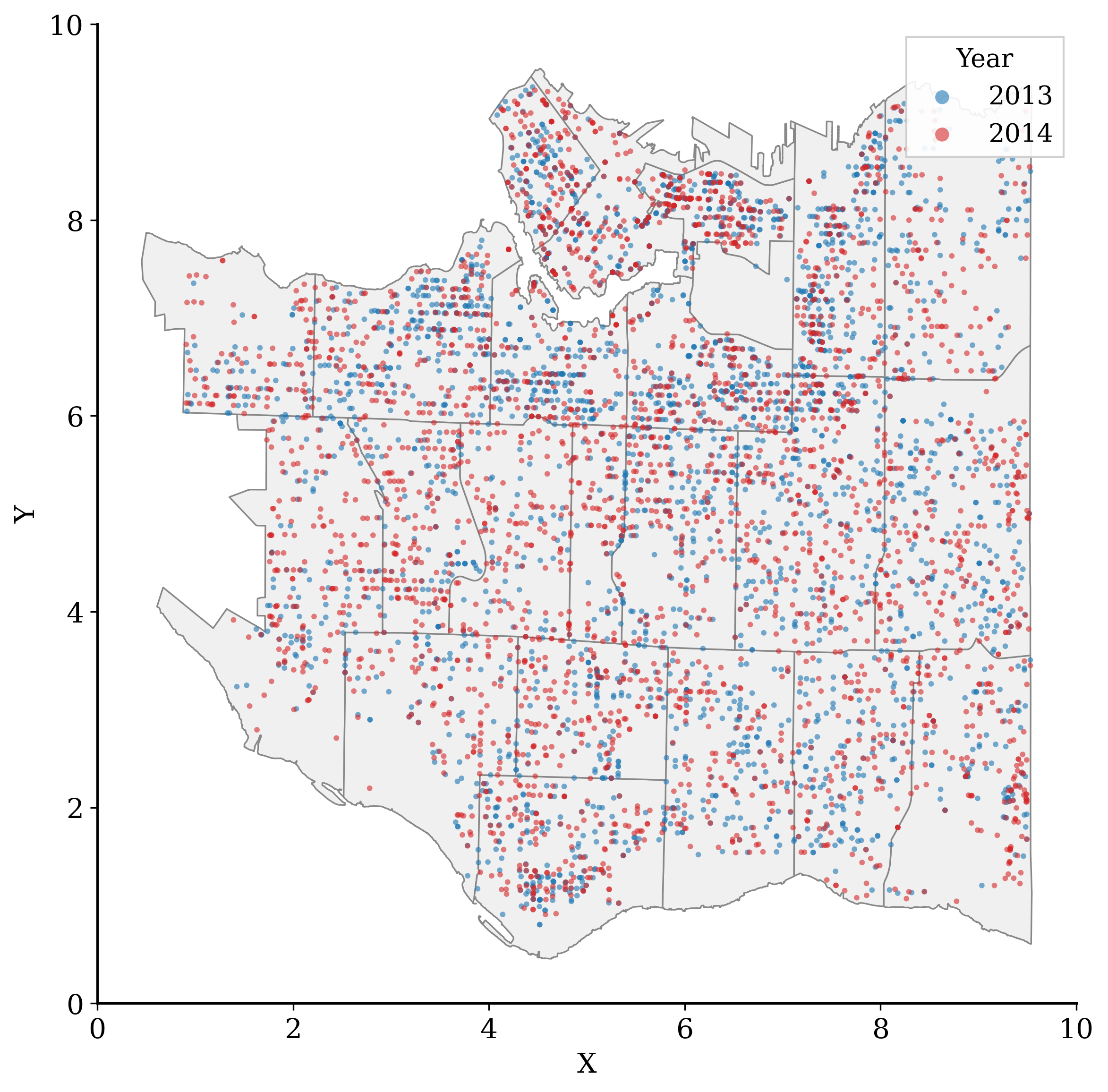}
\end{minipage}

\caption{Spatial distribution of events in the real-world datasets. 
Left: Chicago shooting incidents (2022 for training and 2023 for testing). 
Right: Vancouver break-ins (2013 for training and 2014 for testing).}
\label{fig:real_data_maps}
\end{figure}

We evaluate our nonparametric framework on two real-world spatio-temporal datasets: 
Chicago Shootings~\citep{manring2025bstpp} and Vancouver Break-ins~\citep{zhou2020efficient}. 
Figure~\ref{fig:real_data_maps} illustrates the spatial distribution of events for both datasets.

The Chicago Shootings dataset contains reported shooting incidents with continuous 
timestamps and geographic coordinates. We use the data from 2022 (2,124 events) 
as the training sequence and the data from 2023 (1,809 events) for testing.

The Vancouver Break-ins dataset records break-and-enter incidents in Vancouver. 
We use the 2013 data (3,013 events) for training and the 2014 data (3,033 events) 
for testing.

For both datasets, we rescale the temporal dimension $T$ to $[0,10]$ and the spatial 
coordinates $(X,Y)$ to $[0,10]^2$. For the Chicago Shootings dataset, the triggering kernel window 
is set to $T_\phi = 0.4$ (approximately half a month), with spatial windows 
$X_\phi = Y_\phi = 0.4$ (representing physical distances of approximately 1,278 m East-West and 1,716.4 m North-South). 
For the Vancouver Break-ins dataset, we use $T_\phi = 0.6$ (approximately three weeks), 
with spatial windows $X_\phi = Y_\phi = 0.2$ (representing physical distances of approximately 321.2 m East-West and 233.6 m North-South). 
The selection of these finite window sizes is supported by a sensitivity analysis detailed in Appendix~\ref{app:sensitivity_window}.

The model fit, evaluated using the Expected Log-Likelihood ($ELL$) on the test sets, is presented in Table~\ref{tab:real_world_results}, with more comprehensive results for all tested kernel and link function combinations provided in Appendix~\ref{app:real_world_full}. Alongside standard baselines, we specifically include a Log-Gaussian Cox Process~\citep{moller1998log} to isolate the impact of self-excitation. Because the Log-Gaussian Cox Process models only the background rate without any triggering mechanism, the fact that our proposed model yields a significantly higher $ELL$ across both datasets provides strong evidence for the presence of underlying contagion effects. Furthermore, among our proposed model variants, the additive kernel configurations yield the best overall fit. This suggests that decomposing the spatial and temporal components provides a better empirical fit for these real-world events. 

To evaluate goodness-of-fit, we apply the super-thinning algorithm \citep{reinhart2018review} (details in Appendix~\ref{app:super_thinning}) to the estimated intensity obtained via the variational posterior mean on the background rate and triggering kernel from our additive kernel configurations. After super-thinning, a perfectly specified model yields a thinned process which has the distribution of a homogeneous Poisson process. We assess the distribution of this thinned process using standard temporal and spatial diagnostics (see Appendix~\ref{app:statistical_tests}).

\begin{figure}[h!]
\centering

\begin{minipage}{0.48\textwidth}
\centering
\includegraphics[width=\linewidth]{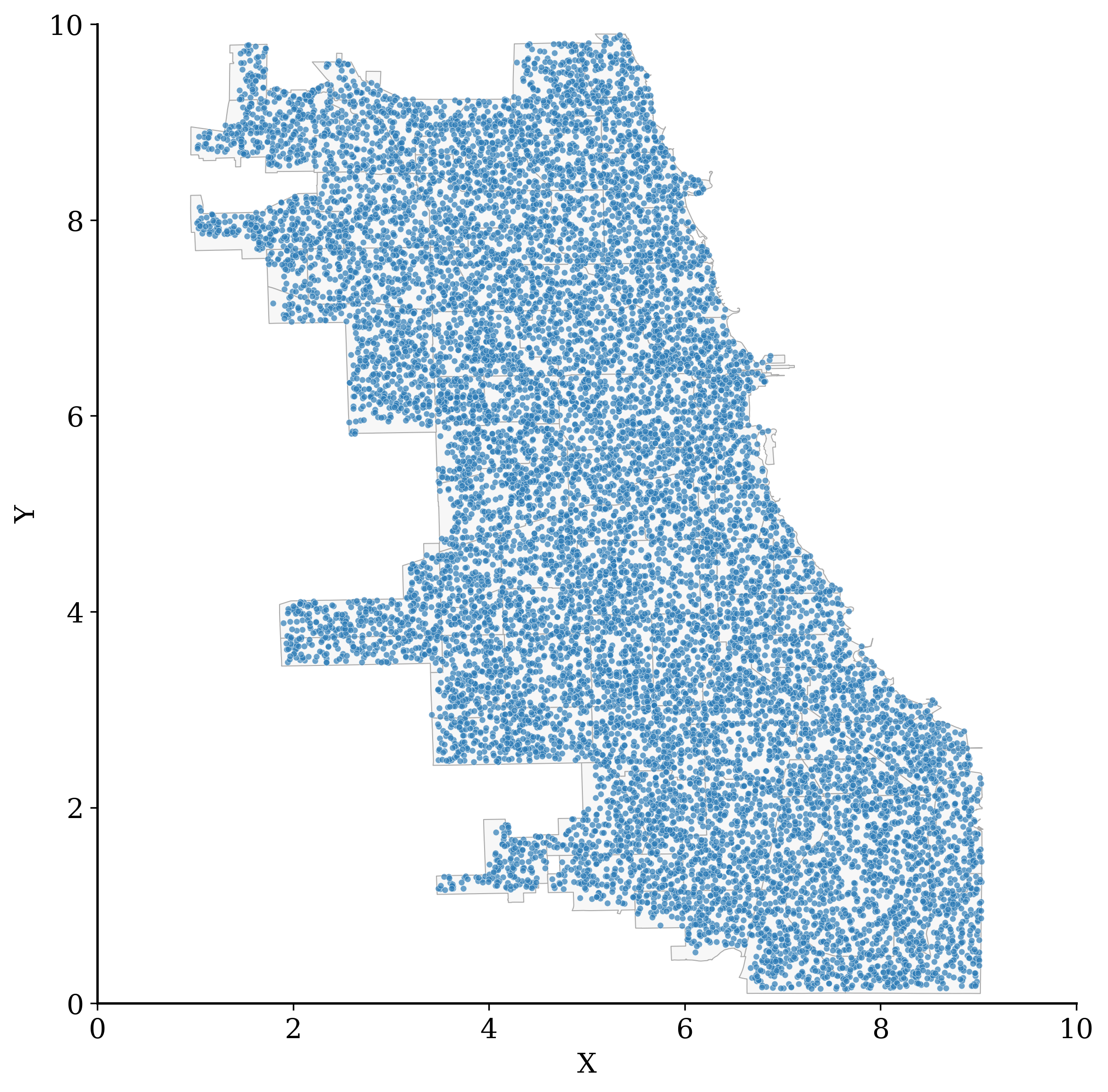}
\end{minipage}
\hfill
\begin{minipage}{0.48\textwidth}
\centering
\includegraphics[width=\linewidth]{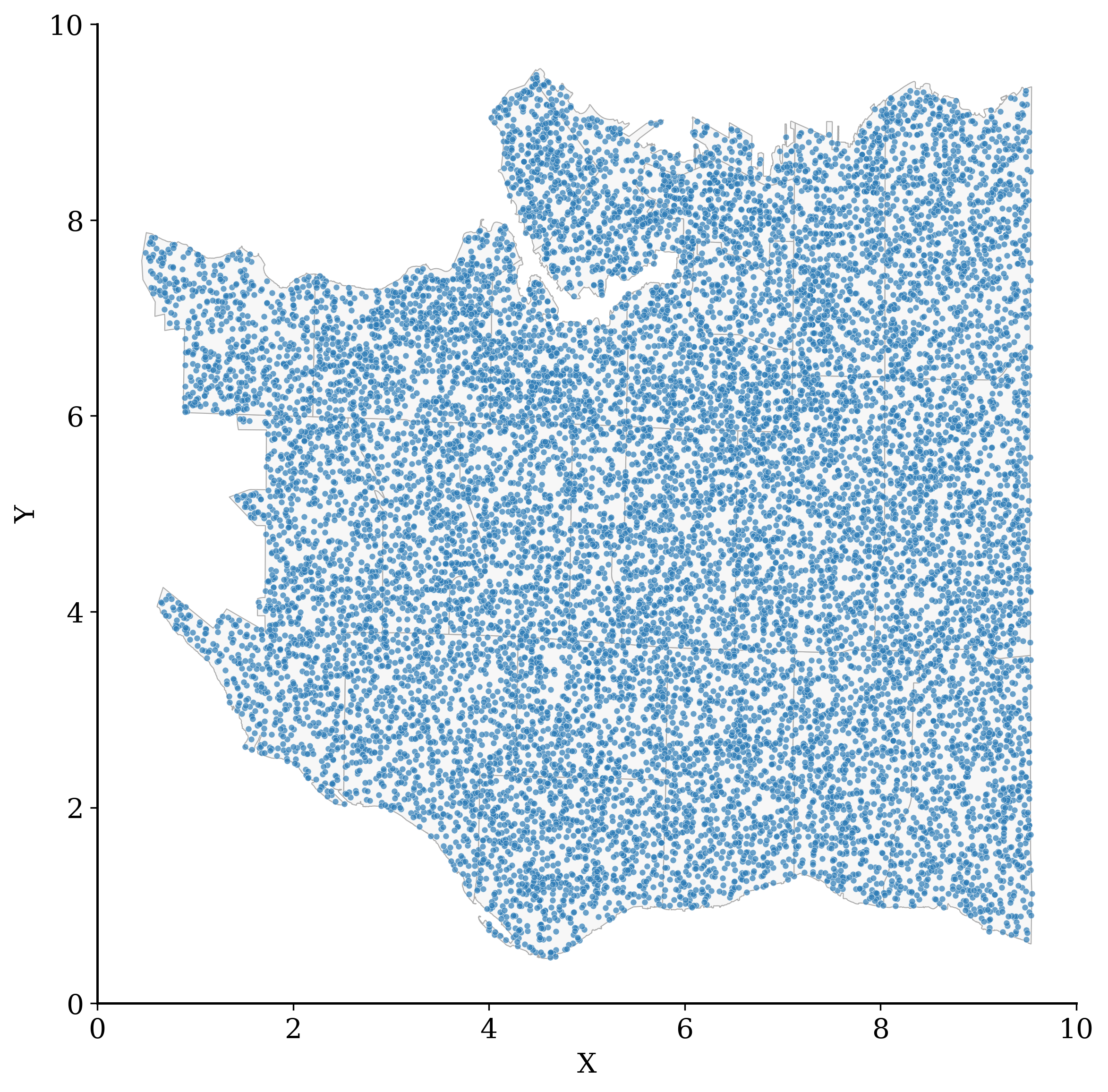}
\end{minipage}

\caption{Spatial distribution of the super-thinned events. 
Left: Chicago shooting incidents. 
Right: Vancouver break-ins.}
\label{fig:thinned_spatial}
\end{figure}

Our temporal diagnostics yield solid results. Kolmogorov-Smirnov (KS) tests confirm that the time between events follows the expected pattern. For the Vancouver dataset, 16 out of 20 random seeds passed the test ($p > 0.05$). For the Chicago dataset, the results were even more consistent, with 19 out of 20 seeds passing the test ($p > 0.05$). This high success rate across multiple independent runs suggests that our temporal modeling is stable and performs reliably.
Spatially, however, visual inspection of the super-thinned events (Figure~\ref{fig:thinned_spatial}) and subsequent quadrat count tests ($p < 0.05$) indicate the presence of residual localised clustering. This suggests that while our model effectively captures the temporal contagion and the primary spatial patterns, certain localised spatial heterogeneities remain unmodeled. This potential remaining clustering is likely driven by external spatial covariates that are not included in our model.

Despite the minor spatial heterogeneities identified in the diagnostics, our Bayesian framework provides the following interpretable insights into the overall generating mechanisms of these events. Note that in what follows, all expected event counts and their CIs are rounded to the nearest integer. For the Chicago Shootings dataset, the variational posterior mean of the expected number of background events is $\|\mu\|_1 = 1398$ (95\% CI: $[1300, 1500]$), and the branching ratio is $\|\phi\|_1 = 0.324$ (95\% CI: $[0.278, 0.382]$). Using the branching process identity $\mathbb{E}[N(\mathcal{W})] \approx \|\mu\|_1 / (1 - \|\phi\|_1)$ (see Appendix~\ref{app:expected_events} for the derivation), the implied expected number of total events is $2073$ (95\% CI: $[1876, 2314]$). This aligns well with both the numerical compensator estimate of $2131$ (95\% CI: $[1992, 2292]$) and the actual observed number of training events (2,124). With an estimated branching ratio of 0.324, we find that nearly a third of these shootings are directly triggered by previous incidents within this window, highlighting a strong short-term contagion in urban gun violence.

Similarly, for the Vancouver Break-ins dataset, the variational posterior mean of the expected number of background events is $\|\mu\|_1 = 2213$ (95\% CI: $[2090, 2346]$), with a branching ratio of $\|\phi\|_1 = 0.210$ (95\% CI: $[0.176, 0.248]$). The branching process identity yields an implied expected number of events of $2804$ (95\% CI: $[2610, 3019]$), while the compensator yields $2904$ (95\% CI: $[2741, 3078]$). Again, the true observed number of training events (3,013) falls well within these CIs. Practically, a 21\% branching ratio indicates that within the chosen spatial-temporal window, localised near-repeat dynamics are responsible for roughly one in five break-ins.

To further understand the behaviour of the additive kernel of the GP, we examine the variational posterior means and 95\% CIs of the hyperparameters for the additive kernel, which decomposes the underlying Gaussian processes into purely temporal, purely spatial, and joint spatio-temporal components.

\begin{table}[h!]
\centering
\caption{Variational posterior means and 95\% CIs (in brackets) for the additive kernel hyperparameters. $\ell$ denotes the length scale and $\sigma^2$ denotes the signal variance for the temporal ($t$), spatial ($s$), and spatio-temporal ($t,s$) components.}
\label{tab:hyperparameters}
\scriptsize 
\setlength{\tabcolsep}{3pt}
\resizebox{\textwidth}{!}{
\begin{tabular}{ll ccc ccc}
\toprule
& & \multicolumn{3}{c}{\textbf{Length Scales} ($\ell$)} & \multicolumn{3}{c}{\textbf{Signal Variance} ($\sigma^2$)} \\
\cmidrule(lr){3-5} \cmidrule(lr){6-8}
\textbf{Dataset} & \textbf{Component} & $t$ & $s$ & $(t,s)$ & $t$ & $s$ & $(t,s)$ \\
\midrule
\multirow{2}{*}{Chicago} & Background $\mu$ & 2.73 [1.46, 4.63] & 2.07 [1.91, 2.23] & 0.28 [0.12, 0.53] & 0.82 [0.15, 2.41] & 6.10 [3.84, 9.20] & 0.47 [0.31, 0.67] \\
\addlinespace[2pt]
& Trigger $\phi$ & 0.62 [0.11, 1.87] & 0.04 [0.03, 0.05] & 0.20 [0.16, 0.24] & 0.26 [0.03, 0.89] & 0.74 [0.43, 1.19] & 0.93 [0.61, 1.37] \\
\midrule
\multirow{2}{*}{Vancouver} & Background $\mu$ & 4.90 [2.24, 10.07] & 1.23 [1.15, 1.31] & 1.44 [1.15, 1.75] & 2.78 [1.05, 6.04] & 7.10 [4.57, 10.95] & 1.44 [0.95, 2.13] \\
\addlinespace[2pt]
& Trigger $\phi$ & 0.50 [0.06, 2.02] & 0.02 [0.01, 0.02] & 0.13 [0.10, 0.16] & 0.39 [0.11, 1.08] & 4.24 [3.12, 5.68] & 2.36 [1.59, 3.32] \\
\bottomrule
\end{tabular}
}
\end{table}

While additive kernels offer interpretable decompositions~\citep{duvenaud2011additive} by measuring the signal variance ($\sigma^2$) of each term, they can suffer from identifiability issues \citep{lu2022additive}. To address this, Orthogonal Additive Kernels (OAK) were proposed \citep{lu2022additive}. However, we do not employ OAK in this work, as the need to specify marginal data densities for each dimension significantly complicates the modeling and increases the computational burden. Instead, we verify the interpretability of our model through empirical testing. Across 20 random seeds, the learned hyperparameters for the Chicago dataset were very similar. For the Vancouver dataset, all hyperparameters remained stable except for one temporal scale ($\ell_{t}$ for $\mu$, which showed two modes). This suggests that our model converges to reliable optima rather than random local optima. Therefore, given the stability of our results, our model's decomposition remains identifiable and interpretable.

\begin{figure}
\centering
\includegraphics[width=1.0\textwidth]{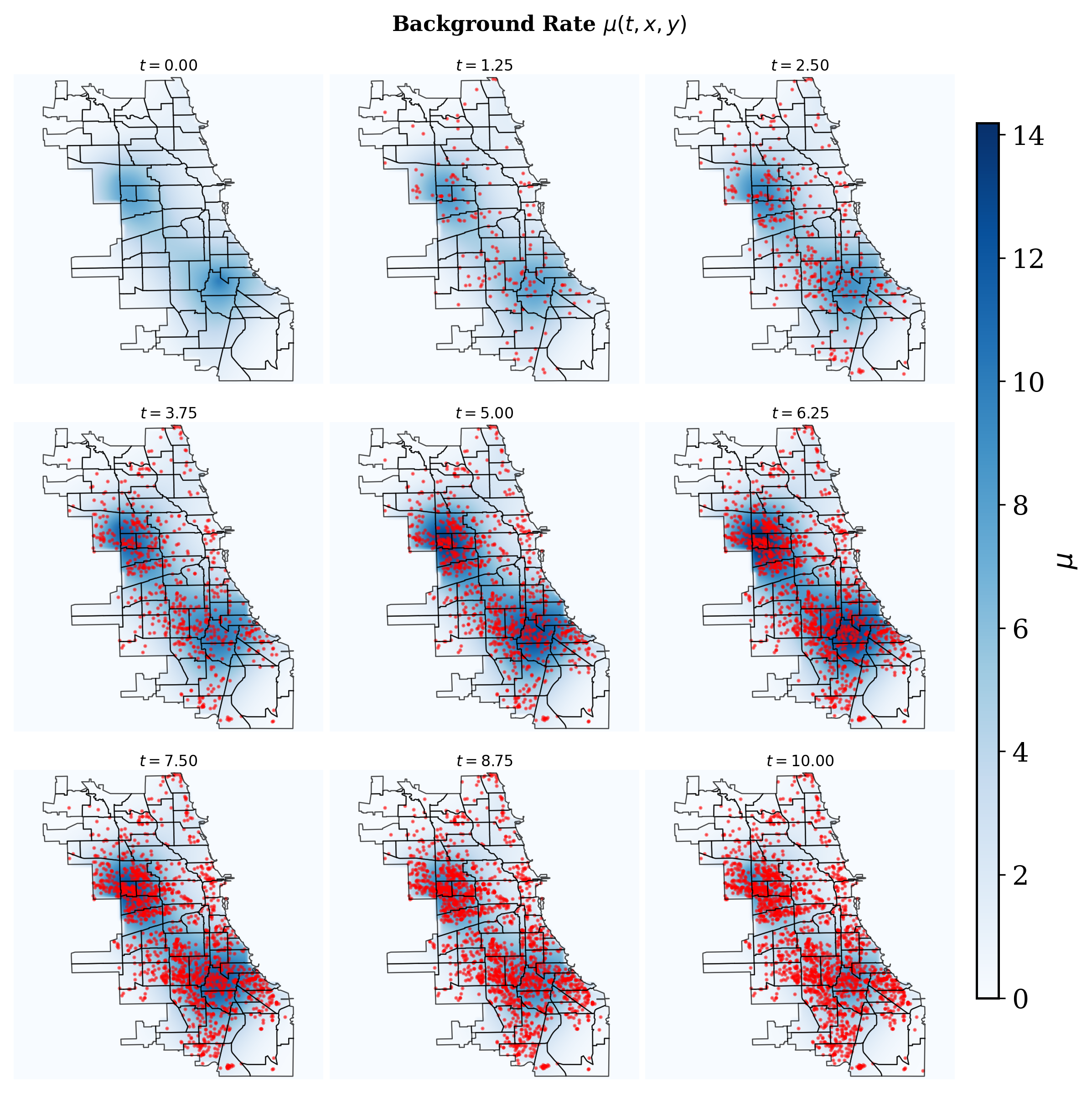}
\caption{Variational posterior mean on the background rate at selected timestamps for the Chicago shootings dataset (2022-2023)
%identifying exogenous spatial hotspots
.}
\label{fig:chicago_u}
\end{figure}

\begin{figure}
\centering
\includegraphics[width=1.0\textwidth]{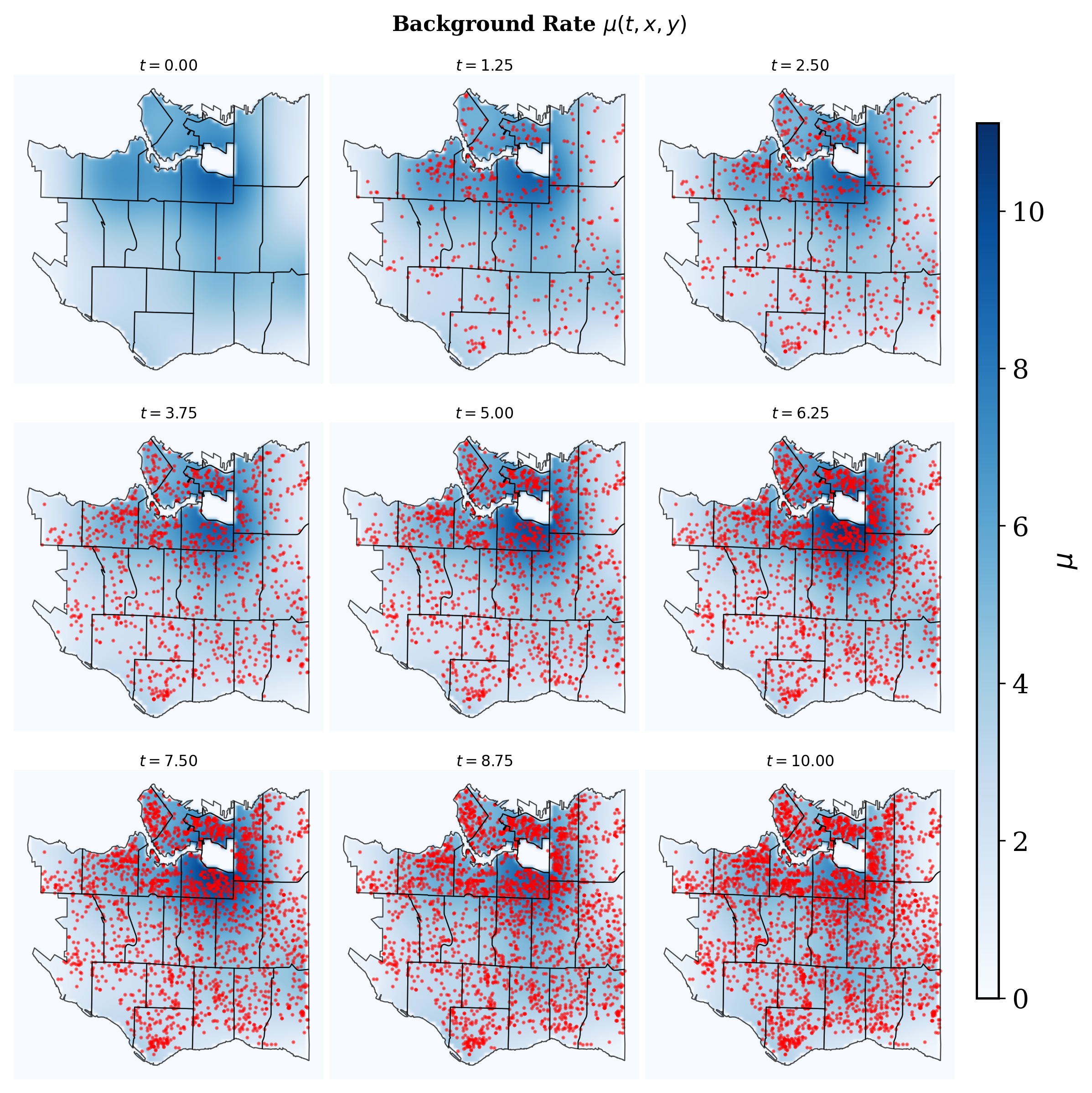}
\caption{Variational posterior mean  on the background rate at selected timestamps for the Vancouver break-ins dataset %, isolating locations with intrinsically high crime bases
.}
\label{fig:vancouver_u}
\end{figure}

\textbf{Background Rate.} When looking at the estimated background rate surfaces ($\mu$) (Figure~\ref{fig:chicago_u} and Figure~\ref{fig:vancouver_u}), we observe distinct spatial hotspots that correspond to areas with persistently high crime baseline rates. For Chicago, these hotspots are concentrated in certain neighbourhoods on the South and West sides, while Vancouver shows elevated background rates in specific downtown area. As shown in Table~\ref{tab:hyperparameters}, In Chicago, the spatial factor is statistically robust; its 95\% CI lower bound ($3.84$) comfortably exceeds the upper bounds of both the temporal ($2.41$) and spatio-temporal ($0.67$) components. This confirms that Chicago's baseline crime risks are largely persistent and tied to long-term locational factors rather than temporal fluctuations. For Vancouver, while the spatial variance remains the largest ($\sigma^2_s = 7.10$), its 95\% CI overlaps with the temporal component ($\sigma^2_t$, upper bound $6.04$). This suggests that while location is the strongest driver, the baseline risk in Vancouver is also influenced by broader temporal trends. Overall, these results indicate that while spatial persistence is a common feature, the temporal stability of the background rate is more noticeable in the Chicago shootings dataset than in Vancouver break-ins.

\textbf{Triggering Kernel.} The posterior means of the self-excitation triggering kernels ($\phi$) for the two datasets are shown in Figure~\ref{fig:chicago_phi} and Figure~\ref{fig:vancouver_phi} in Appendix~\ref{app:triggering_kernel_real_world}. 
We observe that the triggering effects are strongly centered in space and time in both cases. 
Table~\ref{tab:hyperparameters} further highlights clear differences between shootings and break-ins. In Chicago, the triggering effect is mostly driven by the joint space-time component ($\sigma^2_{st} = 0.93$, 95\% CI: $[0.61, 1.37]$). This means that retaliatory violence happens within a very specific window of both time and space. On the other hand, the spread of Vancouver break-ins is mostly driven by the pure spatial component ($\sigma^2_s = 4.24$, 95\% CI: $[3.12, 5.68]$). Notably, despite Vancouver having a smaller physical triggering window than Chicago, it exhibits a significantly smaller spatial length scale ($\ell_s = 0.02$, 95\% CI: $[0.01, 0.02]$ vs. Chicago's $0.04$, 95\% CI: $[0.03, 0.05]$), indicating an even more localised concentration of risk. This reflects the ``near-repeat'' pattern inherent in burglaries: once a property is targeted, the immediate neighbors face a higher risk that is tied more to precise proximity than to a restrictive time window. 

These findings empirically confirm that crime in the aforementioned datasets propagates similarly to epidemic processes, meaning crime triggers more crime. The highly localised spatial length scales for both datasets underscore that crime contagion operates within a very restricted microscopic radius. This suggests that future work focusing on interventions should treat crime spikes not merely as static non-contagious clusters, but as dynamic contagions, aiming to decrease the spread in space and time through rapid, geographically targeted policing.

\begin{table}[ht!]
    \centering
    \tbl{Performance comparison (Expected Log-Likelihood) on real-world datasets.}
    {\small\addtolength{\tabcolsep}{-1.1pt}\begin{tabular}{llccccc} 
        \toprule
        & \multicolumn{2}{c}{\textbf{Chicago Shootings}} & & \multicolumn{2}{c}{\textbf{Vancouver Break-ins}} \\
        \cmidrule(lr){2-3} \cmidrule(lr){5-6}
        Method & Link & $ELL$ & & Link & $ELL$ \\ 
        \midrule
        Parametric Hawkes Process& -- & $1471.17$ && -- & $2203.65$ \\ 
        Log Gaussian Cox Process& exp & $1535.49$ && exp & $2000.43$ \\ 
        Cox-Hawkes Process& exp & $1647.26$ && exp & $2157.32$ \\ 
        \midrule
        Ours (separable RBF) & softplus-softplus & $1615.71$ && softplus-softplus & $2209.45$ \\ 
        Ours (separable RBF) & sigmoid-softplus & $1642.24$ && sigmoid-softplus & $2198.64$ \\ 
        Ours (separable RBF) & exp-softplus & $1643.36$ && exp-softplus & $2188.72$ \\
        \midrule
        Ours (separable Mat\'{e}rn, $\nu=2.5$) & softplus-softplus & $1622.09$ && softplus-softplus & $2220.85$ \\ 
        Ours (separable Mat\'{e}rn, $\nu=2.5$) & sigmoid-softplus & $1645.43$ && sigmoid-softplus & $2205.70$ \\
        Ours (separable Mat\'{e}rn, $\nu=2.5$) & exp-softplus & $1651.15$ && exp-softplus & $2193.20$ \\
        \midrule
        Ours (additive RBF) & softplus-softplus & $1616.90$ && softplus-softplus & $\color{red}2249.02$ \\ 
        Ours (additive RBF) & sigmoid-softplus & $1656.78$ && sigmoid-softplus & $2245.78$ \\ 
        Ours (additive RBF) & exp-exp & $\color{red}1670.68$ && exp-softplus & $2227.90$ \\ 
        \midrule
        Ours (additive Mat\'{e}rn, $\nu=2.5$) & softplus-softplus & $1621.99$ && softplus-softplus & $\color{blue}2247.74$ \\ 
        Ours (additive Mat\'{e}rn, $\nu=2.5$) & sigmoid-softplus & $1655.94$ && sigmoid-softplus & $2241.56$ \\
        Ours (additive Mat\'{e}rn, $\nu=2.5$) & exp-exp & $\color{blue}1670.64$ && exp-softplus & $2215.98$ \\
        \bottomrule
    \end{tabular}}
    \label{tab:real_world_results}
    \begin{tabnote}
    Higher $ELL$ indicates superior performance. Best and second-best results in terms of $ELL$ are highlighted in red and blue, respectively.
    \end{tabnote}
\end{table}

\clearpage
\section{Discussion}
\label{sec:conclusion}

We have proposed a fully Bayesian nonparametric framework for spatio-temporal Hawkes processes, where both the background rate and the triggering kernel are modeled using Gaussian processes with flexible covariance kernels. By leveraging variational inference and sparse GP approximations, our method scales to large spatio-temporal point patterns while retaining uncertainty quantification. Through comprehensive experiments on synthetic data with varying complexity, we demonstrated that our model can accurately recover complex spatial and temporal excitation patterns without imposing restrictive parametric assumptions. On real-world crime datasets (Chicago shootings and Vancouver break-ins), our model consistently outperforms existing parametric baselines, achieving higher $ELL$ and providing interpretable insights into the self-exciting nature of criminal activities.

Our key contributions can be summarised as follows: (i) we introduce a novel fully Bayesian nonparametric spatio-temporal Hawkes process model that flexibly captures both background rate and triggering kernel; (ii) we develop an efficient variational inference algorithm with sparse GP approximations that scales to high-dimensional settings; (iii) we demonstrate superior performance over state-of-the-art baselines on both synthetic and real-world datasets; and (iv) we provide interpretable estimates of branching ratios and spatial hotspots that offer actionable insights for crime analysis.

Despite these strengths, several limitations remain. First, the current model assumes a single event type and may not capture interactions between different event categories. Second, our approach does not explicitly incorporate exogenous covariates that may influence event rates. Thus, there exist several promising directions for future work. First, extending the framework to multivariate Hawkes processes with both self-excitation and cross-excitation between different event types would enable more nuanced modelling of complex phenomena such as the interaction between different crime categories or between crimes and their responses. Second, including external covariates in the model would allow us to better account for environmental factors and provide more detailed insights. Third, exploring alternative nonparametric constructions such as neural network-based intensity estimators could further enhance model flexibility. 

In conclusion, the proposed framework represents a significant advance in flexible and scalable Bayesian nonparametric modelling of spatio-temporal Hawkes processes, offering both methodological contributions and practical insights for crime analysis and beyond.

\section*{Acknowledgements}
We acknowledge access to Piz Daint or Alps at the Swiss National Supercomputing Centre, Switzerland under the Università della Svizzera italiana's share with the project ID u0.

XM has received funding from the European Union's Horizon Europe research and innovation programme under the Marie Sklodowska-Curie grant agreement 101151781 and from the University of Cyprus under
the startup grant programme.

\clearpage
\appendix
\section*{Appendix}
\section{Derivation of the Expected Total Number of Events}
\label{app:expected_events}
To derive the expected total number of events $\mathbb{E}[N(\mathcal{W})]$ using the branching process perspective, we can decompose the Hawkes process into discrete generations of events. Let $N_k$ denote the number of events in the $k$-th generation within the spatial-temporal domain $\mathcal{W} = \mathcal{T} \times \mathcal{S}$.

First, the $0$-th generation consists solely of background events (i.e., immigrants), which are generated independently by the background rate $\mu(t, s)$. The expected number of these background events is the integral of the background rate over the domain:
\begin{equation}
    \mathbb{E}[N_0] = \int_{\mathcal{W}} \mu(t, s) \, \mathrm{d}t \, \mathrm{d}s = \|\mu\|_1.
\end{equation}

Next, each event in the $0$-th generation acts as a parent, generating the $1$-st generation of offspring events. According to the definition of the triggering kernel, the expected number of offspring generated by a single event is given by the branching ratio $\|\phi\|_1$. Thus, the expected number of $1$-st generation events is:
\begin{equation}
    \mathbb{E}[N_1] = \mathbb{E}[N_0] \|\phi\|_1 = \|\mu\|_1 \|\phi\|_1.
\end{equation}

Following this branching mechanism, the total number of events in the $k$-th generation, $N_k$, is the sum of offspring generated by the $N_{k-1}$ parent events from the previous generation. Given $N_{k-1}$, the conditional expectation of $N_k$ is $\mathbb{E}[N_k \mid N_{k-1}] = N_{k-1} \|\phi\|_1$, because each parent independently generates offspring with an expected count of $\|\phi\|_1$. By applying the law of total expectation, we can express the unconditional expectation of $N_k$ as:
\begin{equation}
    \mathbb{E}[N_k] = \mathbb{E} \big[ \mathbb{E}[N_k \mid N_{k-1}] \big] = \mathbb{E} \big[ N_{k-1} \|\phi\|_1 \big] = \mathbb{E}[N_{k-1}] \|\phi\|_1.
\end{equation}

By mathematical induction, applying this recursive relationship starting from $\mathbb{E}[N_0] = \|\mu\|_1$, the expected number of events in the $k$-th generation is:
\begin{equation}
    \mathbb{E}[N_k] = \|\mu\|_1 \|\phi\|_1^k.
\end{equation}

The total number of events in the domain $\mathcal{W}$ is the sum of events across all generations, given by $N(\mathcal{W}) = \sum_{k=0}^{\infty} N_k$. Taking the expectation on both sides yields:
\begin{align}
    \mathbb{E}[N(\mathcal{W})] &= \mathbb{E} \left[ \sum_{k=0}^{\infty} N_k \right] \nonumber \\
    &= \sum_{k=0}^{\infty} \mathbb{E}[N_k] \nonumber \\
    &= \sum_{k=0}^{\infty} \|\mu\|_1 \|\phi\|_1^k \nonumber \\
    &= \|\mu\|_1 \sum_{k=0}^{\infty} \|\phi\|_1^k.
\end{align}

Under the stationarity condition $\|\phi\|_1 < 1$, the infinite geometric series converges to $1 / (1 - \|\phi\|_1)$. Assuming the temporal and spatial boundaries are sufficiently large such that boundary truncation effects are negligible, we arrive at the final approximation:
\begin{equation}
    \mathbb{E}[N(\mathcal{W})] \approx \frac{\|\mu\|_1}{1 - \|\phi\|_1}.
\end{equation}

\vspace*{-10pt}
\section{Kernel Functions}
\label{sec:appendix_kernels}

In this section, we provide the detailed mathematical forms of the covariance kernels. We denote $t, t' \in [0, T]$ as temporal coordinates and $s, s' \in [0, X] \times [0, Y]$ as spatial coordinates, with $s=(x,y)$ and $s'=(x',y')$. Following the definitions in Section~\ref{sec:gp_prior}, each base kernel $k(z, z')$ is parameterised by a signal variance $\sigma^2 > 0$ and lengthscale $\ell > 0$.

\paragraph{Radial Basis Function (RBF) Kernel:}
The RBF kernel between two points $z$ and $z'$ is:
\begin{equation}
    k_{\text{RBF}}(z, z' \mid \sigma^2, \ell) = \sigma^2 \exp\left( - \frac{\|z - z'\|^2}{2\ell^2} \right).
\end{equation}

\paragraph{Mat\'{e}rn Kernels:}
The Mat\'{e}rn kernel with smoothness parameter $\nu \in \{0.5, 1.5, 2.5\}$ is:
\begin{equation}
    k_{\text{Mat}}(z, z' \mid \sigma^2, \nu, \ell) = \begin{cases} 
        \sigma^2 \exp\left( - \frac{\|z - z'\|}{\ell} \right), & \nu = 0.5, \\
        \sigma^2 \left( 1 + \frac{\sqrt{3}\|z - z'\|}{\ell} \right) \exp\left( - \frac{\sqrt{3}\|z - z'\|}{\ell} \right), & \nu = 1.5, \\
        \sigma^2 \left( 1 + \frac{\sqrt{5}\|z - z'\|}{\ell} + \frac{5\|z - z'\|^2}{3\ell^2} \right) \exp\left( - \frac{\sqrt{5}\|z - z'\|}{\ell} \right), & \nu = 2.5.
    \end{cases}
\end{equation}

\subsection{Joint Kernel Forms}

Using the base kernels above, we define the specific joint spatio-temporal structures.

\paragraph{Separable Joint Kernel:}
The separable kernel factorises the dependencies into independent marginal effects, incorporating Automatic Relevance Determination (ARD)~\citep{liu2020gaussian} over the temporal and spatial dimensions:
\begin{equation}
    K_{\text{separable}}\bigl((t,s), (t',s')\bigr)
    = \sigma^2\, k_t(t, t' \mid 1, \ell_t)\, k_s(s, s' \mid 1, \ell_x, \ell_y),
\end{equation}
where $k_t$ and $k_s$ are base kernels defined over the temporal and spatial domains, respectively, and $s=(x,y)$.

A common choice is the squared exponential kernel:
\begin{equation}
K_{\text{separable}}((t,s),(t',s')) = \sigma^2 \exp\!\left(
-\frac{(t-t')^2}{2\ell_t^2}
-\frac{(x-x')^2}{2\ell_x^2}
-\frac{(y-y')^2}{2\ell_y^2}
\right).
\end{equation}

Alternatively, with a Mat\'ern-3/2 kernel, let
\[
s=\left(\frac{x}{\ell_x}, \frac{y}{\ell_y}\right), \qquad
s'=\left(\frac{x'}{\ell_x}, \frac{y'}{\ell_y}\right).
\]
Then
\begin{equation}
K_{\text{separable}}((t,s),(t',s')) = \sigma^2 
\left(1 + \frac{\sqrt{3}|t-t'|}{\ell_t}\right)
\exp\!\left(-\frac{\sqrt{3}|t-t'|}{\ell_t}\right)
\left(1 + \sqrt{3}\,\|s-s'\|\right)
\exp\!\left(-\sqrt{3}\,\|s-s'\|\right).
\end{equation}

\paragraph{Additive Joint Kernel:}
The additive kernel sums marginal temporal ($k_t$), marginal spatial ($k_s$), and joint interacton ($k_{t,s}$) components:
\begin{equation}
    K_{\text{additive}}\bigl((t,s), (t',s')\bigr) = k_t(t, t' \mid \sigma_t^2, \ell_t) + k_s(s, s' \mid \sigma_s^2, \ell_s) + k_{t,s}\bigl((t,s), (t',s') \mid \sigma_{t,s}^2, \ell_{t,s}\bigr),
\end{equation}
where $k_s$ is a 2D isotropic spatial kernel and $k_{t,s}$ is a 3D isotropic spatio-temporal kernel.

\clearpage

\section{Likelihood derivation}
\label{app:LH_derivation}

The likelihood $p(\mathcal{D} \mid \mu, \phi)$ for a spatio-temporal Hawkes process is defined as:
\begin{equation}
L = \exp \left( -\int_{0}^{T} \int_{0}^{X} \int_{0}^{Y} \lambda(t,x,y \mid \mathcal{H}_t) \, \rm{d}t \rm{d}x \rm{d}y \right) \prod_{i=1}^{n} \lambda(t_i,x_i,y_i \mid \mathcal{H}_{t_i}).
\end{equation}
By substituting the conditional intensity $\lambda(t,x,y \mid \mathcal{H}_t) = \mu(t,x,y) + \sum_{j: t_j < t} \phi(t-t_j, x-x_j, y-y_j)$, the product over observed events is expressed as:
\begin{equation}
\prod_{i=1}^{n} \left( \mu(t_i,x_i,y_i) + \sum_{j: t_j < t_i} \phi(\Delta t_{ij}, \Delta x_{ij}, \Delta y_{ij}) \right).
\end{equation}
The compensator term $\Lambda =\iiint_{\mathcal{W}} \lambda(t,x,y) \, \rm{d}t \rm{d}x \rm{d}y$ can be decomposed into background and triggering components, $\Lambda_\mu + \Lambda_\phi$. By swapping the order of summation and integration for the triggering part, and applying the change of variables $\Delta t = t-t_i$, $\Delta x = x-x_i$, and $\Delta y = y-y_i$, we obtain:
\begin{align}
\Lambda_\phi &= \sum_{i=1}^{n} \int_{t_i}^{T} \int_{0}^{X} \int_{0}^{Y} \phi(t-t_i, x-x_i, y-y_i) \, \rm{d}t \rm{d}x \rm{d}y \nonumber \\
&= \sum_{i=1}^{n} \int_{0}^{T-t_i} \int_{-x_i}^{X-x_i} \int_{-y_i}^{Y-y_i} \phi(\Delta t, \Delta x, \Delta y) \, \rm{d}(\Delta t)\, \rm{d}(\Delta x)\, \rm{d}(\Delta y).
\end{align}
Under the triggering kernel support constraints $(T_\phi, X_\phi, Y_\phi)$, the effective integration limits for each event $i$ become:
\begin{itemize}
    \item $\Delta t \in [0, \min(T_\phi, T-t_i)]$
    \item $\Delta x \in [\max(-X_\phi, -x_i), \min(X_\phi, X-x_i)]$
    \item $\Delta y \in [\max(-Y_\phi, -y_i), \min(Y_\phi, Y-y_i)]$
\end{itemize}

Combining these components leads to the final likelihood expression presented in Eq.~\eqref{eq:full_likelihood}.

\vspace*{-10pt}
\section{Variational inference}
\label{app:VI_derivation}

VI approximates the intractable posterior $p(\theta \mid D)$ with a distribution $q(\theta)$ from a tractable family $\mathcal{Q}$ that minimises the KL divergence:
\begin{equation}
    q^*(\theta) = \arg \min_{q \in \mathcal{Q}} \text{KL}(q(\theta) \,\|\, p(\theta \mid D)) = \int q(\theta) \log \frac{q(\theta)}{p(\theta \mid D)} \, d\theta.
\end{equation}

Using Bayes' rule $p(\theta \mid D) = p(D, \theta) / p(D)$, the KL divergence can be expanded as:
\begin{align}
    \text{KL}(q(\theta) \,\|\, p(\theta \mid D)) &= \int q(\theta) \left[ \log \frac{q(\theta)}{p(D, \theta)} + \log p(D) \right] d\theta \nonumber \\
    &= \int q(\theta) \log \frac{q(\theta)}{p(D, \theta)} \, d\theta + \log p(D).
\end{align}
Re-arranging terms yields the decomposition of the marginal log-likelihood:
\begin{equation}
    \log p(D) = \mathcal{L}(q) + \text{KL}(q(\theta) \,\|\, p(\theta \mid D)),
\end{equation}
where $\mathcal{L}(q)$ is the ELBO:
\begin{equation}
    \mathcal{L}(q) = \mathbb{E}_q[\log p(D, \theta)] - \mathbb{E}_q[\log q(\theta)] = \mathbb{E}_q[\log p(D \mid \theta)] - \text{KL}(q(\theta) \,\|\, p(\theta)).
\end{equation}

Since $\text{KL}(q \,\|\, p) \geq 0$, the ELBO provides a rigorous lower bound $\log p(D) \geq \mathcal{L}(q)$. Maximizing $\mathcal{L}(q)$ with respect to $q$ is thus equivalent to minimizing the KL divergence to the posterior.

\section{Derivation of the Variational Marginal Distribution $q(f_\mu)$}
\label{app:sparse_gp_derivation}

Given the conditional distribution $p(f_\mu \mid \mathbf{u}_\mu) = \mathcal{N}(m^{f \mid u}_\mu, \Sigma^{f \mid u}_\mu)$ and the variational posterior over the inducing variables $q(\mathbf{u}_\mu) = \mathcal{N}(\mu_\mu, S_\mu)$, the marginal distribution $q(f_\mu)$ is obtained by integrating out $\mathbf{u}_\mu$:
\begin{equation}
    q(f_\mu) = \int p(f_\mu \mid \mathbf{u}_\mu) q(\mathbf{u}_\mu) \, \mathrm{d}\mathbf{u}_\mu.
\end{equation}
Since this is a linear Gaussian system, the resulting marginal distribution $q(f_\mu)$ is also Gaussian, denoted as $\mathcal{N}(\tilde{m}_\mu, \tilde{\Sigma}_\mu)$. We can derive its mean and covariance using the law of total expectation and the law of total variance, respectively.

\paragraph{Derivation of the Mean $\tilde{m}_\mu$:}
By the law of total expectation, the unconditional mean is the expectation of the conditional mean:
\begin{align}
    \tilde{m}_\mu &= \mathbb{E}_{q(\mathbf{u}_\mu)} \left[ \mathbb{E}[f_\mu \mid \mathbf{u}_\mu] \right] \nonumber \\
    &= \mathbb{E}_{q(\mathbf{u}_\mu)} \left[ K_{ZZ_\mu} K_{Z_\mu Z_\mu}^{-1} \mathbf{u}_\mu \right].
\end{align}
Since $K_{ZZ_\mu} K_{Z_\mu Z_\mu}^{-1}$ is a deterministic matrix independent of $\mathbf{u}_\mu$, we can factor it out of the expectation:
\begin{equation}
    \tilde{m}_\mu = K_{ZZ_\mu} K_{Z_\mu Z_\mu}^{-1} \mathbb{E}_{q(\mathbf{u}_\mu)}[\mathbf{u}_\mu] = K_{ZZ_\mu} K_{Z_\mu Z_\mu}^{-1} \mu_\mu.
\end{equation}

\paragraph{Derivation of the Covariance $\tilde{\Sigma}_\mu$:}
By the law of total variance, the unconditional variance is the sum of the expected conditional variance and the variance of the conditional mean:
\begin{equation}
    \tilde{\Sigma}_\mu = \underbrace{\mathbb{E}_{q(\mathbf{u}_\mu)} \left[ \text{Var}(f_\mu \mid \mathbf{u}_\mu) \right]}_{\text{Term 1}} + \underbrace{\text{Var}_{q(\mathbf{u}_\mu)} \left( \mathbb{E}[f_\mu \mid \mathbf{u}_\mu] \right)}_{\text{Term 2}}.
\end{equation}

For Term 1, the conditional covariance $\Sigma^{f \mid u}_\mu = K_{ZZ} - K_{ZZ_\mu} K_{Z_\mu Z_\mu}^{-1} K_{Z_\mu Z}$ is deterministic and does not depend on the random variable $\mathbf{u}_\mu$. Thus, its expectation is simply itself:
\begin{equation}
    \text{Term 1} = K_{ZZ} - K_{ZZ_\mu} K_{Z_\mu Z_\mu}^{-1} K_{Z_\mu Z}.
\end{equation}

For Term 2, we compute the variance of the linear transformation of $\mathbf{u}_\mu$. Using the property $\text{Var}(A\mathbf{x}) = A\text{Var}(\mathbf{x})A^\top$ and noting that $(K_{ZZ_\mu} K_{Z_\mu Z_\mu}^{-1})^\top = K_{Z_\mu Z_\mu}^{-1} K_{Z_\mu Z}$ due to the symmetry of the kernel matrices, we have:
\begin{align}
    \text{Term 2} &= \text{Var}_{q(\mathbf{u}_\mu)} \left( K_{ZZ_\mu} K_{Z_\mu Z_\mu}^{-1} \mathbf{u}_\mu \right) \nonumber \\
    &= \left( K_{ZZ_\mu} K_{Z_\mu Z_\mu}^{-1} \right) \text{Var}_{q(\mathbf{u}_\mu)}(\mathbf{u}_\mu) \left( K_{ZZ_\mu} K_{Z_\mu Z_\mu}^{-1} \right)^\top \nonumber \\
    &= K_{ZZ_\mu} K_{Z_\mu Z_\mu}^{-1} S_\mu K_{Z_\mu Z_\mu}^{-1} K_{Z_\mu Z}.
\end{align}

Summing Term 1 and Term 2 yields the final expression for the marginal covariance:
\begin{equation}
    \tilde{\Sigma}_\mu = K_{ZZ} - K_{ZZ_\mu} K_{Z_\mu Z_\mu}^{-1} K_{Z_\mu Z} + K_{ZZ_\mu} K_{Z_\mu Z_\mu}^{-1} S_\mu K_{Z_\mu Z_\mu}^{-1} K_{Z_\mu Z}.
\end{equation}
\clearpage

\section{Experimental Settings}
\label{app:experimental_settings}

This section details the evaluation metrics, hyperparameter specifications, optimisation protocol, and numerical configurations used for the sparse GP variational inference framework.

\subsection{Evaluation Metrics}
\label{sec:metrics}

To assess the performance of our model, we employ three metrics targeting point estimates, posterior uncertainty, and overall model fit:

\begin{itemize}
    \item \textbf{Posterior Mean Error ($PM_{\mathrm{mse}}$).} This metric quantifies the accuracy of the point estimates by calculating the average squared distance between the posterior mean and the ground truth:
    \begin{align}
    PM_{\mathrm{mse}}(\mu) &= \frac{1}{N_{\mu}} \sum_{i=1}^{N_{\mu}} (\hat{\mu}_i - \mu_i^\ast)^2, \\
    PM_{\mathrm{mse}}(\phi) &= \frac{1}{N_{\phi}} \sum_{j=1}^{N_{\phi}} (\hat{\phi}_j - \phi_j^\ast)^2,
    \end{align}
    where $\mu^\ast, \phi^\ast$ are ground truth values, and $\hat{\mu} = \mathbb{E}_q(\mu), \hat{\phi} = \mathbb{E}_q(\phi)$ are the posterior mean estimates under the variational posterior $q$.

    \item \textbf{Posterior Expected Error ($PE_{\mathrm{mse}}$).} The expected mean squared error calculated under the variational posterior $q$:
    \begin{align}
    PE_{\mathrm{mse}}(\mu) &= \frac{1}{N_{\mu}} \sum_{i=1}^{N_{\mu}} \mathbb{E}_q \left[ (\mu_i - \mu_i^\ast)^2 \right], \\
    PE_{\mathrm{mse}}(\phi) &= \frac{1}{N_{\phi}} \sum_{j=1}^{N_{\phi}} \mathbb{E}_q \left[ (\phi_j - \phi_j^\ast)^2 \right].
    \end{align}

    \item \textbf{Expected Log-Likelihood ($ELL$).} This metric measures the model's goodness-of-fit by calculating the expectation of the log-likelihood under the variational posterior $q$:
    \begin{equation}
    ELL = \mathbb{E}_{q} \left[ \sum_{(t_i, s_i) \in \mathcal{D}} \log \lambda(t_i, s_i) - \int_{0}^{T} \int_{\mathcal{S}} \lambda(t, s) \, ds \, dt \right],
    \end{equation}
    where $\mathcal{D} = \{(t_i, s_i)\}_{i=1}^n$ denotes the set of observed events.
\end{itemize}

\subsection{Generic Configurations}
\begin{itemize}
    \item \textbf{Gaussian Process Priors:} For the latent processes $f_\mu$ and $f_\phi$, we place Gamma priors on the lengthscales $\ell$ and signal variances $\sigma^2$ to ensure positivity.
    \item \textbf{Optimisation:} We employ Stochastic Variational Inference (SVI) by maximizing the ELBO. Optimisation is performed using the Adam optimiser \citep{kingma2014adam}.
    \item \textbf{Hardware:} All computations are implemented using \texttt{NumPyro} \citep{phan2019composable,bingham2019pyro} and executed on NVIDIA GH200 GPUs.
\end{itemize}

\subsection{Grid Resolution and Numerical Integration}
\label{sec:grid_and_integration}

To maintain consistency and computational stability across all experiments, we adopt a unified grid resolution for evaluating the intensity surfaces and performing the numerical integration of the Hawkes compensator. Specifically, we employ the trapezoidal rule for numerical integration over the following regular grids:
\begin{itemize}
    \item \textbf{Background rate ($\mu$):} The integration grid is set to $n_{\mu,t} = n_{\mu,x} = n_{\mu,y} = 70$.
    \item \textbf{Triggering kernel ($\phi$):} The integration grid is set to $n_{\phi,t} = n_{\phi,x} = n_{\phi,y} = 40$.
\end{itemize}

Regarding the sparse GP approximation, we place the inducing variables on equidistant grids across their respective spatio-temporal domains. For the background component, we use a total of $M_\mu = 4 \times 4 \times 4 = 64$ inducing points. Similarly, for the triggering component, we employ $M_\phi = 3 \times 3 \times 3 = 27$ inducing points.

\subsection{Synthetic Data Experiments}
\begin{itemize}
    \item \textbf{Evaluation Protocol:} We generate 8 independent realisations for each synthetic scenario. Due to the non-convex nature of the ELBO and the stochasticity in initial distributions, we perform 4 independent runs with different random initialisations for each realisation. The model achieving the best training ELBO (lowest final loss) is selected for reporting the final metrics.
    \item \textbf{Metrics:} We report the mean and standard deviation ($\text{mean} \pm \text{std}$) across the 8 realisations for the indicators defined in Section~\ref{sec:baselines_metrics}.
\end{itemize}

\subsection{Real-world Data Experiments}
\begin{itemize}
    \item \textbf{Evaluation Protocol:} For both the Chicago Shootings and Vancouver Break-ins datasets, we perform 20 independent SVI runs with different random seeds. Given the complexity of real-world event dynamics, a higher number of initialisations is used to avoid local optima. The model achieving the best final training ELBO is used for the evaluation on the held-out test set.
    \item \textbf{Metrics:} For real-world datasets, we report the performance metric from the model that achieved the best training loss among all initialisation attempts.
\end{itemize}

\clearpage
\section{Extended Results} \label{app:full_results}

\subsection{Synthetic Experiment Results for Scenario 1}

\begin{table}[ht!]
    \centering
    \tbl{Comparative analysis for Scenario 1 across various link functions and kernel smoothness specifications.}\label{tab:scenario1_full}
    {   \small\addtolength{\tabcolsep}{-1.1pt}\begin{tabular}{crrr|crrr}
        Method & Link & $PM_{\mathrm{mse}}(\mu)$ & $PM_{\mathrm{mse}}(\phi)$ & 
        & $PE_{\mathrm{mse}}(\mu)$ & $PE_{\mathrm{mse}}(\phi)$ & \\ 
        \midrule
        Parametric Hawkes & {} & $\color{red}{1.21 \pm 1.90}$& $\color{blue}4.49 \pm 2.26$ & & $\color{red}{1.90 \pm 1.87}$ & $\color{red}{3.33 \pm 0.47}$ & \\ 
        Cox-Hawkes & {}  & $4.63 \pm 2.76$ & $\color{red}2.92 \pm 1.66$ && $13.84 \pm 4.90$ & $\color{blue}4.42 \pm 1.66$ & \\ 
        \midrule
        Ours (separable RBF kernel) & {softplus-softplus}  & $8.13 \pm 2.50$ & $9.82 \pm 2.31$ && $14.95 \pm 2.73$ & $14.64 \pm 2.42$ &\\ 
        Ours (separable RBF kernel) & {sigmoid-sigmoid}   & $8.03 \pm 2.86$ & $8.16 \pm 2.26$ && $15.96 \pm 3.51$ & $15.23 \pm 2.12$ &\\ 
        Ours (separable RBF kernel) & {exp-exp}   & $8.48 \pm 2.50$ & $8.57 \pm 1.97$ && $16.29 \pm 2.85$ & $14.39 \pm 2.12$ &\\ 
        \midrule
        Ours (separable Mat\'{e}rn, $\nu=0.5$)& {softplus-softplus}    & $8.32 \pm 2.03$ & $9.82 \pm 3.34$ && $17.10 \pm 2.17$ & $14.49 \pm 3.11$ &\\ 
        Ours (separable Mat\'{e}rn, $\nu=0.5$)&{sigmoid-sigmoid}    & $9.54 \pm 2.36$ & $12.33 \pm 2.64$ && $19.78 \pm 2.59$ & $18.39 \pm 2.30$ &
\\ 
        Ours (separable Mat\'{e}rn, $\nu=0.5$)&{exp-exp}    & $11.29 \pm 2.79$ & $15.08 \pm 2.23$ && $22.21 \pm 3.08$ & $21.22 \pm 1.94$ &
\\ 
        \midrule
        Ours (separable Mat\'{e}rn, $\nu=1.5$)& {softplus-softplus}     & $6.75 \pm 1.99$ & $6.89 \pm 2.53$ && $14.15 \pm 2.20$ & $11.68 \pm 2.52$ &\\ 
        Ours (separable Mat\'{e}rn, $\nu=1.5$)&{sigmoid-sigmoid}     & $7.77 \pm 2.63$ & $7.26 \pm 2.30$ && $16.39 \pm 3.16$ & $13.26 \pm 2.19$ &
\\ 
        Ours (separable Mat\'{e}rn, $\nu=1.5$)&{exp-exp}     & $8.62 \pm 2.66$ & $8.79 \pm 2.36$ && $17.60 \pm 3.10$ & $14.62 \pm 2.12$ &
\\ 
        \midrule
        Ours (separable Mat\'{e}rn, $\nu=2.5$)& {softplus-softplus}      & $6.58 \pm 2.08$ & $7.36 \pm 2.69$ && $13.27 \pm 2.24$ & $12.03 \pm 2.64$ &\\ 
        Ours (separable Mat\'{e}rn, $\nu=2.5$)&{sigmoid-sigmoid}      & $7.34 \pm 2.61$ & $7.04 \pm 2.32$ && $15.13 \pm 3.16$ & $13.35 \pm 2.28$ &
\\ 
        Ours (separable Mat\'{e}rn, $\nu=2.5$)&{exp-exp}      & $8.11 \pm 2.69$ & $8.12 \pm 2.09$ && $16.18 \pm 3.19$ & $14.01 \pm 1.99$ &
\\ 
        \midrule
        Ours (additive RBF kernel) & {softplus-softplus}   & $2.88 \pm 0.88$ & $5.01 \pm 1.36$ && $\color{blue}7.63 \pm 0.82$ & $10.16 \pm 1.47$ &\\ 
        Ours (additive RBF kernel) & {sigmoid-sigmoid}    & $3.43 \pm 1.20$ & $7.91 \pm 2.93$ && $9.75 \pm 1.50$ & $15.43 \pm 3.28$ &\\ 
        Ours (additive RBF kernel) & {exp-exp}    & $4.78 \pm 2.22$ & $7.26 \pm 3.46$ && $13.69 \pm 4.23$ & $14.08 \pm 3.78$ &\\ 
        \midrule
        Ours (additive Mat\'{e}rn, $\nu=0.5$)& {softplus-softplus}    & $4.10 \pm 1.49$ & $7.20 \pm 2.05$ && $9.83 \pm 2.16$ & $12.66 \pm 2.06$ &\\ 
        Ours (additive Mat\'{e}rn, $\nu=0.5$)&{sigmoid-sigmoid}     & $4.06 \pm 1.09$ & $11.19 \pm 2.39$ && $9.84 \pm 0.82$ & $17.67 \pm 2.22$ &
\\ 
        Ours (additive Mat\'{e}rn, $\nu=0.5$)&{exp-exp}     & $5.96 \pm 2.58$ & $12.74 \pm 2.38$ && $13.82 \pm 4.91$ & $19.48 \pm 2.26$ &
\\ 
        \midrule
        Ours (additive Mat\'{e}rn, $\nu=1.5$)& {softplus-softplus}      & $3.18 \pm 0.98$ & $5.48 \pm 1.77$ && $8.66 \pm 1.16$ & $11.13 \pm 2.00$ &\\ 
        Ours (additive Mat\'{e}rn, $\nu=1.5$)&{sigmoid-sigmoid}      & $3.31 \pm 0.92$ & $7.55 \pm 2.71$ && $9.40 \pm 0.71$ & $14.11 \pm 3.07$ &
\\ 
        Ours (additive Mat\'{e}rn, $\nu=1.5$)&{exp-exp}      & $4.96 \pm 2.46$ & $8.02 \pm 2.29$ && $13.47 \pm 4.73$ & $14.72 \pm 2.18$ &
\\ 
        \midrule        
        Ours (additive Mat\'{e}rn, $\nu=2.5$)& {softplus-softplus}       & $\color{blue}3.14 \pm 1.10$ & $5.28 \pm 1.77$ && $8.44 \pm 1.45$ & $10.83 \pm 1.93$ &\\ 
        Ours (additive Mat\'{e}rn, $\nu=2.5$)&{sigmoid-sigmoid}       & $3.76 \pm 1.51$ & $7.55 \pm 2.93$ && $10.45 \pm 2.06$ & $14.62 \pm 3.41$ &
\\ 
        Ours (additive Mat\'{e}rn, $\nu=2.5$)&{exp-exp}       & $5.11 \pm 2.50$ & $7.36 \pm 2.31$ && $13.82 \pm 4.84$ & $13.97 \pm 2.31$ &
\\ 
        \bottomrule
    \end{tabular}}
    \begin{tabnote}
        Results are averaged over 8 independent realisations. Mean values and standard deviations ($\pm$) are reported. Red and blue indicate the best and second-best performance, respectively. All values are scaled by $10^{2}$.
\end{tabnote}
\end{table}
\clearpage
\subsection{Synthetic Experiment Results for Scenario 2}

\begin{table}[ht!]
    \centering
    \tbl{Comparative analysis for Scenario 2 across various link functions and kernel smoothness specifications.}\label{tab:scenario2_full}
    {   \small\addtolength{\tabcolsep}{-1.1pt}\begin{tabular}{crrr|crrr}
        Method & Link & $PM_{\mathrm{mse}}(\mu)$ & $PM_{\mathrm{mse}}(\phi)$ & 
        & $PE_{\mathrm{mse}}(\mu)$ & $PE_{\mathrm{mse}}(\phi)$ & \\ 
        \midrule
        Parametric Hawkes & {} &$152.64 \pm 3.24$& $16.78 \pm 7.56$ & & $	153.37 \pm 3.26$ & $\color{blue}11.58 \pm 6.34$ & \\ 
        Cox-Hawkes & {}  & $16.08 \pm 4.72$ & $\color{red}4.21 \pm 2.33$ && $30.57 \pm 6.09$ & $\color{red}5.68 \pm 2.27$ & \\ 
        \midrule
        Ours (separable RBF kernel) & {softplus-softplus}   & $22.09 \pm 7.26$ & $13.24 \pm 1.68$ && $37.01 \pm 6.70$ & $18.04 \pm 1.94$ &
\\ 
        Ours (separable RBF kernel) & {sigmoid-sigmoid}    & $19.56 \pm 5.97$ & $11.76 \pm 2.07$ && $36.59 \pm 5.88$ & $19.24 \pm 2.48$ &
\\ 
        Ours (separable RBF kernel) & {exp-exp}   & $22.34 \pm 6.55$ & $11.58 \pm 1.74$ && $38.29 \pm 6.21$ & $17.71 \pm 2.21$ &
\\ 
        \midrule
        Ours (separable Mat\'{e}rn, $\nu=0.5$)& {softplus-softplus}     & $19.79 \pm 6.30$ & $11.59 \pm 1.64$ && $34.26 \pm 5.79$ & $16.96 \pm 1.87$ &
\\ 
        Ours (separable Mat\'{e}rn, $\nu=0.5$)&{sigmoid-sigmoid}     & $19.98 \pm 5.46$ & $15.67 \pm 1.48$ && $37.52 \pm 4.93$ & $22.18 \pm 1.71$ &
\\ 
        Ours (separable Mat\'{e}rn, $\nu=0.5$)&{exp-exp} & $21.98 \pm 6.45$ & $18.26 \pm 1.46$ && $39.91 \pm 6.14$ & $24.89 \pm 1.85$ &
\\ 
        \midrule
        Ours (separable Mat\'{e}rn, $\nu=1.5$)& {softplus-softplus}      & $17.78 \pm 6.43$ & $9.24 \pm 1.43$ && $31.20 \pm 5.73$ & $14.65 \pm 1.87$ &
\\ 
        Ours (separable Mat\'{e}rn, $\nu=1.5$)&{sigmoid-sigmoid}      & $17.27 \pm 5.92$ & $10.46 \pm 1.58$ && $32.90 \pm 5.65$ & $16.91 \pm 1.96$ &
\\ 
        Ours (separable Mat\'{e}rn, $\nu=1.5$)&{exp-exp} & $18.45 \pm 6.14$ & $11.56 \pm 1.37$ && $33.73 \pm 5.64$ & $17.93 \pm 1.79$ &
\\ 
        \midrule
        Ours (separable Mat\'{e}rn, $\nu=2.5$)& {softplus-softplus}       & $18.16 \pm 6.48$ & $9.16 \pm 1.43$ && $31.73 \pm 5.80$ & $14.58 \pm 1.89$ &
\\ 
        Ours (separable Mat\'{e}rn, $\nu=2.5$)&{sigmoid-sigmoid} & $17.28 \pm 5.91$ & $10.14 \pm 1.66$ && $32.90 \pm 5.69$ & $16.91 \pm 2.08$ &
\\ 
        Ours (separable Mat\'{e}rn, $\nu=2.5$)&{exp-exp}       & $18.62 \pm 6.08$ & $10.87 \pm 1.36$ && $33.69 \pm 5.60$ & $17.27 \pm 1.80$ &
\\ 
        \midrule
        Ours (additive RBF kernel) & {softplus-softplus}   & $\color{red}11.14 \pm 6.21$ & $\color{blue}7.67 \pm 1.37$ && $\color{red}20.22 \pm 6.14$ & $13.44 \pm 2.74$ &
\\ 
        Ours (additive RBF kernel) & {sigmoid-sigmoid}    & $\color{blue}11.95 \pm 6.18$ & $11.96 \pm 2.58$ && $22.88 \pm 6.12$ & $20.20 \pm 3.56$ &
\\ 
        Ours (additive RBF kernel) & {exp-exp}    & $14.12 \pm 5.72$ & $9.87 \pm 1.91$ && $25.52 \pm 5.75$ & $17.37 \pm 3.16$ &\\ 
        \midrule
        Ours (additive Mat\'{e}rn, $\nu=0.5$)& {softplus-softplus}     & $16.37 \pm 6.12$ & $10.32 \pm 1.83$ && $27.42 \pm 6.18$ & $16.21 \pm 2.02$ &
\\ 
        Ours (additive Mat\'{e}rn, $\nu=0.5$)&{sigmoid-sigmoid}     & $17.17 \pm 5.52$ & $14.42 \pm 1.56$ && $28.87 \pm 5.77$ & $21.37 \pm 1.89$ &
\\ 
        Ours (additive Mat\'{e}rn, $\nu=0.5$)&{exp-exp}     & $19.66 \pm 6.45$ & $16.07 \pm 1.80$ && $31.81 \pm 7.46$ & $23.69 \pm 2.06$ &
\\ 
        \midrule
        Ours (additive Mat\'{e}rn, $\nu=1.5$)& {softplus-softplus}      & $12.94 \pm 5.98$ & $8.36 \pm 1.58$ && $22.89 \pm 6.09$ & $14.80 \pm 2.12$ &
\\ 
        Ours (additive Mat\'{e}rn, $\nu=1.5$)&{sigmoid-sigmoid} & $13.36 \pm 5.71$ & $10.10 \pm 1.37$ && $24.23 \pm 5.64$ & $16.82 \pm 1.74$ &
\\ 
        Ours (additive Mat\'{e}rn, $\nu=1.5$)&{exp-exp} & $15.03 \pm 6.31$ & $10.90 \pm 1.97$ && $26.03 \pm 6.49$ & $18.42 \pm 2.22$ &
\\ 
        \midrule
        Ours (additive Mat\'{e}rn, $\nu=2.5$)& {softplus-softplus}       & $11.81 \pm 6.14$ & $8.08 \pm 1.26$ && $\color{blue}21.00 \pm 5.86$ & $14.48 \pm 1.71$ &
\\ 
        Ours (additive Mat\'{e}rn, $\nu=2.5$)&{sigmoid-sigmoid}       & $12.38 \pm 5.80$ & $10.02 \pm 1.85$ && $23.27 \pm 5.79$ & $16.98 \pm 2.55$ &
\\ 
        Ours (additive Mat\'{e}rn, $\nu=2.5$)&{exp-exp} & $14.43 \pm 6.22$ & $10.50 \pm 1.97$ && $25.50 \pm 6.44$ & $18.26 \pm 2.01$ &
\\ 
        \bottomrule
    \end{tabular}}
    \begin{tabnote}
Results are averaged over 8 independent realisations. Mean values and standard deviations ($\pm$) are reported. Red and blue indicate the best and second-best performance, respectively. All values are scaled by $10^{2}$.
\end{tabnote}
\end{table}
\clearpage
\subsection{Synthetic Experiment Results for Scenario 3}

\begin{table}[ht!]
\centering
\label{tab:scenario3_full}
\tbl{Comparative analysis for Scenario 3 across various link functions and kernel smoothness specifications.}
{   \small\addtolength{\tabcolsep}{-1.1pt}\begin{tabular}{crrr|crrr}
        Method & Link & $PM_{\mathrm{mse}}(\mu)$ & $PM_{\mathrm{mse}}(\phi)$ & 
        & $PE_{\mathrm{mse}}(\mu)$ & $PE_{\mathrm{mse}}(\phi)$ & \\ 
        \midrule
        Parametric Hawkes & {} & $169.55 \pm  5.35$ & $32.50 \pm  4.42$ & & $170.35 \pm  5.31$ & $31.11 \pm  3.81$ & \\ 
        Cox-Hawkes & {} & $24.39 \pm 5.60$ & $24.97 \pm 1.46$ && $47.32 \pm 5.61$ & $26.02 \pm 1.44$ &\\ 
        \midrule
Ours (separable RBF) & {softplus-softplus}    & $22.92 \pm 3.97$ & $10.72 \pm 5.46$ && $41.49 \pm 3.82$ & $18.92 \pm 4.71$ &
\\ 
        Ours (separable RBF) & {sigmoid-sigmoid}     & $25.24 \pm 5.24$ & $11.61 \pm 5.07$ && $47.67 \pm 4.93$ & $20.31 \pm 4.94$ &
\\ 
        Ours (separable RBF) & {exp-exp} & $28.08 \pm 5.25$ & $12.41 \pm 5.29$ && $48.23 \pm 5.34$ & $20.75 \pm 4.68$ &
\\ 
        \midrule
Ours (separable Mat\'{e}rn, $\nu=0.5$)& {softplus-softplus} & $22.58 \pm 3.49$ & $12.14 \pm 4.82$ && $40.55 \pm 3.42$ & $19.52 \pm 4.33$ &
\\ 
        Ours (separable Mat\'{e}rn, $\nu=0.5$)&{sigmoid-sigmoid}      & $25.75 \pm 5.40$ & $12.94 \pm 4.47$ && $47.83 \pm 5.75$ & $20.81 \pm 4.28$ &
\\ 
        Ours (separable Mat\'{e}rn, $\nu=0.5$)&{exp-exp}  & $28.26 \pm 4.94$ & $13.74 \pm 4.25$ && $51.09 \pm 5.11$ & $21.46 \pm 3.82$ &
\\ 
        \midrule
        Ours (separable Mat\'{e}rn, $\nu=1.5$)& {softplus-softplus} & $20.54 \pm 3.79$ & $10.53 \pm 4.54$ && $37.82 \pm 3.76$ & $18.83 \pm 3.93$ &
\\ 
        Ours (separable Mat\'{e}rn, $\nu=1.5$)&{sigmoid-sigmoid} & $23.04 \pm 5.37$ & $10.89 \pm 5.17$ && $43.86 \pm 5.32$ & $18.43 \pm 4.79$ &
\\ 
        Ours (separable Mat\'{e}rn, $\nu=1.5$)&{exp-exp} & $24.47 \pm 5.27$ & $11.70 \pm 4.43$ && $44.36 \pm 5.34$ & $19.91 \pm 3.98$ &
\\ 
        \midrule
        Ours (separable Mat\'{e}rn, $\nu=2.5$)& {softplus-softplus}      & $20.75 \pm 3.86$ & $10.23 \pm 4.39$ && $38.07 \pm 3.80$ & $18.74 \pm 3.77$ &
\\ 
        Ours (separable Mat\'{e}rn, $\nu=2.5$)&{sigmoid-sigmoid}      & $23.00 \pm 5.34$ & $10.57 \pm 5.20$ && $43.84 \pm 5.06$ & $18.39 \pm 4.84$ &
\\ 
        Ours (separable Mat\'{e}rn, $\nu=2.5$)&{exp-exp}      & $24.79 \pm 5.32$ & $11.75 \pm 4.36$ && $44.17 \pm 5.43$ & $20.18 \pm 3.93$ &
\\ 
        \midrule
Ours (additive RBF) & {softplus-softplus}   & $19.17 \pm 5.28$ & $10.14 \pm 4.27$ && $\color{red}37.11 \pm 4.42$ & $18.54 \pm 4.23$ &\\ 
        Ours (additive RBF) & {sigmoid-sigmoid}    & $25.50 \pm 4.14$ & $\color{blue}9.31 \pm 4.61$ && $40.57 \pm 8.14$ & $\color{red}17.49 \pm 4.42$ &\\ 
        Ours (additive RBF) & {exp-exp}    & $24.88 \pm 6.60$ & $\color{red}9.26 \pm 4.24$ && $43.98 \pm 6.32$ & $\color{blue}17.84 \pm 3.96$ &
\\ 
        \midrule
Ours (additive Mat\'{e}rn, $\nu=0.5$)& {softplus-softplus}     & $21.08 \pm 3.50$ & $12.18 \pm 5.48$ && $38.85 \pm 3.60$ & $20.14 \pm 5.12$ &
\\ 
        Ours (additive Mat\'{e}rn, $\nu=0.5$)&{sigmoid-sigmoid}     & $26.62 \pm 6.17$ & $12.38 \pm 5.21$ && $49.02 \pm 6.70$ & $20.51 \pm 4.67$ &
\\ 
        Ours (additive Mat\'{e}rn, $\nu=0.5$)&{exp-exp} & $29.36 \pm 5.89$ & $13.31 \pm 4.60$ && $52.82 \pm 6.02$ & $21.00 \pm 4.14$ &
\\ 
        \midrule
        Ours (additive Mat\'{e}rn, $\nu=1.5$)& {softplus-softplus}      & $\color{blue}18.58 \pm 3.51$ & $11.06 \pm 5.78$ && $\color{blue}37.57 \pm 3.68$ & $18.80 \pm 5.36$ &
\\ 
        Ours (additive Mat\'{e}rn, $\nu=1.5$)&{sigmoid-sigmoid}      & $26.55 \pm 4.19$ & $10.21 \pm 4.82$ && $42.19 \pm 8.40$ & $18.16 \pm 4.60$ &
\\ 
        Ours (additive Mat\'{e}rn, $\nu=1.5$)&{exp-exp} & $25.71 \pm 6.77$ & $11.05 \pm 5.25$ && $46.25 \pm 7.67$ & $18.87 \pm 4.93$ &
\\ 
        \midrule
        Ours (additive Mat\'{e}rn, $\nu=2.5$)& {softplus-softplus}      & $\color{red}18.27 \pm 3.56$ & $10.91 \pm 5.89$ && $37.76 \pm 3.77$ & $18.75 \pm 5.45$ &
\\ 
        Ours (additive Mat\'{e}rn, $\nu=2.5$)&{sigmoid-sigmoid}       & $26.22 \pm 4.27$ & $9.70 \pm 4.78$ && $41.47 \pm 8.80$ & $17.94 \pm 4.50$ &
\\ 
        Ours (additive Mat\'{e}rn, $\nu=2.5$)&{exp-exp}       & $25.32 \pm 6.85$ & $10.50 \pm 4.39$ && $45.91 \pm 7.83$ & $19.47 \pm 4.47$ &
\\ 
        \bottomrule
    \end{tabular}}
    \begin{tabnote}
Results are averaged over 8 independent realisations. Mean values and standard deviations ($\pm$) are reported. Red and blue indicate the best and second-best performance, respectively. All values are scaled by $10^{2}$.
\end{tabnote}
\end{table}

\clearpage

\subsection{Detailed Results for Real-world Data} \label{app:real_world_full}

\begin{table}[htbp]
    \centering
    \tbl{Performance comparison (Expected Log-Likelihood) on real-world datasets.}
    {\small\addtolength{\tabcolsep}{-1.1pt}
    \begin{tabular}{llc @{\hspace{2em}} lc} 
        \toprule
        & \multicolumn{2}{c}{\textbf{Chicago Shootings}} & \multicolumn{2}{c}{\textbf{Vancouver Break-ins}} \\
        \cmidrule(lr){2-3} \cmidrule(lr){4-5}
        Method & Link & $ELL$ & Link & $ELL$ \\ 
        \midrule
        Parametric Hawkes & --- & $1471.17$ & --- & $2203.65$ \\ 
        Log Gaussian Cox Process& exp & $1535.49$ & exp & $2000.43$ \\ 
        Cox-Hawkes & exp & $1647.26$ & exp & $2157.32$ \\ 
        \midrule
        Ours (separable RBF) & softplus-softplus & $1615.71$ & softplus-softplus & $2209.45$ \\ 
        Ours (separable RBF) & sigmoid-softplus & $1642.24$ & sigmoid-softplus & $2198.64$ \\ 
        Ours (separable RBF) & exp-softplus & $1643.36$ & exp-softplus & $2188.72$ \\ 
        \midrule
        Ours (separable Mat\'{e}rn, $\nu=0.5$) & softplus-softplus & $1626.44$ & softplus-softplus & $2233.63$ \\ 
        Ours (separable Mat\'{e}rn, $\nu=0.5$) & sigmoid-softplus & $1639.41$ & sigmoid-softplus & $2217.16$ \\
        Ours (separable Mat\'{e}rn, $\nu=0.5$) & exp-softplus & $1645.53$ & exp-softplus & $2203.97$ \\
        \midrule
        Ours (separable Mat\'{e}rn, $\nu=1.5$) & softplus-softplus & $1624.10$ & softplus-softplus & $2226.73$ \\ 
        Ours (separable Mat\'{e}rn, $\nu=1.5$) & sigmoid-softplus & $1648.11$ & sigmoid-softplus & $2212.76$ \\
        Ours (separable Mat\'{e}rn, $\nu=1.5$) & exp-softplus & $1652.56$ & exp-exp & $2197.42$ \\ 
        \midrule
        Ours (separable Mat\'{e}rn, $\nu=2.5$) & softplus-softplus & $1622.09$ & softplus-softplus & $2220.85$ \\ 
        Ours (separable Mat\'{e}rn, $\nu=2.5$) & sigmoid-softplus & $1645.43$ & sigmoid-softplus & $2205.70$ \\
        Ours (separable Mat\'{e}rn, $\nu=2.5$) & exp-softplus & $1651.15$ & exp-softplus & $2193.20$ \\
        \midrule
        Ours (additive RBF) & softplus-softplus & $1616.90$ & softplus-softplus & $\color{red}2249.02$ \\ 
        Ours (additive RBF) & sigmoid-softplus & $1656.78$ & sigmoid-softplus & $2245.78$ \\ 
        Ours (additive RBF) & exp-softplus & $\color{red}1670.68$ & exp-softplus & $2227.90$ \\ 
        \midrule
        Ours (additive Mat\'{e}rn, $\nu=0.5$) & softplus-softplus & $1623.12$ & softplus-softplus & $2239.53$ \\ 
        Ours (additive Mat\'{e}rn, $\nu=0.5$) & sigmoid-softplus & $1642.45$ & sigmoid-softplus & $2221.36$ \\
        Ours (additive Mat\'{e}rn, $\nu=0.5$) & exp-softplus & $1653.71$ & exp-softplus & $2207.75$ \\
        \midrule
        Ours (additive Mat\'{e}rn, $\nu=1.5$) & softplus-softplus & $1623.46$ & softplus-softplus & $2244.67$ \\ 
        Ours (additive Mat\'{e}rn, $\nu=1.5$) & sigmoid-softplus & $1655.61$ & sigmoid-softplus & $2238.62$ \\
        Ours (additive Mat\'{e}rn, $\nu=1.5$) & exp-softplus & $1669.86$ & exp-softplus & $2214.18$ \\ 
        \midrule
        Ours (additive Mat\'{e}rn, $\nu=2.5$) & softplus-softplus & $1621.99$ & softplus-softplus & $\color{blue}2247.74$ \\ 
        Ours (additive Mat\'{e}rn, $\nu=2.5$) & sigmoid-softplus & $1655.94$ & sigmoid-softplus & $2241.56$ \\
        Ours (additive Mat\'{e}rn, $\nu=2.5$) & exp-softplus & $\color{blue}1670.64$ & exp-softplus & $2215.98$ \\
        \bottomrule
    \end{tabular}}
    \label{tab:real_world_full_}
    \begin{tabnote}
        Higher $ELL$ indicates superior performance. Best and second-best results are highlighted in red and blue, respectively.
    \end{tabnote}
\end{table}

\clearpage
\subsection{Estimated Triggering Kernels for Real-World Datasets}
\label{app:triggering_kernel_real_world}
\begin{figure}[h!]
\centering
\includegraphics[width=1.0\textwidth]{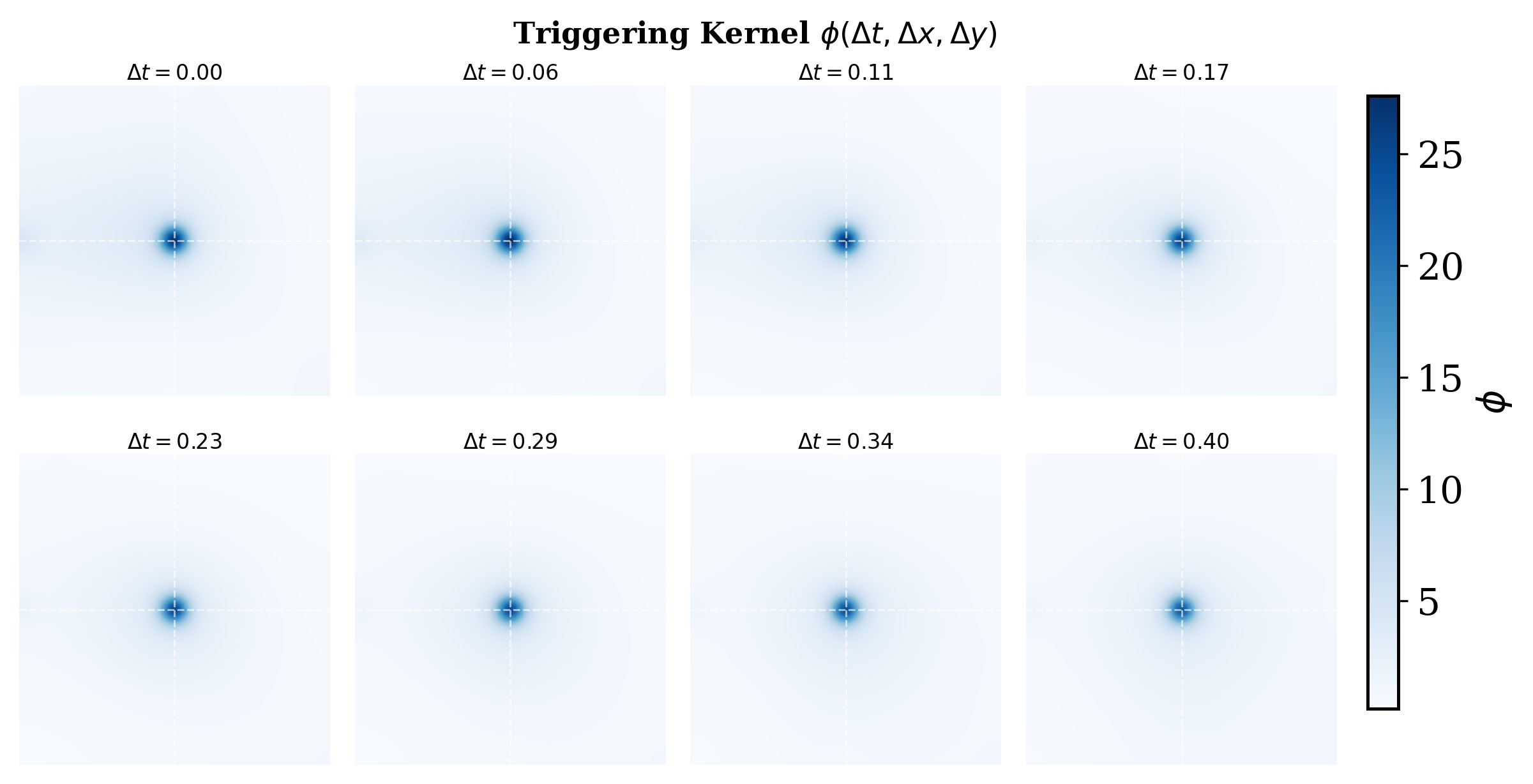}
\caption{Variational posterior mean on the self-excitation triggering kernel at selected timestamps for the Chicago shootings dataset.}
\label{fig:chicago_phi}
\end{figure}
\begin{figure}[h!]
\centering
\includegraphics[width=1.0\textwidth]{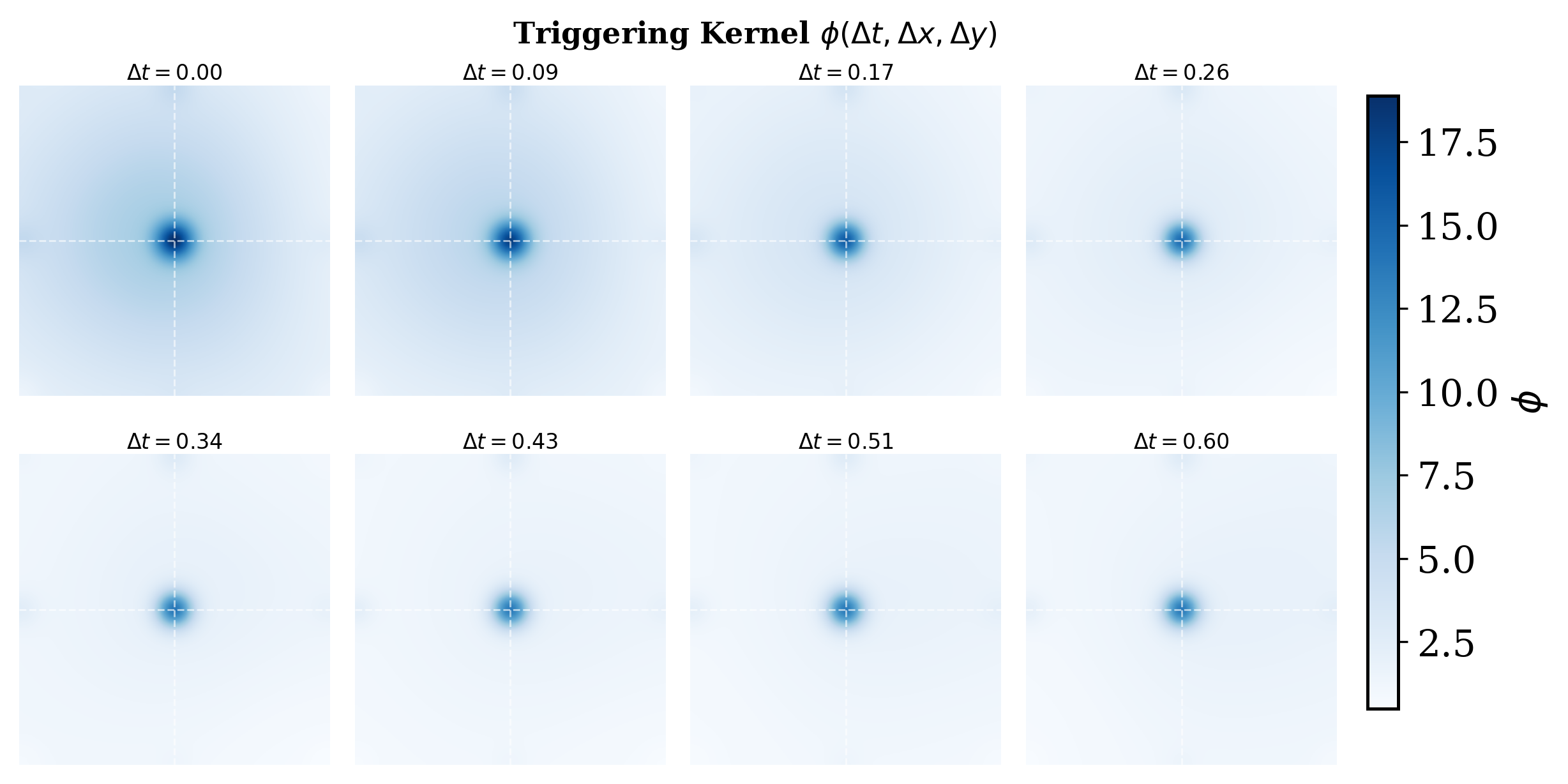}
\caption{Variational Posterior mean on the  self-excitation triggering kernel at selected timestamps  for the Vancouver break-ins dataset %, capturing how a break-in elevates the immediate risk in the surrounding neighbourhood
.}
\label{fig:vancouver_phi}
\end{figure}
\clearpage

\section{Sensitivity Analysis of the Triggering Kernel Window}
\label{app:sensitivity_window}
In this section, we study how sensitive our nonparametric framework is to the choice of the finite support window $(T_\phi, X_\phi, Y_\phi)$ used for the triggering kernel. 
Choosing this window involves a trade-off. If the window is too small, part of the true self-exciting effect may be truncated, and some contagion information will be lost. 
If the window is too large, distant events that should belong to the background may be incorrectly included in the triggering component, and the computational cost also increases.

To examine the robustness of the model, we vary the temporal range 
$T_\phi \in \{0.4, 0.6, 0.8\}$ 
and the spatial ranges 
$X_\phi = Y_\phi \in \{0.2, 0.4, 0.6, 0.8, 1.0\}$. 
The $ELL$ values for our proposed model, as well as for the Parametric Hawkes and Cox-Hawkes baselines, are reported in Tables~\ref{tab:window_ell} and~\ref{tab:window_ell_vancouver}.

The results show different behaviors in the spatial and temporal dimensions. 
For the spatial range, the $ELL$ follows a clear unimodal pattern: performance improves when the window increases from small to moderate size, but decreases when the window becomes too large. 
This suggests that large spatial windows include too much distant background noise, which weakens the local triggering signal. 
For the temporal range, the $ELL$ slightly increases as $T_\phi$ becomes larger, but this improvement is small and likely comes from increased model flexibility rather than a better description of the true crime dynamics. 
Using a very long temporal window may also lead to overfitting and higher computational cost.

Overall, a moderate window size provides a good balance between capturing the main triggering effects and keeping the model simple. 
More importantly, under most window settings, our nonparametric model
achieves higher $ELL$ than the parametric baselines. This overall
pattern across different window sizes suggests that the proposed
nonparametric framework is generally more robust than the parametric
and semi-parametric alternatives.

\begin{table}[htbp]
\centering
\caption{Sensitivity analysis for different triggering kernel windows on the Chicago Shootings dataset. We report $ELL$ under various temporal and spatial truncation thresholds ($T_\phi, X_\phi, Y_\phi$). The specific link function used for our proposed model is indicated for each configuration.}
\label{tab:window_ell}

\begin{tabular}{ccc @{\hspace{2em}} cccc}
\toprule
\multicolumn{3}{c}{\textbf{Window Size}} & \multicolumn{4}{c}{\textbf{Expected Log-Likelihood ($ELL$)}} \\
\cmidrule(lr){1-3} \cmidrule(lr){4-7}
$T_\phi$ & $X_\phi$ & $Y_\phi$ & Parametric & Cox-Hawkes & Link (Ours) & Ours \\
\midrule

\multirow{5}{*}{0.4} 
& 0.2 & 0.2 & 1280.65 & 1644.08 & exp-softplus & 1637.28 \\
& 0.4 & 0.4 & 1472.31 & 1659.15 & exp-softplus & 1660.30 \\
& 0.6 & 0.6 & 1535.65 & 1670.45 & exp-softplus & 1675.61 \\
& 0.8 & 0.8 & 1540.65 & 1671.34 & exp-softplus& 1674.27 \\
& 1.0 & 1.0 & 1539.09 & 1671.37 & exp-softplus & 1673.08 \\

\midrule

\multirow{5}{*}{0.6} 
& 0.2 & 0.2 & 1375.93 & 1673.33 & exp-softplus& 1679.37 \\
& 0.4 & 0.4 & 1569.82 & 1686.78 & exp-softplus& 1693.48 \\
& 0.6 & 0.6 & 1603.39 & 1699.08 & exp-softplus & 1723.94 \\
& 0.8 & 0.8 & 1604.72 & 1699.77 & exp-softplus& 1719.28 \\
& 1.0 & 1.0 & 1603.71 & 1699.76 & exp-softplus& 1712.97 \\

\midrule

\multirow{5}{*}{0.8} 
& 0.2 & 0.2 & 1463.18 & 1708.16 & exp-softplus & 1720.70 \\
& 0.4 & 0.4 & 1621.08 & 1717.60 & exp-softplus & 1735.83 \\
& 0.6 & 0.6 & 1643.70 & 1728.81 & exp-softplus & 1761.38 \\
& 0.8 & 0.8 & 1643.95 & 1729.21 & exp-softplus& 1757.17 \\
& 1.0 & 1.0 & 1643.63 & 1729.20 & exp-softplus& 1751.01 \\

\bottomrule
\end{tabular}
\end{table}

\begin{table}[htbp]
\centering
\caption{Sensitivity analysis for different triggering kernel windows on the Vancouver Break-ins dataset. We report $ELL$ under various temporal and spatial truncation thresholds ($T_\phi, X_\phi, Y_\phi$). The specific link function used for our proposed model is indicated for each configuration.}
\label{tab:window_ell_vancouver}

\begin{tabular}{ccc @{\hspace{2em}} cccc}
\toprule
\multicolumn{3}{c}{\textbf{Window Size}} & \multicolumn{4}{c}{\textbf{Expected Log-Likelihood ($ELL$)}} \\
\cmidrule(lr){1-3} \cmidrule(lr){4-7}
$T_\phi$ & $X_\phi$ & $Y_\phi$ & Parametric & Cox-Hawkes & Link (Ours) & Ours \\
\midrule

\multirow{5}{*}{0.4} 
& 0.2 & 0.2 & 2189.07 & 2151.20 & exp-softplus & 2178.71 \\
& 0.4 & 0.4 & 2195.06 & 2157.38 & exp-exp & 2280.83 \\
& 0.6 & 0.6 & 2193.22 & 2149.07 & exp-exp & 2249.51 \\
& 0.8 & 0.8 & 2193.15 & 2147.06 & exp-exp & 2222.47 \\
& 1.0 & 1.0 & 2193.16 & 2146.55 & exp-exp & 2206.10 \\

\midrule

\multirow{5}{*}{0.6} 
& 0.2 & 0.2 & 2200.49 & 2168.45 & exp-softplus & 2224.25\\
& 0.4 & 0.4 & 2208.15 & 2180.96 & exp-exp & 2314.55 \\
& 0.6 & 0.6 & 2208.02 & 2170.62 & exp-exp & 2283.94 \\
& 0.8 & 0.8 & 2208.09 & 2168.49 & exp-exp & 2251.14 \\
& 1.0 & 1.0 & 2208.11 & 2169.26 & exp-exp & 2229.18 \\

\midrule

\multirow{5}{*}{0.8} 
& 0.2 & 0.2 & 2215.27 & 2200.45 & exp-softplus & 2264.96 \\
& 0.4 & 0.4 & 2227.84 & 2216.78 & exp-exp & 2350.82 \\
& 0.6 & 0.6 & 2228.86 & 2196.50 & exp-exp & 2311.60 \\
& 0.8 & 0.8 & 2228.95 & 2194.79 & exp-exp & 2255.38 \\
& 1.0 & 1.0 & 2228.97 & 2192.95 & exp-exp & 2263.91 \\

\bottomrule
\end{tabular}
\end{table}

\clearpage

\section{Model Diagnostics}
\label{app:diagnostics}

\subsection{Super-Thinning Algorithm}
\label{app:super_thinning}
To assess the goodness-of-fit of our proposed model, we employ the super-thinning algorithm \citep{clements2012evaluation}. This method evaluates a fitted spatio-temporal point process by transforming the observed events into a residual process that, if the model is correctly specified, is distributed as a homogeneous Poisson process with rate $k$.

Let $\hat{\lambda}(t,x,y)$ be the estimated conditional intensity from our model. We choose the target rate $k$ such that $\inf \hat{\lambda} \leq k \leq \sup \hat{\lambda}$. The procedure involves two steps:

\begin{enumerate}
    \item \textbf{Thinning:} Each observed event $i$ at location $(x_i,y_i)$ and time $t_i$ in the original dataset is retained with probability
    $$ p_i = \min\left\{ \frac{k}{\hat{\lambda}(t_i,x_i,y_i)}, 1 \right\}. $$
    \item \textbf{Superposition:} We simulate a new inhomogeneous Poisson process with rate $\lambda_{\text{sim}}(t, x,y) = \max\{k - \hat{\lambda}(t, x,y), 0\}$, and add these simulated points to the thinned observed process.
\end{enumerate}

If the estimated intensity $\hat{\lambda}(t, x,y)$ accurately reflects the true underlying data-generating mechanism, the resulting combined process is equivalent to a homogeneous Poisson process with a constant rate $k$. 

\subsection{Statistical Goodness-of-Fit Tests}
\label{app:statistical_tests}

We evaluate the homogeneity of the super-thinned process by examining its spatial and its temporal distributions. Let $N_{thinned}$ be the total number of events in the resulting super-thinned process.

\medskip
\noindent \textbf{Temporal Analysis (Inter-arrival Time Test)} \\
To assess the temporal fit, we analyze the inter-arrival times $\Delta t_i = t_{(i)} - t_{(i-1)}$ of the sorted events. For a homogeneous Poisson process, these intervals follow an exponential distribution with rate $\Lambda = N_{thinned}/T_{max}$. We perform a Kolmogorov-Smirnov (KS) test against the cumulative distribution function:
\begin{equation}
    F(\Delta t) = 1 - \exp(-\Lambda \Delta t)
\end{equation}
A non-significant $p$-value ($p > 0.05$) indicates that the model has successfully captured the temporal triggering effects.

\medskip
\noindent \textbf{Spatial Analysis (Quadrat Count Test)} \\
To assess spatial fit, the domain $X \times Y$ is partitioned into an $n_{grid} \times n_{grid}$ regular grid. The expected count per quadrat is denoted by $E = N_{thinned}/n_{grid}^2$. We then perform a $\chi^2$ test:
\begin{equation}
    \chi^2 = \sum_{j=1}^{n_{grid}^2} \frac{(O_j - E)^2}{E}
\end{equation}
where $O_j$ is the observed count in the $j$-th quadrat. A significant result ($p < 0.05$) indicates residual spatial clustering not captured by the model.
\clearpage

\bibliographystyle{plainnat}
\bibliography{paper-ref}

\end{document}